\documentclass[aps,epsfigure,twocolumn,superscriptaddress,showkeys,nofootinbib]{revtex4-1}
\usepackage{url}

\usepackage[colorlinks=true,linkcolor=blue,urlcolor=blue,citecolor=blue,pdfusetitle]{hyperref}
\usepackage[utf8]{inputenc}
\usepackage[english]{babel}
\usepackage{amsmath}
\usepackage[caption = false]{subfig}
\usepackage{graphicx,epstopdf}
\usepackage{blindtext}
\usepackage[table,xcdraw]{xcolor}
\usepackage{lipsum}
\usepackage{amsfonts}
\usepackage{bbm}
\usepackage{amssymb}
\usepackage{enumerate}
\usepackage{comment}
\usepackage{color}
\usepackage{latexsym}
\usepackage{multirow}
\usepackage{threeparttable}
\usepackage{hyperref}
\usepackage{tabularx}
\usepackage{array}
\usepackage{float}

\usepackage{times,txfonts}
\usepackage[normalem]{ulem}
\graphicspath{ {./figures/} }
\usepackage{subcaption}
\usepackage{algorithm}
\usepackage{algpseudocode}
\usepackage{tikz}
\usetikzlibrary{quantikz}
\usepackage{array}
\usepackage{caption}
\usepackage{booktabs}
\usepackage{ragged2e}
\usepackage{comment}
\usepackage{orcidlink}
\usepackage{enumitem}

\newtheorem{definition}{Definition}

\usepackage[T1]{fontenc}
\usepackage{lmodern}  
\usepackage{titlesec}
\usepackage{titletoc}
\usepackage{hyperref}
\usepackage{xcolor}

\titleformat{\section}
  {\normalfont\Large\bfseries\sffamily}
  {\thesection}{1em}{}

\titleformat{\subsection}
  {\normalfont\large\bfseries\sffamily}
  {\thesubsection}{1em}{}

\titlecontents{section}
  [1.8em]
  {\vspace{0.5\baselineskip}\bfseries\sffamily}
  {\contentslabel{2em}}
  {\hspace*{-2em}}
  {\titlerule*[0.5pc]{.}\contentspage}

\titlecontents{subsection}
  [4.5em]
  {\vspace{0.2\baselineskip}\normalfont\sffamily}
  {\contentslabel{2.0em}}
  {\hspace*{-2.5em}}
  {\titlerule*[0.5pc]{.}\contentspage}

\titlecontents{subsubsection}
  [6.5em]
  {\vspace{0.1\baselineskip}\normalfont\sffamily\small}
  {\contentslabel{2em}}
  {\hspace*{-3em}}
  {\titlerule*[0.5pc]{.}\contentspage}

\usepackage[most]{tcolorbox}
\tcbuselibrary{theorems,skins}

\newtcbtheorem[auto counter]{example}{Example}{%
  enhanced,
  colback=blue!4!white,
  colframe=blue!75!black,
  fonttitle=\bfseries,
  coltitle=white,
  colbacktitle=blue!75!black,
  sharp corners,
  boxrule=0.8pt,
  title style={left color=blue!100!black,right color=blue!80!black},
  separator sign=:\ %
}{ex}

\begin{document}

\title{Quantum Random Number Generators: principles, practical challenges and mainstream implementations}

\author{Lucas Q. Galvão \orcidlink{0009-0001-2682-250X}}
\email{Corresponding author: Lucas Q. Galvão \newline Email: lucas.queiroz@fieb.org.br | lqgalvao3@gmail.com}
\affiliation{Universidade SENAI CIMATEC}
\affiliation{QuIIN - Quantum Industrial Innovation, Centro de Competência Embrapii Cimatec. SENAI CIMATEC, Av. Orlando Gomes, 1845, Salvador, BA, Brazil CEP 41850-010}

\author{Gubio G. de Lima \orcidlink{https://orcid.org/0000-0002-4551-3251}}
\affiliation{QuIIN - Quantum Industrial Innovation, Centro de Competência Embrapii Cimatec. SENAI CIMATEC, Av. Orlando Gomes, 1845, Salvador, BA, Brazil CEP 41850-010}

\author{Rubicely Francisco do Nascimento \orcidlink{?}}
\affiliation{QuIIN - Quantum Industrial Innovation, Centro de Competência Embrapii Cimatec. SENAI CIMATEC, Av. Orlando Gomes, 1845, Salvador, BA, Brazil CEP 41850-010}

\author{José Batista de Souza Júnior \orcidlink{?}}
\affiliation{Universidade SENAI CIMATEC}
\affiliation{QuIIN - Quantum Industrial Innovation, Centro de Competência Embrapii Cimatec. SENAI CIMATEC, Av. Orlando Gomes, 1845, Salvador, BA, Brazil CEP 41850-010}

\author{Davi Juvêncio G. de Sousa\orcidlink{0009-0009-1946-5573}}
\affiliation{QuIIN - Quantum Industrial Innovation, Centro de Competência Embrapii Cimatec. SENAI CIMATEC, Av. Orlando Gomes, 1845, Salvador, BA, Brazil CEP 41850-010}

\author{Hugo Santos da Costa Bessa\orcidlink{0009-0003-9388-9995}}
\affiliation{QuIIN - Quantum Industrial Innovation, Centro de Competência Embrapii Cimatec. SENAI CIMATEC, Av. Orlando Gomes, 1845, Salvador, BA, Brazil CEP 41850-010}

\author{Vitor L. Sena \orcidlink{0000-0002-8787-2009}}
\affiliation{QuIIN - Quantum Industrial Innovation, Centro de Competência Embrapii Cimatec. SENAI CIMATEC, Av. Orlando Gomes, 1845, Salvador, BA, Brazil CEP 41850-010}

\author{Nelson Alves Ferreira Neto \orcidlink{0000-0003-2278-1082}}
\affiliation{QuIIN - Quantum Industrial Innovation, Centro de Competência Embrapii Cimatec. SENAI CIMATEC, Av. Orlando Gomes, 1845, Salvador, BA, Brazil CEP 41850-010}

\author{Valéria Loureiro da Silva \orcidlink{0000-0001-5466-7933}}
\affiliation{Universidade SENAI CIMATEC}
\affiliation{QuIIN - Quantum Industrial Innovation, Centro de Competência Embrapii Cimatec. SENAI CIMATEC, Av. Orlando Gomes, 1845, Salvador, BA, Brazil CEP 41850-010}

\begin{abstract}

\noindent  Quantum random number generators have become a cornerstone technology for modern digital cryptography, including key generation, quantum key distribution, and post-quantum cryptographic systems. While existing reviews typically concentrate on specific entropy sources or implementation platforms, a comprehensive perspective encompassing the complete QRNG pipeline remains largely absent. This work addresses this gap by providing a unified review of the theoretical foundations, practical methodologies, and technological developments that underpin modern QRNGs. We examine randomness from both classical and quantum perspectives and survey the principal discrete-variable and continuous-variable entropy sources, together with entropy estimation, randomness certification, extraction techniques, randomness enhancement protocols, and statistical validation methods. Beyond the fundamental components, we review representative implementations reported in the literature, comparing performance metrics, architectural trade-offs, integration strategies, and attainable generation rates across different technological platforms. Practical deployment challenges, including scalability, trust assumptions, hardware integration, and cryptographic adoption, are also discussed. By complementing the discussion with a broad survey of state-of-the-art implementations and applications, this review provides researchers with a comprehensive reference on the current landscape, open challenges, and future directions of quantum random number generation.

\end{abstract}

\maketitle

\tableofcontents

\section{Introduction}

Random numbers have become an important resource in modern science and technology \cite{Hayes2001}. The security of communications, validation of simulations or fundamental science concepts critically depends on the availability of high-quality randomness. Indeed, randomness is present from the initial stages of applications such as quantum communications \cite{RevModPhys.74.145, Usenko2026}, post-quantum cryptography \cite{Bernstein2017, chen2016report}  and Monte-Carlo simulations \cite{testealeat, Metropolis01091949}. At present, its importance has gained so much prominence that Entropy-as-a-Service (EaaS) has recently been proposed as a novel Internet architecture \cite{Apostol2016} designed to provide on-demand access to high-quality entropy.

In the past few decades, pseudorandom number generators (PRNGs) have been used to generate random numbers due to their simplicity \cite{JAMES1990329, BHATTACHARJEE2022100471, goldreich1999modern}. Cryptographically secure pseudorandom number generators (CSPRNG) are preferable because they generate a bit set that approximates a uniform distribution, passing statistical security tests such as NIST and Diehard  \cite{math11234812, rock2005pseudorandom, luby1996pseudorandomness}. However, their generation process is provided by a deterministic algorithm, resulting in weak randomness valid only if an adversary, commonly called Eve, does not have access to its seed.  In contrast, quantum random number generators (QRNG) are devices based on the impossibility of obtaining deterministic results from quantum processes \cite{Bera_2017}. Considering that {QRNG} are based on a fundamental imposition of nature, their adoption is preferable for guaranteeing unconditional security. 

The construction of QRNG devices entails the measurement of a quantum process through dedicated entropy source implementations. Most QRNGs operate with accessible optical devices \cite{RevModPhys.89.015004}, which makes them particularly attractive for practical deployment. The challenge, however, lies in extracting the amount of pure quantum entropy, since the measurement apparatus inevitably introduces noise of classical nature \cite{frauchiger2013truerandomnessrealisticquantum}. The development of post-processing methods for randomness extraction makes it possible to distill high-quality random bits from raw data while suppressing any classical bias or correlations introduced \cite{PhysRevA.87.062327}. This combination of strategies has rendered QRNG the most mature technology in the second quantum revolution scenario \cite{Dowling2003}. In fact, commercial applications are already widespread \cite{idquantique_quantis_chips, cryptalabs_qrng, quside_ruby_s1}, with applications ranging from secure communications \cite{Martin2015, Crampton2025} to consumer electronics \cite{Vinay2024}. 

In this review, we survey the substantial body of literature on QRNG. Our main goal is to provide a comprehensive overview that facilitates the identification of the most promising directions, taking into account principles, mainstream implementations and fundamental technical challenges. Previous reviews have addressed specific aspects of QRNGs. Entropy sources with a focus on optical devices are discussed in the reviews \cite{RevModPhys.89.015004, ma2016quantum, mannalatha2023review}, which also include a brief overview of post-processing. Randomness certification techniques are discussed in refs. \cite{Pironio2010, PRXQuantum.2.010201, Acin2016}. A comparison between techniques of post-processing is presented in \cite{PhysRevA.87.062327}. Our review covers both practical challenges and systematic performance benchmarking in QRNG, presenting principles, modern trends, and recent developments.

We start our review in Section \ref{sec:Randomness} with a brief overview of randomness generation across a wide range of scientific and technological domains. In Section \ref{sec:entropy_sources}, we discuss the entropy sources classified as discrete-variable (DV) or continuous-variable (CV) sources. We then discuss the different trusted levels for entropy estimation (Section \ref{sec:Est_Ent}) and randomness certification (Section \ref{sec:Cert_al}). Section \ref{sec:post} presents the main techniques used in post-processing for randomness extraction. In the following, we discuss techniques for randomness enhancement, briefly detailing the process of amplification and expansion of randomness in Section \ref{sec:enhance}. In Section \ref{sec:Test}, the main randomness testing methods are detailed, focusing on NIST and DieHard. The typical currently-achievable implementations are detailed in Section \ref{sec:mainstream}, including high generation speed, photonic chip integrability, dedicated hardware and cryptographic applications. Finally, we conclude the review with a summary and future outlook in Sec. \ref{sec:summary}. 

\section{Randomness in Real World} \label{sec:Randomness}

Randomness plays an important role in scientific and practical applications, yet its operational definition remains notably nontrivial, diverging markedly across different fields of science \cite{Bera_2017, schneier2007applied}. In computation, for example, a bit sequence is called random if its Kolmogorov complexity approaches maximality for infinitely long sequences, signifying incompressibility \cite{LiVitanyi2019}. This asymptotic criterion, however, renders the metric unsuitable for practical finite-length applications. In physics, the situation is fundamentally different. Bell's theorem shows that some correlations obtainable from local measurements on entangled systems cannot be reproduced by any local hidden-variable model \cite{bell1964einstein}. More generally, nonclassicality witnesses in prepare-and-measure scenarios demonstrate that the observed statistics are incompatible with classical hidden-variable models satisfying specified constraints, such as a bound on the dimension of the communicated system or an assumption of generalized noncontextuality \cite{poderini2020criteria}. These results collectively indicate that quantum randomness is not merely epistemic but rooted in the intrinsic structure of the theory.

These definitions naturally lead to a central question: what exactly qualifies as a sequence sufficiently random for a given application? This section addresses this issue by examining the concept of randomness generation across disparate fields. We begin by discussing pseudorandomness produced by algorithmic generators, followed by an analysis of apparent randomness in classical deterministic systems. We then contrast these epistemic forms with the intrinsic randomness posited by quantum theory.

\subsection{Pseudorandomness}

The need for randomness in science and communication is fundamental, underpinning both statistical inference \cite{Metropolis01091949} and cryptographic security \cite{marton2010randomness}. For decades, pseudorandom number generators have fulfilled this role by expanding a short random seed into a long sequence through deterministic recursive transformations \cite{luby1996pseudorandomness, 8672854}. Security in this setting rests on the assumption that an adversary lacking knowledge of the seed and internal state cannot distinguish the generated sequence from a truly random one, as doing so would require computational resources far beyond practical feasibility to detect statistical patterns \cite{rock2005pseudorandom}.

To ensure the generated sequence meets these operational demands and resists such statistical analysis, the underlying algorithm relies on specific mathematical and practical properties. Accordingly, a PRNG is typically evaluated based on the following criteria \cite{BHATTACHARJEE2022100471}:

\begin{enumerate}[label=(\arabic*)]
    \item \textbf{Uniformity}\label{crit:uniformity}: The generated values follow a uniform probability distribution across the target space. This indicates that each possible value or bit sequence has an equal probability of occurrence, avoiding statistical bias towards specific outputs.
    
    \item \textbf{Independence}\label{crit:independence}: The sequence exhibits correlation between successive numbers. Consequently, observing previous outputs provides no computational advantage in predicting future values, approximating the behavior of independent identically distributed variables.
    
    \item \textbf{Long Period}\label{crit:long_period}: Because PRNGs operate as finite state machines, their output sequences eventually repeat. A suitable generator possesses a cycle length large enough so that the repetition period is not exhausted during practical applications, mitigating structural predictability.
    
    \item \textbf{Reproducibility}\label{crit:reproducibility}: The algorithm relies on a deterministic state transition function. Providing the same initial seed initializes the internal state identically, yielding the corresponding output sequence, which allows for controlled simulations and system debugging.
    
    \item \textbf{Consistency and Disjoint Subsequences}\label{crit:consistency}: The statistical quality of the generated numbers does not depend on the specific choice of the initial seed. Furthermore, subsequences derived from distinct seeds present minimal cross-correlation, allowing deployment in parallel computing environments where multiple independent streams are required.
    
    \item \textbf{Efficiency and Portability}\label{crit:efficiency}: The generation process requires low computational overhead and memory footprint, facilitating integration into resource-constrained systems. The algorithm implementation produces identical output streams for a given seed regardless of the underlying hardware architecture or operating system.
\end{enumerate}

These criteria make PRNG often preferred over True Random Number Generators (TRNGs) due to their high generation speed and implementation simplicity. While TRNGs rely on unpredictable but slow physical processes \cite{SHARMA2026100138, YUFEIANDLI}, PRNGs achieve high throughput and hardware efficiency \cite{https://doi.org/10.1002/wics.1404}. They require low computational resources, often utilizing basic logic gates and flip-flops, to provide consistent data streams. Based on their underlying mathematical recurrence, these generators are classified into distinct categories, each with specific applications. Henceforth, we review these categories and main implementations.

Linear Congruential Generators (LCGs) represent an early structural category, evaluating linear recurrence sequences within modular arithmetic \cite{knuth2005art}. The state transition is defined by $x_{n+1} = (a x_n + c) \pmod{m}$, where $m$ is the modulus (defining the maximum period), $a$ is the multiplier, and $c$ is the increment. Common implementations include Knuth's MMIX and \textit{minstd\_rand}. To address the statistical limitations of basic LCGs without sacrificing speed, the Permuted Congruential Generator (PCG) family, such as PCG-32 \cite{draper2025}, applies a randomized permutation function to the LCG output \cite{knuth2005art}. This approach maintains high throughput while improving statistical quality \cite{draper2025}.

Linear Feedback Shift Registers (LFSRs) operate on modulo-2 arithmetic, relying primarily on exclusive-OR (XOR) gates and cascaded memory elements. Requiring no complex arithmetic operations, this architecture is well-suited for hardware, offering high speed and low area overhead \cite{Salehi2024}. Notable algorithms based on this concept include the \textit{Mersenne Twister} (MT19937) \cite{matsumoto1998mersenne}, which features a long period, and the \textit{SIMD-oriented Fast Mersenne Twister} (SFMT), designed for parallel processor pipelines. The \textit{Xorshift} family, including Xorshift128+ and Xorshift64*, achieves high throughput by using only successive bitwise shifts and XOR operations.

Cellular Automata (CAs) consist of a discrete network of finite state machines updated simultaneously via local transition rules \cite{wolfram1986random}. Because cells communicate only with direct neighbors, this structure offers high hardware scalability. Cellular Automata based algorithms range from the standard Rule 30 to Decimal CAs, which exhibit low correlation and high statistical quality \cite{wolfram1986random}. For secure applications, Cryptographically Secure PRNGs   (CSPRNGs) rely on computationally intractable problems and one-way functions. The Blum Blum Shub (BBS) generator, for instance, bases its security on integer factorization \cite{Divyanjali2014}. CSPRNGs ensure next-bit unpredictability in polynomial time; however, the computational cost of modular exponentiation results in lower data throughput compared to LFSR or CA architectures \cite{Divyanjali2014}. An example of 3-bit LFSR PRNG can be see in Ex. \ref{ex:prng}. 

\begin{example}{Linear Feedback Shift Register}{prng}

Let the model below be a 3-bit LFSR PRNG 
\begin{center}
        \includegraphics[width=0.8\textwidth]{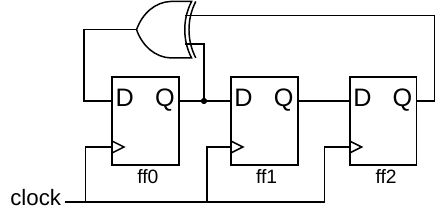}
        \label{fig:diagrama_lfsr}
    \end{center}
    with three D flip-flops ($ff_0, ff_1, ff_2$) and an XOR gate. At each clock pulse, bits shift right, and the new input for $ff_0$ is computed by XORing the outputs of $ff_0$ and $ff_2$. \\ 
    
    Using a sequence generation example with a seed = 101, we have
    \begin{itemize}[nosep, leftmargin=*]
        \item \textbf{Init:} $(ff_2,ff_1,ff_0) = 101$
        \item \textbf{Clock cycle 1:} $1\oplus 1 \rightarrow 010$
        \item \textbf{Clock cycle 2:} $0\oplus 0 \rightarrow 001$
        \item \textbf{Clock cycle 3:} $0\oplus 1 \rightarrow 100$
        \item \textbf{Clock cycle 4:} $1\oplus 0 \rightarrow 110$
        \item \textbf{Clock cycle 5:} $1\oplus 0 \rightarrow 111$
        \item \textbf{Clock cycle 6:} $1\oplus 1 \rightarrow 011$
        \item \textbf{Clock cycle 7:} $0\oplus 1 \rightarrow 101$ (Repeats)
    \end{itemize}
    
\end{example}

For software implementations requiring high throughput and strong statistical properties without the need for cryptographic security, SFMT, the Xorshift family, and PCG-32 are suitable options \cite{bhattacharjee2022search}. In dedicated hardware environments demanding high parallelism and area efficiency, Decimal CAs and standard LFSR architectures are suitable options \cite{Salehi2024}.

Despite their computational speed and efficiency, standard PRNGs are fundamentally insecure in cryptographically sensitive or non-deterministic contexts. Because PRNGs operate via deterministic algorithms, their output is entirely bounded by the initial seed, i.e., if an adversary discovers or guesses a weakly initialized seed, they can reconstruct the internal state and fully predict both past and future outputs, violating both forward and backward security \cite{lenstra2012ron, Divyanjali2014, markoff2012flaw}. Furthermore, because a PRNG relies on a finite memory space, its output inevitably cycles \ref{crit:long_period} and exhibits inherent mathematical correlations, allowing adversaries to deduce the full internal state from a short sequence of observed outputs without ever knowing the seed \cite{marton2010randomness}. Consequently, applications requiring genuine unpredictability or physical entropy cannot rely on standard PRNGs and must instead employ  alternative random generators.

\subsection{Randomness in classical theory}

The use of statistical theory in classical physics allows for an efficient description of many-body systems by focusing on average properties rather than on the precise details of individual particles \cite{landau2013statistical, Penrose_1979}. Such probabilistic descriptions intuitively introduce the conceptualization of randomness. In this context, randomness is typically understood as an apparent property: it arises not from any intrinsic indeterminism in nature, but from our incomplete knowledge of a deterministic system whose microscopic dynamics are too complex to track in full detail \cite{Bera_2017}. The stochasticity thus emerges from the practical necessity, and this does not imply that randomness is a fundamental feature of the system.

Apparent randomness, also termed epistemic randomness, presupposes an underlying deterministic framework. This determinism is rooted in the classical worldview, according to which the state of a physical system at any given moment, together with the laws governing its evolution, uniquely determines all future states  \cite{Volovich2011}. In this view, the apparent stochasticity of macroscopic observations merely reflects our ignorance of the precise microscopic configuration, rather than any fundamental absence of causality. These physical phenomena can be incorporated in the devices such that form the class of TRNGs \cite{stipcevic2014true}. A widely used example is electronic noise present in electronic circuits. Random number generators based on these elements typically employ resistors or photodiodes as entropy sources \cite{somlo1975zener, stipcevic2007quantum}. The fluctuations in voltage or current arise from thermal agitation of charge carriers, a phenomenon well described by statistical mechanics \cite{johnson1928thermal, Nyquist32, VANDERZIEL1978313}.  Another examples include chaotic systems, which are governed by deterministic equations, yet their extreme sensitivity to initial conditions leads to behaviour that is unpredictable in practice \cite{gleick2008chaos, ivancevic2008complex}. 

This deterministic perspective is historically encapsulated by Laplace's notion of a hypothetical intellect that, knowing all positions and velocities of every particle in the universe, could predict the future with absolute certainty \cite{marquis1819essai}. It is precisely this deterministic underpinning that distinguishes epistemic randomness from the intrinsic, ontological randomness of quantum mechanics, where even complete knowledge of the state does not guarantee deterministic predictions \cite{Volovich2011}. However, if an adversary finds finer methods to predict outcomes the security can be seriously compromised. Consequently, randomness that is merely a manifestation of our practical ignorance cannot guaranty security unless it is accompanied by a rigorous proof that no feasible prediction strategy exists, given the computational resources available to an attacker. 

\subsection{Randomness in quantum theory}

The source of randomness in quantum processes admits multiple interpretations \cite{RevModPhys.64.339}. According to the Born rule, the outcome of a measurement on a quantum system is probabilistic, with probabilities given by the squared modulus of the wave function's projection onto the eigenstates of the measured observable. Complementing this description, the theory of quantum measurement attributes this probabilistic character to the interaction between the system, including its measurement apparatus, and the surrounding environment \cite{wheeler2014quantum}. This coupling generically leads to decoherence, which effectively precludes the possibility of measuring the system in a fully deterministic manner \cite{zurek1991decoherence}. Indeed, this holds for the vast majority of quantum systems, with the notable exception of those prepared in an eigenstate of the observable being measured, for which the outcome is deterministic.

This discussion can be formalized through the postulates of quantum mechanics, as stated in the textbook \cite{nielsen2010quantum}, which we summarize below:

\begin{enumerate} [label=(\arabic*)]
    \item \textbf{State space}: Associated to any isolated physical system is a complex vector space with inner product (that is, a Hilbert space) known as the state space of the system. The system is completely described by its state vector, which is a unit vector in the system’s state space.
    \item \textbf{Evolution}: The evolution of a closed quantum system is described by a unitary transformation. That is, the state $\ket{\psi}$ of the system at time $t_1$ is related to the state $\ket{\psi}$ of the system at time $t_2$ by a unitary operator $U$ which depends only on the times $t_1$ and $t_2$,
    \begin{equation}
        \ket{\psi' } = U \ket \psi 
    \end{equation}

    \item \label{post:3} \textbf{Quantum measurement}: Quantum measurements are described by a collection $\{M_m\}$ of measurement operators. These are operators acting on the state space of the system being measured. The index $m$ refers to the measurement outcomes that may occur in the experiment. If the state of the quantum system is $\ket{\psi}$ immediately before the measurement then the probability that result $m$ occurs is
    \begin{equation}
        p_m(m) = \bra{\psi}M_m^\dagger M_m \ket{\psi} \,,
    \end{equation}

    and the state of the system after the measurement is $M_m \ket\psi/\sqrt{p_m(m)}$.

    \item \label{post:4} \textbf{Composite systems}: The state space of a composite physical system is the tensor product of the state spaces of the component physical systems. Moreover, if we have systems numbered $1$ through $n$, and system number i is prepared in the state $\ket{\psi_1}$, then the joint state of the total system is $\ket{\psi_1}\otimes\ket{\psi_2}\otimes \dots \otimes \ket{\psi_n}$.
\end{enumerate}

From postulate \ref{post:3}, it is evident that the inherent randomness of quantum theory is intimately connected to the measurement process. Indeed, the measurement interpretation implies that we have access only to a probabilistic description of outcomes. It is worth noting, however, that the postulates do not preclude the existence of deterministic cases \cite{Bera_2017}. For instance, if the system is prepared in the pure state corresponding to eigenstate \(m_0\) of the spin z-component operator, a measurement along that axis yields the outcome \(m_0\) with certainty. If, instead, we measure the spin x-component, the result becomes non-deterministic and can only be described probabilistically. This example illustrates a more general feature of quantum mechanics: the existence of non-commuting observables that cannot be measured simultaneously with arbitrary precision.

This feature has a particular relevance in the continuous-variable scenario, where the Heisenberg uncertainty relation can be used to describe the randomness inherent to continuous-measurement \cite{serafini2023quantum}. They quantify the fundamental limits on the simultaneous measurement precision of two non-commuting observables.

\begin{definition} \label{def:heis}
    (Heisenberg uncertainty relation) For a given state $\rho$ and observables $X$ and $Y$, the relation provides a lower bound on the product of their variances when measured jointly, expressed as
    
    \begin{equation} \label{eq:uncertainty}
        \delta X^2\delta Y^2 \geq \frac{1}{4} |\mathrm{Tr}\rho[X, Y]|^2\,,
    \end{equation}
    where \(\delta X^2 = \operatorname{Tr}\rho X^2 - \left(\operatorname{Tr}\rho X\right)^2\) denotes the variance and \([X,Y] = XY - YX\) is the commutator. 
\end{definition}

From Def. \ref{def:heis}, a non-vanishing \(\delta X\) signifies randomness in the measurement outcome, which may originate from non-commutativity between the state and the observable, or alternatively from classical uncertainty present in the state itself, independent of quantum superposition \cite{wheeler2014quantum}. Equation \eqref{eq:uncertainty} allows either \(\delta X\) or \(\delta Y\) to vanish individually for a given state \(\rho\) with \([X,Y] \neq 0\), but not simultaneously. However, when \(\delta X\) vanishes, the bound provides no meaningful information about \(\delta Y\), and vise versa.

Beyond the quantitative constraints imposed by uncertainty relations, two deeper foundational features play an important role in the discussion about inherent randomness in quantum measurements: contextuality and nonlocality \cite{Bera_2017}. Formalized by the Kochen-Specker theorem \cite{kochen1967}, contextuality establishes that the outcome of a quantum measurement cannot be interpreted as revealing a pre-existing value independent of the measurement context. In other words, no noncontextual hidden variable theory can reproduce the predictions of quantum mechanics. Nonlocality, as demonstrated by Bell's theorem \cite{bell1964einstein}, goes further by showing that quantum correlations between spatially separated systems cannot be accounted for by any local hidden variable theory. Both results preclude the existence of an underlying noncontextual and local deterministic description that would assign definite values to all observables prior to measurement. Consequently, the randomness observed in quantum measurements is not merely epistemic, as in classical or pseudorandom settings, but genuinely intrinsic, rooted in the algebraic structure of quantum observables and the non-classical correlations they exhibit \cite{RevModPhys.38.447}.

The inherent randomness in quantum process renders quantum mechanics the only empirically established physical theory capable of providing true randomness. This unique property has catalyzed the development of quantum technologies that exploit this feature to generate truly random numbers, known as quantum random number generators \cite{RevModPhys.89.015004, ma2016quantum, mannalatha2023review}. These devices have found significant application in cryptographic protocols and secure communication scenarios \cite{Martin2015, Crampton2025}, where the unpredictability of the generated sequences is paramount \cite{marton2010randomness, Gennaro2006}. Quantum random number generators currently stand as one of the most mature technologies of the second quantum revolution \cite{Dowling2003}, with commercial devices already available in the market \cite{idquantique_quantis_chips, cryptalabs_qrng, quside_ruby_s1}. Their practical realization often relies on the intrinsic randomness of quantum measurements and can be certified at most cases \cite{Pironio2010, Acin2016}, providing a degree of security unattainable by classical or pseudorandom alternatives. 

\begin{figure*}
    \centering
    \includegraphics[width=0.85\linewidth]{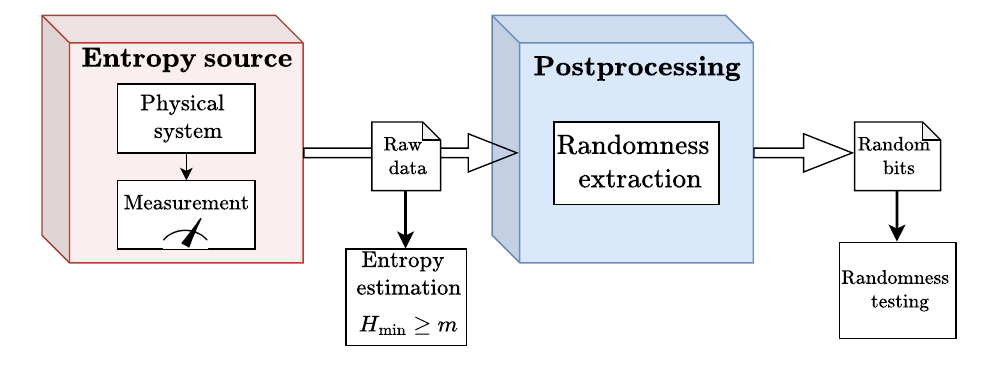}
    \caption{\justifying Conceptual scheme of a practical QRNG. The entropy source (red block) comprises a quantum physical system and its measurement stage, which together generate raw bits. This raw data is subjected to entropy estimation to evaluate the min-entropy. Subsequently, the raw bits are sent to the post-processing stage (blue block) to extract randomness, yielding final random bits that are then validated through randomness testing.}
    \label{fig:QRNG_Bulding_blocks}
\end{figure*}

Figure \ref{fig:QRNG_Bulding_blocks} illustrates the general scheme of a {QRNG}. The process begins with the preparation and measurement of the system known as an entropy source. In this stage, an experimental setup capable of preparing and performing measurements on a quantum system is conducted, ensuring a certain degree of randomness in the resulting data. Thus, various quantum phenomena can be exploited, such as radioactive decay, vacuum noise, and photon counting, among others \cite{RevModPhys.89.015004, ma2016quantum, mannalatha2023review}. In general, the output of the entropy source must be a sequence of raw bits formed by $n$-bit words, derived from discrete event detections (discrete-variables) or from analog signal conversions (continuous-variables) via an  analog-to-digital converter (ADC). However, practical measurements are susceptible to classical noise, which introduces impurities into the results \cite{frauchiger2013truerandomnessrealisticquantum}.


From a security perspective, the distinction between quantum and classical noise is crucial \cite{DAriano_2005}. If part of the classical noise affecting the device can be controlled or predicted by an eavesdropper then the raw data may contain side information accessible to an adversary \cite{PhysRevA.100.052318, Gehring2021, frauchiger2013truerandomnessrealisticquantum, PhysRevApplied.3.054004}. Therefore, a secure QRNG requires both a reliable characterization of the available entropy and a post-processing method capable of extracting the useful random component from the raw data. Randomness extraction provides precisely this functionality: it distills nearly perfect randomness from imperfect experimental outputs, producing final bits with negligible bias and correlations \cite{NISAN1999148}.

\section{Entropy Sources} \label{sec:entropy_sources}

Entropy quantifies the amount of extractable randomness available from a given entropy source \cite{cover1999elements}. In the context of random number generation, higher entropy corresponds to greater uncertainty in the measurement outcomes, whereas deterministic or predictable processes contain little or no usable randomness \cite{shannon1948mathematical, von195113, eastlake1994randomness}. An entropy source is therefore the physical mechanism from which raw random data are generated before any post-processing is applied \cite{stipcevic2014true}. 

In this review, we classify entropy sources into two categories: discrete-variable (DV) and continuous-variable (CV) sources. This classification reflects the nature of the measured quantum observable and determines both the detection architecture and the subsequent entropy extraction procedure. The following subsections review the principal entropy sources belonging to each category, emphasizing their underlying physical mechanisms, statistical properties, experimental implementations, and practical performance. 

\subsection{Discrete-variable sources}
\label{subsec:discrete_sources}

Discrete-variable sources exploit quantum observables whose measurement outcomes belong to a finite or countable set. Each elementary quantum measurement produces a discrete symbol, most commonly a binary digit, whose value is determined by an irreducibly probabilistic measurement outcome. Consequently, entropy extraction is performed directly from the measured events without requiring the digitization of an analog signal. The following subsections review the principal classes of these sources, including radioactive decay, photon-path measurement, photon time-of-arrival, photon counting, attenuated optical pulses, and Raman scattering. Although these implementations rely on different physical mechanisms and detection architectures, they share a common principle: randomness originates from the projective measurement of quantum systems whose outcomes are fundamentally discrete and intrinsically unpredictable.
 
\subsubsection{Radioactive decay }
\label{subsec:radioactive_decay}

The radioactive decay of unstable atomic nuclei is among the earliest and most paradigmatic quantum phenomena exploited as an entropy source for QRNG \cite{schmidt1970quantum, walker1996hotbits}. The nature of the phenomenon is related to the spontaneous transformation of an unstable atomic nucleus into a more stable daughter nucleus, accompanied by the emission of ionizing particles and/or radiation. In this process, there are three principal decay modes: $\alpha$ decay, characterized by the emission of a helium-4 nucleus; $\beta$ decay, involving the emission of an electron or positron together with an (anti)neutrino; and $\gamma$ emission, consisting of high-energy photon radiation subsequent to nuclear de-excitation \cite{knoll2010radiation}.

The quantum mechanical foundation of randomness in radioactive decay is well established: the transition of an unstable nucleus to its daughter state is governed exclusively by quantum tunneling and nuclear matrix elements, which render the precise instant of disintegration fundamentally unpredictable \cite{gurney1929quantum, zafar2025randomness, silverman1999tests}. Even in the hypothetical case of complete knowledge of the nuclear quantum state, classical mechanics provides no mechanism by which the exact moment of decay could be determined; the process is irreducibly stochastic \cite{park2020qec}.  Additionally, individual nuclear decay events are statistically independent and exhibit no memory of past disintegrations; the process is therefore Markovian \cite{Bisadi2017, silverman1999tests}. These characteristics constitute the origin of the genuine quantum entropy exploited by radioactive QRNG devices.

Under the Markovian hypothesis, each nucleus is characterized by a constant probability of decay per unit time, quantified by the decay constant $\lambda_m$ (related to the half-life $t_{1/2}$ by $\lambda_m = \ln 2 / t_{1/2}$). The nuclear lifetime $t$ follows an exponential distribution with probability density function
 
\begin{equation}
    p(t) = \lambda_m \, e^{-\lambda_m t}, \quad t \geq 0,
    \label{eq:exponential_lifetime}
\end{equation}
 
\noindent and the number of decay events $N$ observed within a fixed time window $\Delta t$ follows a Poisson distribution with mean $\mu = \lambda_m \Delta t$ \cite{Bisadi2017}. These statistical properties constitute the fundamental entropy source exploited by radioactive-decay QRNGs, in which the random occurrence of nuclear disintegrations is converted into binary sequences. The independence, stationarity, and memorylessness of the Poisson process have been rigorously verified by statistical analyses of long temporal sequences of alpha, beta, and electron-capture decay events across nuclides whose half-lives span twelve orders of magnitude, confirming the complete absence of correlations at any run length \cite{silverman1999tests}. Note that the intrinsic emission rate $\lambda_m$ must be distinguished from the effective detection rate $\lambda$, which depends on the efficiency and geometry of the radiation detector. The effective detection rate is therefore given by $\lambda = k \lambda_m$, where $0 < k \leq 1$ is a factor that accounts for the detector's quantum efficiency, the solid angle subtended by the sensitive volume, signal attenuation in the intervening medium, and read-out channel losses \cite{park2020qec}. Thus, the value of $k$ is a critical experimental parameter, since it determines the raw bit rate achievable by the generator.
 
The experimental realization of a radioactive QRNG requires a detection system capable of resolving individual decay events and converting them into electronic signals suitable for digital processing \cite{schmidt1970quantum, walker1996hotbits, alkassar2005true, park2020qec,schmidt1970quantum, vincent1970generation,  alkassar2005obtaining, park2020lightweight}. The classical implementation employs a Geiger-Müller (GM) tube, a gas-filled cylindrical detector maintained at a high voltage bias \cite{zafar2025randomness, knoll2010radiation}. 
When an ionizing particle enters the active volume of the GM tube, it initiates a gas avalanche that produces a macroscopic current pulse, which is subsequently detected and processed by the read-out electronics  \cite{zafar2025randomness, lutz2007semiconductor}. The detector is operated within the Geiger plateau, where the count rate becomes nearly insensitive to small bias-voltage variations, ensuring stable and reproducible operation  \cite{zafar2025randomness}.

\begin{figure*}
    \centering
    \includegraphics[width=1\linewidth]{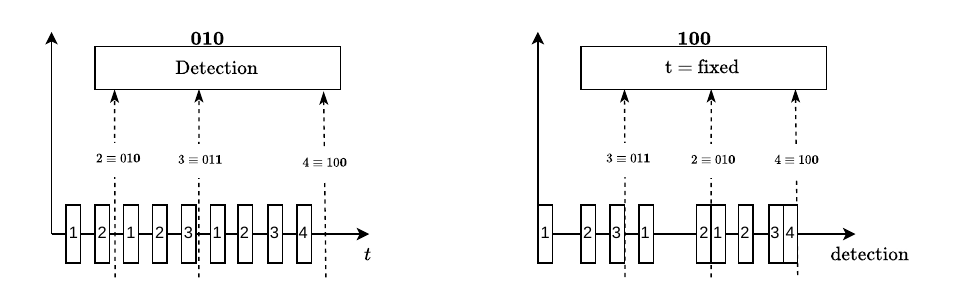}
    \caption{\justifying \textit{Fast clock} (left) and \textit{slow clock} (right) bit extraction schemes. In the \textit{fast clock}, the number of clock cycles elapsed between consecutive detector pulses is counted, and the least significant bit is extracted. In the \textit{slow clock}, time is divided into fixed windows of duration $\Delta t$ and the number of detected pulses $k$ per window is mapped to bits.}
    \label{fig:Fast-slow_clock}
\end{figure*}

Bit extraction from radioactive decay is typically realized through two canonical schemes: the \textit{fast clock} and the \textit{slow clock} (see Fig. \ref{fig:Fast-slow_clock}). In the first scheme, a reference clock operating at frequency \(f_c \gg \lambda\) timestamps each detector pulse; the number of clock cycles elapsed between consecutive pulses is recorded, and bits are extracted from this integer value \cite{park2020qec, alkassar2005true}. In the second scheme, time is partitioned into fixed windows of duration \(\Delta t\), and the number of pulses \(k\) per window is mapped to bits via parity (\(k \bmod 2\)) or modular reduction (\(k \bmod 2^n\)) \cite{park2020qec, alkassar2005true}. The entropy per observation window is maximized near \(\mu = \lambda \Delta t \approx 1\), where the Poisson distribution exhibits its greatest relative spread. In both cases, the output bits retain the independence and memorylessness of the underlying Poisson process, provided the detector operates below its dead-time limit \cite{zafar2025randomness}.

An important limitation inherent to GM-based detectors is the dead time $\tau_d$, the finite recovery interval following each detected pulse during which the detector is insensitive to new events, typically of order $10^{-4}$--$10^{-3}\,\mathrm{s}$.Dead time is conventionally described by two idealized models, the paralyzable and non-paralyzable responses, as well as by hybrid models combining both time constants \cite{muller1973dead, lee2000new}. For high-activity sources, dead-time effects lead to a systematic undercount of true decay events and introduce correlations in the detected sequence, necessitating appropriate dead-time corrections during post-processing \cite{knoll2010radiation,vincent1970generation, vincent1971precautions}.


The first generators, based on $^{60}$Co, $^{90}$Sr and $^{137}$Cs coupled to GM tubes, operated at rates of only $1-0.8$ bits/s \cite{schmidt1970quantum, walker1996hotbits}, and the count rate remains capped in the MHz range by the detector dead time discussed above, so that decay-based devices typically stay below  a few hundred kilobits per second \cite{RevModPhys.89.015004}. 
Recent work has pushed against this ceiling by raising the source activity, from the kbit range of $^{241}$Am to the $\sim$100~Mbit accessible with pure $\beta$ emitters such as $^{63}$Ni \cite{park2020lightweight, park2020high}, and by replacing bulk GM tubes with Complementary Metal-Oxide-Semiconductor (CMOS) integrated detectors \cite{park2020qec}. Radioactive decay survives today mainly as a benchmark of entropy quality and as a practical option for low-power, low-duty-cycle applications such as key seeding in IoT devices 
\cite{park2020lightweight}.

\subsubsection{Branching path}

\begin{figure}
    \centering
    \includegraphics[width=1.\linewidth]{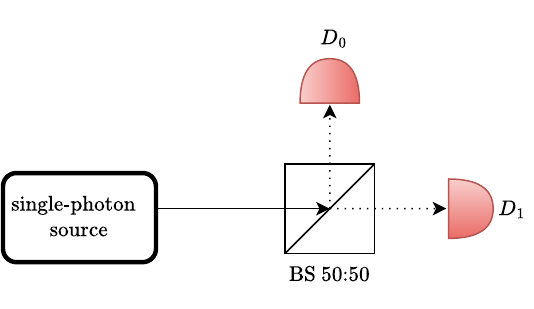}
    \caption{\justifying Conceptual schematic of a photon-path QRNG. A single photon prepared in a spatial superposition impinges on a balanced 50:50 beam splitter (BS), yielding equal probabilities of being detected by detectors $D_0$ and $D_1$. Detection events at $D_0$ and $D_1$ are encoded as bits 0 and 1, respectively, forming the quantum entropy source used for quantum random number generation.}
    \label{fig:qrng_path}
\end{figure}

Branching path QRNG constitute one of the simplest and most widely adopted photonic implementations of quantum randomness \cite{liu_2017_high, Jennewein2000, soubusta2001quantum, Francis2017PlasmonicQRNG, an_2017_proof, Sarkar2019, white_2019_quantum, Li2011, leone_2022_certified, argillander_2022_tunable, lin_2022_imperfection, leone2023generation, du2025source, pewkhom2025enhanced, miszczak_generating, gavrylko_physical}. In this scheme, a single photon incident on a balanced 50:50 beam splitter is coherently transformed into a superposition of two spatial modes, which subsequently propagate along two possible output paths, as illustrated in Fig.~\ref{fig:qrng_path}. The two output modes are represented by the orthogonal Fock states $\ket{10}$ and $\ket{01}$, corresponding to the photon emerging from the first or the second output port, respectively \cite{kwon2009quantum, dudek2025optical, pewkhom2025enhanced, Jennewein2000, haider2025quantum, Francis2017PlasmonicQRNG}.

In Dirac notation, the photon state after the balanced beam splitter is
\begin{equation}
    \ket{\psi} =
    \frac{1}{\sqrt{2}}
    \left(  \ket{10} + \ket{01} \right),
\label{eq:path_state}
\end{equation}
which, according to the Born rule, yields
\begin{equation}
    P(D_0) = |\langle10|\psi\rangle|^2
    = P(D_1)
    = |\langle01|\psi\rangle|^2
    = \frac12\,,
\label{eq:path_probability}
\end{equation}
for a perfectly balanced beam splitter. 

Upon measurement, the superposition collapses onto one of the two output modes, yielding equal probabilities of detecting the photon at either output port \cite{pewkhom2025enhanced, Jennewein2000, haider2025quantum}.
This measurement-induced randomness is the fundamental entropy source exploited by photon path QRNGs. Unlike a classical particle with a predetermined trajectory, the output port is not defined prior to measurement, since it is established only when the projective measurement collapses the spatial superposition.

Each measurement projects the quantum state onto one of the two output modes. A click registered by detector $D_0$ is encoded as bit 0, whereas a click registered by detector $D_1$ is encoded as bit 1. Ideally, each incident photon produces exactly one detector click, yielding one unbiased random bit per measurement. The detector outputs are therefore converted directly into a binary sequence that inherits the intrinsic unpredictability of the underlying quantum measurement \cite{kwon2009quantum, dudek2025optical, Jennewein2000, haider2025quantum, Francis2017PlasmonicQRNG}.

The bit generation rate can be further increased by adding additional output paths with equal splitting probabilities. In a more general representation, the state $\ket{\psi_n}$, with $n$ possible paths, can be written as
\begin{equation}
\label{eq.n_caminhos}
    \ket{\psi_n}=\frac{1}{\sqrt{n}}\sum_{i=1}^{n}\ket{i},
\end{equation}
for an ideal symmetric splitter, i.e., each output port is detected with probability $1/n$. In this architecture, each output mode is monitored by an independent photodetector, allowing each detection to encode up to $\log_2(n)$ bits under ideal uniform splitting \cite{Stefanov2000}.

Practical implementations are affected by detector dark counts, optical losses, finite detection efficiency, and multiphoton events \cite{mannalatha2023review}. Multiphoton pulses may produce simultaneous clicks in both detectors, whereas optical losses and finite efficiency lead to discarded events (see ref. \cite{frauchiger2013truerandomnessrealisticquantum} for a detailed model of a non-ideal polarizing beam splitter). These nonidealities reduce the extractable entropy and must be mitigated. Consequently, practical photon-path QRNGs typically incorporate randomness extraction to suppress residual bias introduced by experimental imperfections.


\subsubsection{Time of arrival}


The arrival time of individual photons constitutes one of the most fundamental temporal entropy sources employed in DV-QRNG.
For weak continuous wave optical fields, each photon emission and detection event is intrinsically probabilistic, so that the precise instant at which a photon is detected cannot be predicted, even with complete knowledge of the experimental conditions.
This irreducible unpredictability originates directly from quantum mechanics and forms the physical basis of photon arrival time QRNGs \cite{nie2014practical, yan2014multi, bisadi2016generation, lin2024xray, dynes2008high, wayne2009photon, wayne2010low, yan2015high, kravtsov2015minimalist, tomasi_2018_model, iavich_2020_novel, ferreira_2021_timeinterleaved,zhang_2023_realization_journal, Schranz2023Stochastic, strydom2024quantum, fan2024metalens, hegazy2025quantum, wang2025efficient, 7038196, Stefanov2000, massari_isscc, Tontini2019, massari_icecs, solymos_secure, yan_optical}.

For a coherent optical source, photon detections constitute a Poisson point process with statistically independent events \cite{mandel1995optical, loudon2000quantum, saleh2019fundamentals}. Consequently, the waiting time $\Delta t$ between successive detections follows an exponential probability density,
\begin{equation}
p(\Delta t)=\lambda e^{-\lambda\Delta t},
\qquad
\Delta t\ge0,
\label{eq:waiting_time}
\end{equation}
where $\lambda$ denotes the mean photon detection rate. These statistical properties provide the temporal entropy exploited by time-of-arrival QRNGs, while different extraction schemes process the same underlying stochastic process in distinct ways \cite{stipcevic2007quantum, wayne2009photon}.

A typical time-of-arrival QRNG comprises a weak continuous-wave laser or an attenuated single-photon source, a single-photon detector, typically a Single-Photon Avalanche Diode (SPAD), high-resolution timing electronics such as a Time-to-Digital Converter (TDC), and a digital post-processing unit \cite{nie2014practical, tomasi_2018_model}. Each photon detection event is assigned a timestamp, from which random bits are subsequently extracted according to the adopted timing protocol. In the following, we provide a detailed description of some of these techniques.

The waiting time method extracts entropy from the interval $\Delta t$ between two consecutive detection events \cite{wayne2009photon, wayne2010low}. Since $\Delta t$ follows the exponential distribution given by Eq.~\eqref{eq:waiting_time}, high-resolution counters or TDCs discretize the interval into time bins, allowing several raw bits to be extracted from a single photon detection. Because short intervals occur more frequently than long ones, the raw sequence is intrinsically biased and therefore requires randomness extraction based on hash functions \cite{wayne2009photon, wayne2010low}.

An alternative approach measures each photon's arrival time relative to an external periodic reference clock rather than comparing consecutive detection events \cite{nie2014practical}. The clock period $T$ is divided into $N$ equally spaced bins, and the detection bin defines the raw output. Ideally, $ P_i=1/N,$ for every bin, resulting in substantially lower statistical bias than waiting time schemes. 


When discrete time intervals are measured using a continuous (non-restartable) clock, discretization introduces a systematic positive autocorrelation between successive bits \cite{stipcevic2007quantum}. As illustrated in Fig.~\ref{fig:Fast-slow_clock}, timing extraction under continuous clocks is typically structured around two operational regimes fast-clock limit and slow-clock limit. 


The restartable clock method overcomes these limitations by synchronizing (restarting) the reference clock at the beginning of every measurement interval \cite{stipcevic2007quantum}. By ensuring that each time measurement depends strictly on its own interval length, it completely eliminates autocorrelation regardless of the clock frequency. Random bits are generated by comparing two independent waiting times,
\begin{equation}
b=
\begin{cases}
    0, & t_1 < t_2, \\
    1, & t_1 > t_2,
\end{cases}
\end{equation}
thereby suppressing systematic timing bias while requiring only a single-photon detector \cite{stipcevic2007quantum}.

The performance of time-of-arrival QRNGs is primarily limited by detector dead time, afterpulsing, timing jitter, and the finite resolution of the timing electronics \cite{Tontini2019, wayne2009photon, lin_2022_imperfection}. These nonidealities introduce correlations and bias into the raw timestamps, reducing the available min-entropy. Consequently, practical implementations incorporate randomness extraction procedures to suppress residual deterministic contributions and recover uniformly distributed random bits.

Time-of-arrival constitute one of the most mature implementations of DV-QRNG. Their entropy originates directly from the intrinsic unpredictability of photon detection times, while modern timing electronics enable efficient extraction of multiple random bits from individual detection events. Owing to their conceptual simplicity, excellent statistical performance, and compatibility with integrated photonic technologies \cite{acerbi2018robust}, time-of-arrival QRNGs remain among the most widely adopted architectures for practical high-speed QRNG \cite{ma2016quantum, stipcevic2007quantum}.

\subsubsection{Photon-counting}

Photon-counting QRNG exploits the intrinsically probabilistic generation and detection of individual photons when a weak coherent optical source, such as an attenuated continuous wave laser or a light-emitting diode (LED), is monitored by a single-photon detector. Each detection event results from an irreversible quantum measurement whose outcome cannot be predicted in advance \cite{Khanmohammadi2015, Bisadi2017, haider2025quantum, Chen2019, Sarkar2019}. 
This measurement-induced randomness is the fundamental source of entropy in photon-counting QRNGs and distinguishes them from classical intensity-fluctuation generators. While classical generators rely on complex or chaotic physical processes, photon-counting QRNG provides irreducible, fundamentally non-deterministic randomness derived directly from the collapse of the optical quantum state, rendering the output unpredictable even to an adversary with full knowledge of the setup \cite{Khanmohammadi2015}.

For coherent optical fields with a constant mean photon detection rate $\lambda_d$, successive detection events constitute a Poisson point process, so that the probability of detecting $n$ photons within an acquisition window of duration $T$ is
\begin{equation}
P(n)=
\frac{(\lambda_d T)^n}{n!}
e^{-\lambda_d T},
\label{eq:Poisson}
\end{equation}
where $\lambda_d T$ corresponds to the mean number of detected photons during the acquisition interval \cite{Aungskunsiri2023, soubusta2001quantum, Applegate2015, 10.1063/1.3613952, Tisa2015, ma2016quantum}. These Poissonian counting statistics constitute the entropy source exploited by photon-counting QRNGs. At sufficiently high count rates, however, detector dead time modifies the measured statistics, requiring careful optimization of the detection regime to preserve maximum entropy \cite{Furst:10, Tisa2015}.

A typical photon-counting QRNG consists of a weak optical source, a single-photon detector, such as a SPAD or photomultiplier tube (PMT), counting electronics synchronized by an acquisition window, and a digital post-processing unit \cite{Tisa2015, balygin_2017_quantum, balygin_2017_implementation, coleman_2020_paritybased, Aungskunsiri2023}. During each acquisition interval, the number of detected photons is recorded and subsequently converted into random bits according to the selected extraction protocol.


The simplest bit extraction protocol is the parity method, in which the number of detected photons within an acquisition interval $\tau_s$ is converted directly into one random bit \cite{Bisadi2017, Furst:10}. Even photon counts are mapped to bit 0, whereas odd counts are mapped to bit 1. For sufficiently large mean photon numbers, typically $\lambda>10$, the least significant bit of the photon count approaches an unbiased distribution, allowing high-quality random sequences to be generated with little or no post-processing \cite{Tisa2015, mannalatha2023review}. Photomultiplier-based implementations further benefit from dead-time effects that partially compensate statistical bias under appropriate operating conditions \cite{Furst:10}.

An alternative extraction protocol compares the photon counts measured in two consecutive acquisition windows, denoted by $n_1$ and $n_2$ \cite{Khanmohammadi2015}. Thus, random bits are assigned according to
\begin{equation}
b=
\begin{cases}
1,&n_1>n_2,\\
0,&n_1<n_2,
\end{cases}
\end{equation}
while equal counts are discarded. This procedure effectively acts as a deterministic randomness extractor analogous to the von Neumann algorithm (see Section \ref{post_sub_B}), suppressing bias caused by slow variations in optical power or detector efficiency. The price paid for this robustness is a reduced bit generation efficiency owing to the discarded events \cite{Bisadi2017, Applegate2015}.

More advanced photon-counting architectures exploit additional information contained in the detection process. Photon-number-resolving ({PNR}) detectors distinguish the exact number of photons contained in each optical pulse, allowing multiple random bits to be extracted from a single detection event through appropriate mapping of the Poisson distribution into equiprobable subsets \cite{Applegate2015, 10.1063/1.3613952}. Likewise, integrated SPAD arrays perform massive parallel photon detection, where both the photon count and the address of the triggered pixel contribute to the available entropy, substantially increasing the overall generation rate \cite{Marangon2016, wang2025efficient}.

The performance of photon-counting QRNGs is primarily limited by detector dead time, afterpulsing, and dark counts. Afterpulsing introduces temporal correlations through spurious avalanches generated by trapped charge carriers, whereas dark counts contribute classical detection events unrelated to the optical source \cite{Khanmohammadi2015, Bisadi2017, Ma2016}. These nonidealities reduce the extractable quantum entropy and therefore require careful detector calibration together with randomness extraction based on the estimated min-entropy of the raw data \cite{Ma2016, Marangon2016}.

Recent advances in integrated detector technologies have significantly increased the performance of photon-counting QRNGs. Large SPAD arrays enable massively parallel photon acquisition, with demonstrated generation rates exceeding hundreds of megabits per second \cite{Tisa2015}, while recent architectures combining multi-pixel arrays with high-resolution time-to-digital converters have achieved rates up to 2.1~Gbps \cite{wang2025efficient}. These improvements mitigate the bandwidth limitations imposed by detector dead time while preserving the Poissonian statistics of the detected photons through appropriate optimization of the illumination level.


\subsubsection{Attenuated pulse}

Attenuated-pulse QRNG, also known as weak coherent-state QRNG, exploits the fundamentally random photon-number statistics of weak optical pulses. In practice, these sources are typically implemented using pulsed laser diodes or silicon nanocrystal (Si-nc) LEDs, whose optical pulses are strongly attenuated to approach the single-photon regime \cite{Bisadi2017, Wei:09, Bisadi2015_SiliconNC, bisadi2016generation, balygin_2017_quantum, argillander_2023_quantum}. The quantum entropy originates from the spontaneous and stimulated emission processes governing photon generation, making the number of photons emitted in each pulse intrinsically probabilistic \cite{siegman1986lasers, loudon2000quantum, garrison2008quantum}.

For weak coherent optical pulses, the photon number follows a Poisson distribution,
\begin{equation}
P(n)=\frac{\mu^n e^{-\mu}}{n!},
\label{eq:Poisson1}
\end{equation}
where $\mu$ denotes the mean photon number per pulse \cite{Wei:09, Furst:10}. To obtain an unbiased binary source, the effective mean photon number at the detector must satisfy $\eta\mu=\ln2$, where $\eta$ is the overall detection efficiency \cite{Wei:09, Furst:10}. Under this operating condition, the probabilities of obtaining a no-click event,
\[
P(n=0)=e^{-\eta\mu},
\]
and a click event,
\[
P(n\ge1)=1-e^{-\eta\mu},
\]
are both equal to $1/2$, providing an ideal unbiased raw bit source \cite{Wei:09, Applegate2015}.

In the ideal single-photon approximation, the optical field may be approximated as an equal superposition of the vacuum state $\ket{0}$ and the single-photon state $\ket{1}$,
\begin{equation}
\ket{\psi}
=
\frac{1}{\sqrt2}
\left(
\ket0+\ket1
\right),
\end{equation}
which provides equal probabilities for click and no-click events. In realistic implementations, however, weak coherent pulses also contain multi-photon components, and the optical state is more accurately described by
\begin{equation}
\ket{\psi}
=
\frac1{\sqrt2}\ket0
+
\sum_{k=1}^{\infty}\alpha_k\ket{k},
\end{equation}
subject to
\begin{equation}
\sum_{k=1}^{\infty}
|\alpha_k|^2
=
\frac12,
\end{equation}
thereby preserving the equal probabilities of vacuum and non-vacuum detection events \cite{garrison2008quantum, loudon2000quantum, mandel1995optical}.

Experimentally, attenuated pulses are detected using SPADs operating in Geiger mode \cite{Wei:09, Furst:10, bisadi2016generation, balygin_2017_quantum, bisadi2016robust, argillander_2023_quantum}. Each optical pulse produces either a click or a no-click event, which are directly encoded as binary values. Under the operating condition $\eta\mu=\ln2$, both outcomes occur with equal probability, yielding an unbiased raw random sequence without requiring balanced optical splitting. The detector output is subsequently processed by threshold electronics that assign click and no-click events to binary values.

The performance of attenuated-pulse QRNG is primarily limited by detector afterpulsing, dead time, and fluctuations in the optical pulse intensity \cite{Cova2004, Hadfield2009}. Afterpulsing introduces temporal correlations between consecutive detection events through trapped charge carriers released after an avalanche, whereas source-intensity drift may alter the click probability from its ideal value. Consequently, practical implementations employ controlled detector dead times, active source stabilization, or randomness extraction algorithms such as the von Neumann procedure to suppress residual bias and preserve statistical independence \cite{Bisadi2017, Wei:09}.

Recent developments have introduced self-testing protocols capable of certifying randomness even when the internal operation of the devices is only partially trusted \cite{Bisadi2017, PhysRevA.100.062338}. These protocols require only an upper bound on the mean photon number of the emitted optical pulses, allowing continuous monitoring of the source while guaranteeing that the extracted entropy remains genuinely quantum in origin.


\subsubsection{Optical parametric oscillators}
\label{subsubsec:parametric_oscillators}

Optical parametric oscillators (OPOs) constitute a class of optical entropy sources in which microscopic quantum fluctuations are converted into a macroscopic optical field by parametric gain inside an optical resonator.
An OPO consists, in its simplest form, of a nonlinear medium placed inside a cavity and pumped by a coherent optical field \cite{Marandi2012, Zhang_2017, Okawachi2017Microresonator, Zhang2020DualEntropy, Okawachi2021KerrOscillator, Okawachi2021PhotonLifetime, Li2023FiveBit, Marah2024SecureComms}. 
The cavity provides optical feedback for the down-converted modes and imposes a threshold condition: only when the parametric gain exceeds the round-trip losses can the microscopic seed field be amplified into a stable macroscopic oscillation  \cite{Breunig2011}.

The elementary physical process underlying an OPO is spontaneous parametric down-conversion (SPDC). 
In a nonlinear medium $\chi^{(2)}$, one pump photon at angular frequency $\omega_p$ can be annihilated while two lower-frequency photons, conventionally called the signal($\omega_s$) and idler($\omega_i$), are created.

Energy conservation requires:
\begin{equation}
    \omega_p = \omega_s + \omega_i.
\end{equation}

\begin{figure}
    \subfloat[Energy Conservation\label{fig:energy_conservation}]{%
        \includegraphics[width=0.48\linewidth]{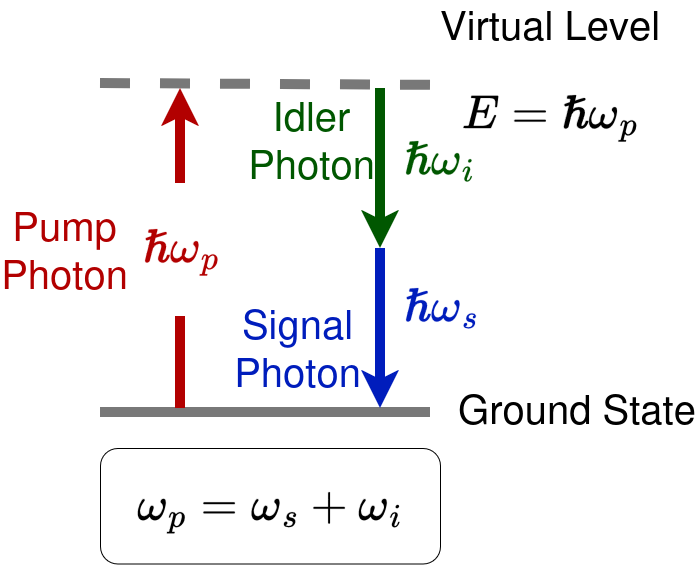}%
    }
    \hfill
    \subfloat[Momentum Conservation\label{fig:momentum_conservation}]{%
        \includegraphics[width=0.48\linewidth]{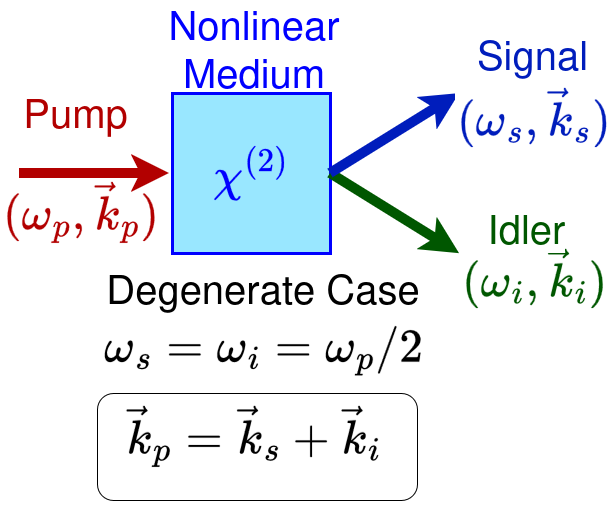}%
    }
    \caption{\justifying Conservation laws in the SPDC process within a $\chi^{(2)}$ medium. 
    (a) Energy conservation diagram ($\omega_p = \omega_s + \omega_i$). 
    (b) Phase matching and wavevector conservation ($\vec{k}_p = \vec{k}_s + \vec{k}_i$).}
    \label{fig:spdc_conservation}
\end{figure}

Momentum conservation, or phase matching ($\vec{k}_p = \vec{k}_s + \vec{k}_i$), determines which optical modes are efficiently amplified in the nonlinear medium \cite{Harris1967OpticalParametricFluorescence}, as illustrated in Fig.~\ref{fig:spdc_conservation}.

In the special case of a degenerate OPO, the signal and idler modes become spectrally and polarization-indistinguishable, oscillating at exactly half the pump frequency ($\omega_s = \omega_i = \omega_p/2$). The dynamics of the emitted field depend critically on the cavity pump power relative to the oscillation threshold:
\begin{itemize}
    \item \textbf{Below threshold ($g < g_{\mathrm{th}}$):} The parametric interaction is insufficient to overcome round-trip cavity losses, yielding squeezed vacuum fluctuations at the output.
    \item \textbf{Above threshold ($g > g_{\mathrm{th}}$):} Optical gain exceeds total losses, and the optical resonator amplifies these initially microscopic zero-point vacuum fluctuations into a macroscopic, bright oscillating coherent field \cite{Louisell1961QuantumFluctuations}.
\end{itemize}

The role of spontaneous emission in an OPO is analogous, in an operational sense, to the role of spontaneous emission in amplified spontaneous emission (ASE) sources (Section \ref{subsubsec:ampl_spont_emiss}). 
In ASE-based QRNGs, spontaneously emitted photons generated in a gain medium are amplified by stimulated emission until they form a macroscopic random-amplitude signal \cite{Williams2010FastPhysical}. 
In an OPO, however, the gain medium is not an inverted atomic medium but a nonlinear parametric medium \cite{Boyd2020Nonlinear, WallsMilburn2008Quantum}. The initiating fluctuations are the zero-point vacuum fluctuations of the signal and idler modes. 
These fluctuations enter the nonlinear interaction through SPDC and are amplified by the cavity gain. 
Thus, the randomness is not produced by detector noise or by a classical instability, but by the quantum uncertainty of the field that seeds the parametric oscillation.

For a degenerate OPO, the gain is phase sensitive. Defining the relative signal phase as
$\theta_s = \phi_s - \phi_p/2 + \pi/4$, the small-signal amplitude gain has period $\pi$ and can be written phenomenologically as $g(\theta_s)=g_{\max}\cos(2\theta_s)$.
Maximum amplification occurs for $\theta_s=0$ and $\theta_s=\pi$, whereas maximum deamplification occurs near $\theta_s=\pi/2$. 
If the maximum gain exceeds the round-trip cavity losses ($g_{\max}>g_{\mathrm{th}}$\footnote{The subscript ``$\mathrm{th}$'' denotes the oscillation threshold.}), the system evolves toward one of two stable phase states separated by $\pi$ \cite{Nabors1990Coherence, Marandi2012}.

The entropy source in OPO-based QRNGs is therefore the random selection of the above-threshold phase state. 
In each generation cycle, the pump is switched off or the cavity is detuned so that the intracavity field decays back to the quantum-noise level.
When the pump is restored, the oscillation grows again from vacuum fluctuations, and the final phase state is newly selected. 
This reset step is crucial: if residual photons from the previous cycle remain in the cavity, they act as a classical seed and force the next oscillation to inherit the previous phase, producing correlations and reducing the available quantum entropy \cite{Marandi2012}.

The binary random bit can be obtained by converting the two possible phase relations into two distinguishable intensity levels. A common architecture uses two independent degenerate OPOs pumped by the same source. Let their output fields be
$E_j = E_0 e^{i\phi_j}$, with $\phi_j \in \{\phi_0,\phi_0+\pi\}$ and $j=1,2$.
After interference at a balanced beam splitter, one output intensity is
\begin{equation}
    I_{\mathrm{out}}
    =
    \frac{1}{2}|E_1+E_2|^2
    =
    I_0\left[1+\cos(\Delta\phi)\right],
    \qquad
    \Delta\phi=\phi_1-\phi_2 .
    \label{eq:opo_interference}
\end{equation}

If the two OPOs choose the same phase, $\Delta\phi=0$, the interference is constructive, and the detected voltage is high. If they choose opposite phases, $\Delta\phi=\pi$, the interference is destructive, and the detected voltage is low. The raw bit is then assigned by thresholding:
\begin{equation}
    b_i =
    \begin{cases}
        1, & V(t_i) > V_{\mathrm{th}},\\
        0, & V(t_i) < V_{\mathrm{th}}.
    \end{cases}
    \label{eq:opo_bit_extraction}
\end{equation}

In the ideal symmetric case, $P(b_i=0)=P(b_i=1)=1/2$, and one random bit is obtained per independent OPO restart. In the all-optical implementation of Marandi \textit{et al.} \cite{Marandi2012}, the random variable is the relative phase between two degenerate OPO pulse trains, and photodetection only reads out the already amplified binary state rather than creating the randomness itself.

The main advantage of OPO QRNGs is that the quantum fluctuation is amplified before electronic detection. The detector therefore does not need to resolve single photons or weak quadrature fluctuations; it only distinguishes between two macroscopic optical states. 
The main limitation is the reset time required to erase the previous intracavity field before the next independent phase selection. 
Shorter cavities, high-repetition-rate pumping, operation closer to threshold, and integrated microresonator implementations are possible routes toward higher bit rates and compact all-optical random number generation \cite{RevModPhys.89.015004, Marandi2012, Steinle2017Unbiased}.

\subsubsection{Raman scattering}
\label{subsubsec:raman_scattering}


Raman scattering is an inelastic light-matter interaction involving the exchange of energy between an optical field and the vibrational modes of a molecular or crystalline medium \cite{raman1928, long2002raman,10.1021/j100207a018, Boyd2020Nonlinear}.
An incident pump photon with angular frequency $\omega_{p}$ interacts with the medium through a virtual intermediate state. If the scattered photon has the same energy as the incident photon, no vibrational excitation is created or annihilated, and the process is referred to as elastic, or Rayleigh, scattering \cite{long2002raman, Boyd2020Nonlinear}.
Conversely, when energy is exchanged with a vibrational mode of the medium, the scattered photon has a frequency different from that of the pump, and the process is classified as Raman scattering \cite{raman1928, long2002raman, 10.1021/j100207a018}.

The vibrational excitations of a solid-state medium are commonly described in terms of phonons \cite{Loudon1964, Agrawal2019Nonlinear}. For a vibrational mode with angular frequency $\Omega_{\mathrm{v}}$, the frequencies of the Stokes and anti-Stokes photons are, respectively,
\begin{equation}
    \omega_{\mathrm{S}} = \omega_{p}-\Omega_{\mathrm{v}}, \quad \omega_{\mathrm{AS}} = \omega_{p}+\Omega_{\mathrm{v}}.
\end{equation}

In the Stokes process, the scattered photon loses energy and a phonon is created in the medium. In the anti-Stokes process, a pre-existing phonon is annihilated, and its energy is transferred to the scattered photon \cite{long2002raman, Bustard2011}. Consequently, the anti-Stokes process requires an initially populated vibrational mode and is generally less probable than the Stokes process under thermal equilibrium conditions \cite{long2002raman}.

Raman scattering can occur either spontaneously or through stimulated emission \cite{long2002raman, Bustard2011, 10.1063/1.4864095}. In spontaneous Raman scattering, the pump field interacts with the medium at an optical power below the stimulated Raman threshold, and individual Raman photons are generated probabilistically \cite{10.1063/1.4864095, Bustard2011, long2002raman}.
The emission time and spectral mode of each detected photon are therefore not predetermined and can be employed as sources of quantum entropy.
For example, Collins \textit{et al.} \cite{Collins2014, Collins2015} generated random bits by separating spontaneous Raman photons into two spectral bins and assigning a binary value according to the detected frequency interval.
Other implementations have simultaneously exploited the wavelength and arrival-time distributions of backward spontaneous Raman photons generated in optical fibers \cite{nie2014practical, Hu2020}.
Stimulated Raman scattering occurs when a pump field and a Stokes seed field satisfy the approximate resonance condition $\omega_{p}-\omega_{\mathrm{S}}\simeq\Omega_{\mathrm{v}}$ \cite{long2002raman, Bustard2011}.

Under this condition, the Stokes field experiences Raman gain: pump photons are converted into additional Stokes photons while vibrational excitations are created in the medium \cite{Bustard2011, 10.1063/1.4864095}. Unlike spontaneous emission into multiple optical modes, stimulated amplification predominantly occurs in the spatial and spectral mode defined by the interacting fields \cite{Bustard2011}.  The resulting Stokes radiation is therefore more intense and directional, allowing its fluctuations to be measured using high-speed photodetectors rather than single-photon detectors \cite{Bustard2011, 10.1063/1.4864095}.

In a spontaneously initiated stimulated Raman process, the initial Stokes field is seeded by quantum vacuum fluctuations \cite{Bustard2011, raymer1990quantum}. These microscopic fluctuations are amplified by the Raman interaction and appear as measurable random variations in the phase or energy of the generated Stokes pulses \cite{raymer1990quantum, Bustard2011, 10.1063/1.4864095}. Quantum random-number generators based on this principle have extracted entropy from the optical phase \cite{Bustard2011, 10.1063/1.4864095} and from pulse-energy fluctuations \cite{Bustard2013}. Thus, depending on the experimental architecture, Raman QRNGs can encode quantum entropy in spectral detection outcomes \cite{Collins2014, Collins2015}, photon arrival times \cite{nie2014practical, Hu2020}, optical phase \cite{Bustard2011, 10.1063/1.4864095}, or pulse energy \cite{Bustard2013}.

In general, the contribution of thermal phonons is determined by the mean thermal occupation number
\begin{equation}
    \overline{n}_{\mathrm{th}} = \frac{1}{\exp\left(\hbar\Omega_{\mathrm{v}}/k_{\mathrm{B}}T\right)-1} \,,
\end{equation}
where for high-frequency optical phonons satisfying $\hbar\Omega_{\mathrm{v}}\gg k_{\mathrm{B}}T$, the thermal occupation is small, where $k_{\mathrm{B}}$ is the Boltzmann constant \cite{long2002raman, Boyd2020Nonlinear}. In this regime, the fluctuations that initiate the Stokes field are dominated by the zero-point fluctuations of the electromagnetic field rather than by thermally populated vibrational modes \cite{raymer1990quantum, Bustard2011, Bustard2013, long2002raman}. These vacuum-seeded fluctuations provide the quantum entropy underlying Raman scattering QRNG architectures \cite{Bustard2011, 10.1063/1.4864095}.

\subsection{Continuous-variable sources}
\label{subsec:continuous_sources}

Continuous-variable QRNG exploit quantum observables whose measurement outcomes belong to a continuous domain \cite{serafini2023quantum}. Instead of producing intrinsically discrete symbols, each quantum measurement yields an analog value determined by the quantum fluctuations of an optical field. Consequently, the measured signal must first be digitized by an ADC, after which suitable randomness extraction techniques are applied to remove residual classical contributions and generate uniformly distributed random bits. This distinguishes CV-QRNG from DV implementations, where each elementary quantum measurement directly produces digital outcomes.

The following subsections review the principal classes of CV-sources employed in QRNG, including vacuum fluctuations, laser phase noise, and amplified spontaneous emission. Although these implementations rely on different physical mechanisms and experimental architectures, they all exploit continuous quantum observables whose irreducible fluctuations originate from the quantum nature of the electromagnetic field. Their high detection bandwidth and compatibility with standard photonic components have made CV architectures the dominant approach for modern high-speed quantum random number generation.

\subsubsection{Vacuum noise}

In modern physics, the notion of the vacuum has undergone a profound conceptual transformation: far from representing an inert ``empty space'', the quantum vacuum is recognized as the lowest-energy eigenstate of the electromagnetic field and, more generally, of every bosonic quantum field, constituting a dynamically active medium endowed with well-defined physical properties \cite{milonni2013quantum}. 
This includes the electric permittivity $\epsilon_{0}$ and magnetic permeability $\mu_{0}$, which together determine the speed of light and fix the vacuum refractive index at $n = 1$~\cite{Battesti2012Magnetic}.

Within quantum field theory, the vacuum behaves as an effectively nonlinear medium under sufficiently intense electromagnetic fields, with predictions of observable optical effects such as vacuum birefringence and photon-photon scattering \cite{Battesti2012Magnetic, King2015Measuring}. Although these effects have not yet been directly confirmed experimentally due to their extreme weakness, they constitute unambiguous predictions of quantum electrodynamics. More directly observable consequences of the non-trivial vacuum structure are the zero-point fluctuations of the field modes and the perpetual creation and annihilation of virtual particle-antiparticle pairs, which underlie several experimentally confirmed phenomena, including spontaneous emission \cite{Dirac1927}, the Lamb shift of atomic energy levels \cite{LambRetherford1947, Bethe1947}, and the Casimir-Polder attraction between closely spaced conducting surfaces \cite{CasimirPolder1948, Lamoreaux1997}. These vacuum fluctuations, spontaneous and unpredictable oscillations of quantized fields arising directly from the probabilistic nature of quantum mechanics, cannot be attributed to any external thermal or classical stochastic source; they are an intrinsic feature of the quantum ground state itself \cite{Riek2015Direct, Appugliese2021Breakdown}.

The single-mode vacuum state, denoted as $\ket{0}$, is formally defined as the eigenstate annihilated by the photon annihilation operator $\hat{a}$, satisfying $\hat{a}\ket{0} = 0$, with an energy expectation value $\langle\hat{H}\rangle = \hbar\omega/2$ that persists even in the complete absence of photonic excitations. This zero-point energy is the physical origin of the vacuum noise directly observable in quantum-optical experiments \cite{Loudon2000, ScullyZubairy1997, WallsMilburn2008}.
The quantum state of each field mode is fully characterized by a pair of canonically conjugate quadrature operators, the amplitude quadrature $\hat{x}_{k}$ and the phase quadrature $\hat{p}_{k}$, defined through the annihilation and creation operators $\hat{a}_{k}$ and $\hat{a}_{k}^{\dagger}$ of mode $k$ as

\begin{equation}
    \hat{x}_{k} = \frac{\hat{a}_{k} + \hat{a}_{k}^{\dagger}}{\sqrt{2}},
    \qquad
    \hat{p}_{k} = \frac{\hat{a}_{k} - \hat{a}_{k}^{\dagger}}{i\sqrt{2}},
    \label{eq:quadratures}
\end{equation}

\noindent
where $\hat{a}_{k}$ and $\hat{a}_{k}^{\dagger}$ satisfy the bosonic canonical commutation relations $[\hat{a}_{k},\hat{a}_{l}^{\dagger}] = \delta_{kl}$ and $[\hat{a}_{k},\hat{a}_{l}] = [\hat{a}_{k}^{\dagger},\hat{a}_{l}^{\dagger}] = 0$, from which the fundamental phase-space commutation relation $[\hat{x}_{k},\,\hat{p}_{l}] = i\,\delta_{kl}$, immediately follows, constituting a direct expression of the Heisenberg uncertainty principle and imposing the lower bound $\Delta\hat{x}_{k}^{\,2}\,\Delta\hat{p}_{k}^{\,2} \geq 1/4$ on the product of quadrature variances for any physical state (see Def. \ref{def:heis}).
For the vacuum state $\ket{0}$, this bound is saturated with the uncertainty equally distributed between both quadratures, $\Delta\hat{x}^{2} = \Delta\hat{p}^{2} = 1/2$, establishing the standard quantum limit below which no classical description can account for the measured fluctuations. This irreducible, symmetrically distributed quantum noise is precisely what is designated \emph{vacuum shot noise}.

\begin{figure}
    \centering
    \includegraphics[width=1\linewidth]{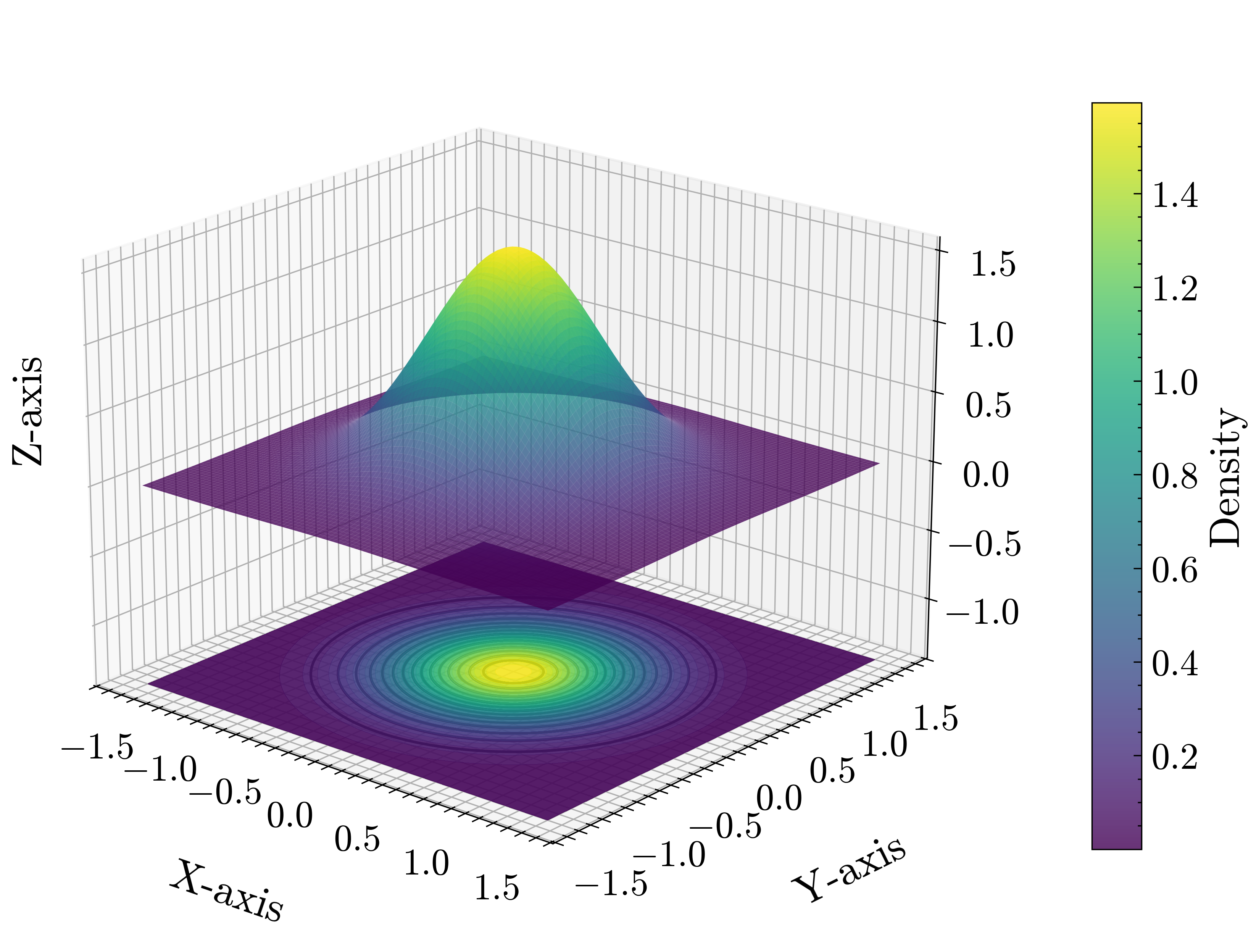}
    \caption{\justifying Combined three-dimensional and two-dimensional representation of the Wigner function Eq.\eqref{eq:marginal} of the electromagnetic vacuum state.}
    \label{fig:wignerfuction}
\end{figure}

The complete phase-space description of the vacuum state is captured by its Wigner quasi-probability distribution described in Def. \ref{def:Wig_quase}. A visual representation of this distribution is illustrated in Fig.~\ref{fig:wignerfuction}.

\begin{definition} \label{def:Wig_quase}
    (Wigner quasi-probability distribution for vacuum state) For any quantum state is defined as
$W(x,p) = \frac{1}{\pi}\int_{-\infty}^{+\infty}\psi^{*}(x+y)\psi(x-y)
e^{2ipy}\,dy$, and which for the vacuum evaluates to the bivariate Gaussian

\begin{equation}
    W_{\ket{0}}(x,p)
      = \frac{1}{\pi}\,\exp\!\bigl[-(x^{2}+p^{2})\bigr],
    \label{eq:wigner_vacuum}
\end{equation}

\noindent
centered at the phase-space origin, with vanishing mean quadrature values $\langle\hat{x}\rangle = \langle\hat{p}\rangle = 0$.
\end{definition}

The vacuum state $\ket{0}$ is, accordingly the canonical example of a pure Gaussian state, defined as any state whose characteristic function and quasi-probability distributions in quantum phase space are Gaussian-valued functions~\cite{Weedbrook2012, adesso2014continuous, bachor2019guide}.
The positive-definiteness of $W_{\ket{0}}(x,p)$ everywhere in phase space indicates classical simulability when restricted to Gaussian measurements, yet the state is genuinely quantum: it saturates the Robertson-Schrodinger uncertainty relation and its zero-point noise is irreducible, as confirmed experimentally through direct electro-optic sampling of vacuum fluctuations~\cite{Riek2015Direct, Appugliese2021Breakdown}.
Integrating Eq.~\eqref{eq:wigner_vacuum} over the momentum quadrature $p$ yields the marginal probability density for any amplitude quadrature measurement outcome

\begin{equation}
    |\psi_{\ket{0}}(x)|^{2}
      = \int_{-\infty}^{+\infty} W_{\ket{0}}(x,p)\,dp
      = \frac{1}{\sqrt{\pi}}\,\exp(-x^{2}),
    \label{eq:marginal}
\end{equation}

\noindent
which is a zero-mean Gaussian distribution with variance $\sigma^{2} = 1/2$ ~\cite{Shen2010,Huang2020}. By rotational symmetry of $W_{\ket{0}}(x,p)$, the identical distribution is obtained for any phase angle $\phi$, \textit{i.e.}, for the generalized quadrature $\hat{x}_{\phi} = \cos\phi\,\hat{x} + \sin\phi\,\hat{p}$.
This result is the cornerstone of vacuum noise QRNGs: any quadrature measurement performed on the vacuum state yields outcomes governed by a continuous, zero-mean Gaussian distribution whose spread is dictated solely by the quantum zero-point fluctuations of the field and is therefore provably of quantum origin by the Robertson-Schrodinger relation~\cite{Gabriel2010, haider2025quantum, Huang2020}.
\begin{figure*}
    \centering
    \includegraphics[width=0.9\linewidth]{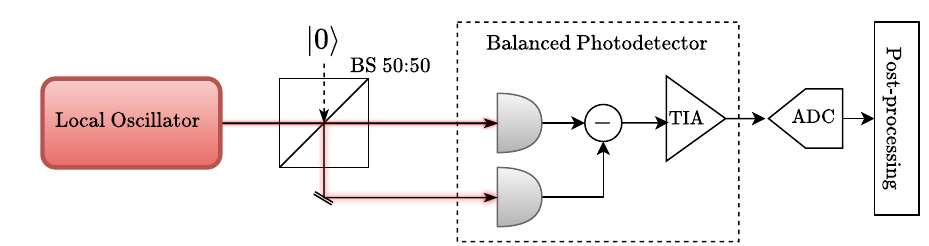}
    \caption{\justifying Conceptual experimental schematic of a vacuum-noise QRNG. A continuous-wave laser acting as a LO (local oscillator) injects a coherent beam into one input port of a balanced $50/50$ BS (beam splitter), while the second port is held in the vacuum state $\ket{0}$. Both output ports are monitored by matched photodiodes whose photocurrents are subtracted, thereby suppressing common-mode classical noise and isolating the vacuum shot-noise component. The difference signal is amplified by a TIA, digitized by an ADC, and forwarded to a classical post-processing module that extracts and certifies the final random-bit sequence.}
    \label{fig:qrng_vacuum}
\end{figure*}

The standard experimental realization of a vacuum noise QRNG, depicted in Fig.~\ref{fig:qrng_vacuum}, employs balanced homodyne detection: A continuous-wave laser acting as a LO injects a strong coherent beam into one input port of a balanced $50:50$ BS, while the second port is held in the vacuum state $\ket{0}$. The two output fields, each carrying an equal superposition of the LO and vacuum contributions, are directed onto a matched pair of photodiodes whose photocurrents are electronically subtracted~\cite{bachor2019guide}.
This subtraction operation rejects all common-mode classical noise present on the LO while constructively reinforcing the vacuum shot-noise contribution, so that the resulting difference photocurrent is dominated exclusively by the quantum fluctuations described by Eq.~\eqref{eq:marginal}.
The photocurrent is amplified by a TIA, band-pass filtered to isolate the shot noise-dominated spectral window, and digitized by an ADC that maps each continuous outcome to an $n$-bit binary word, forming the raw random sequence~\cite{haider2025quantum, Shen2010}.

The total measured signal may be decomposed as $V_{\text{out}} = V_{\text{vac}} + V_{\text{class}}$, where $V_{\text{vac}}$ and $V_{\text{class}}$ denote the vacuum-fluctuation and residual classical-noise components, respectively. Finally, a classical post-processing stage, typically based on universal hashing or Toeplitz matrix extraction (see Section \ref{sec:post}), removes the residual classical contribution $V_{\text{class}}$ and inter-sample correlations introduced by the finite electronic bandwidth, yielding a final bit sequence whose randomness is information-theoretically certified and whose extraction rate is determined by the shot-noise clearance and ADC resolution.

\subsubsection{Phase noise of lasers}

Quantum phase fluctuations in semiconductor lasers originate from spontaneous emission inside the laser cavity, where each spontaneously emitted photon is added to the coherent intracavity field with a random phase, instantaneously perturbing the phase of the total field \cite{Henry1982, Xu2012}. 
Whereas amplitude perturbations are damped by the gain dynamics of the laser, the optical phase is not subject to any restoring force and therefore performs an unbounded random walk, a process known as phase diffusion \cite{Henry1982, Yang2026}. 
The output field of a single-mode laser can be written as
$$E(t) = E_{0}\,\exp\!\big\{ i\left[\omega_{0} t + \theta(t)\right]\big\},$$ 
where $E_{0}$ is the field amplitude, $\omega_{0}$ the central angular frequency, and $\theta(t)$ the fluctuating phase. 
Mathematically, $\theta(t)$ is well described by a Wiener process, whose increments over a time interval $\tau$ are zero-mean Gaussian random variables, $\theta(t+\tau)-\theta(t)\sim\mathcal{N}(0,\,2\pi\Delta\nu\,\tau)$, with a variance that grows linearly in time.

\begin{equation}
    \big\langle \Delta\theta^{2}(\tau) \big\rangle
    = \frac{R_{sp}}{2S}\left(1+\alpha^{2}\right)\tau
    = \frac{2\tau}{\tau_{c}},
    \label{eq:phase_variance}
\end{equation}
where $R_{sp}$ is the spontaneous emission rate, $S$ the mean intracavity photon number, $\alpha$ the linewidth enhancement factor, and $\tau_{c}=1/(\pi\Delta\nu)$ the laser coherence time \cite{Henry1982, Yang2026}. 
In the frequency domain, this diffusion manifests itself as a Lorentzian broadening of the emission spectrum with linewidth $\Delta\nu = R_{sp}(1+\alpha^{2})/(4\pi S)$ \cite{Henry1982}. 
Since $S$ is proportional to the optical output power $P$, the quantum contribution to the phase fluctuations scales as $1/P$, while technical (classical) phase noise is power independent. The variance of the total measured fluctuation can thus be decomposed as
\begin{equation}
    \big\langle \Delta\theta^{2} \big\rangle = \frac{Q}{P} + C,
    \label{eq:quantum_classical}
\end{equation}
where $Q/P$ and $C$ denote the quantum and classical contributions, respectively. 
Operating the laser slightly above threshold therefore maximizes the quantum-to-classical noise ratio and ensures that the detected signal is dominated by fluctuations of quantum origin \cite{Xu2012}.

\begin{figure*}
    \centering
    \includegraphics[width=0.9\linewidth]{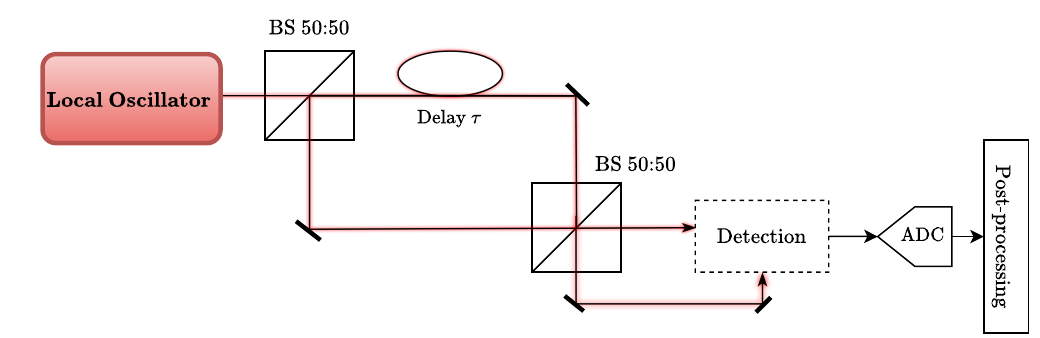}
    \caption{\justifying Conceptual experimental schematic of a phase-noise QRNG. A continuous-wave laser feeds an unbalanced MZI, which converts the phase diffusion accumulated over the delay $T_{d}$ into intensity fluctuations. After recombination, the optical outputs are directed to a balanced detector composed of two photodiodes, whose difference photocurrent is amplified by a TIA. The electrical signal is then digitized by an ADC and delivered to the post-processing stage, where the conditional min-entropy is estimated and a randomness extractor distills the final random bits associated with the quantum phase fluctuations.}
    \label{fig:qrng_Phase_noise}
\end{figure*}

The phase fluctuations cannot be detected directly by a photodiode and must first be mapped onto intensity fluctuations. Two main detection configurations are employed: delayed self-heterodyne interferometry with a single laser and the beating of two independent lasers \cite{Sun2017, Li2024}. In the self-heterodyne scheme illustrated in Fig.~\ref{fig:qrng_Phase_noise}, the continuous-wave beam is split by a BS 50:50 into the two arms of an unbalanced Mach-Zehnder interferometer (MZI); one arm introduces a time delay $T_{d}$, and the fields are recombined at a second BS \cite{Xu2012}. 
The interferometer thereby compares the laser phase at two different times, and, with the phase bias locked at quadrature, the photodetected signal is  $v(t) \propto P \sin\!\big[\Delta\theta(t, \tau)\big] \approx P\,\Delta\theta(t, \tau)$, such that $ \Delta\theta(t, \tau) = \theta(t+\tau)-\theta(t)$ \cite{Xu2012, Yang2026}. 
Increasing $T_{d}/\tau_{c}$ improves the uniformity of the sampled distribution at the expense of the spectral flatness (and hence the bandwidth) of the entropy source, so the delay and the sampling rate must be jointly optimized \cite{Yang2026}.

In the two-laser configuration, two independent distributed-feedback laser diodes emit continuous waves at closely spaced wavelengths, variable optical attenuators set the optical power at the detector and the fields interfere at a 50:50 BS \cite{Sun2017}. 
The photodetected beat signal is $v(t) = A\cos\!\big[2\pi\Delta f\, t + \theta_{1}(t)-\theta_{2}(t)\big],$ where $\Delta f$ is the beat frequency and $\theta_{1}(t)-\theta_{2}(t)$ is the difference of two independent Wiener processes. 
When the sampling period greatly exceeds the coherence times, the phase difference is uniformly distributed over $(-\pi,\pi] $ and the sampled voltage follows an arcsine distribution \cite{Sun2017}. 
This architecture eliminates the long delay line and the active phase stabilization required by the interferometric scheme, which makes it particularly attractive for photonic chip integration, spectral filtering of the beat note has recently extended the entropy source bandwidth of this scheme to 20~GHz \cite{Yang2026}. 
Balanced detection at the BS outputs additionally suppresses common-mode classical intensity noise, so that the differential photocurrent is dominated by the quantum phase fluctuations \cite{RevModPhys.89.015004, Xu2012}.
 
\subsubsection{Amplified spontaneous emission}
\label{subsubsec:ampl_spont_emiss}

Amplified spontaneous emission ({ASE}) originates from the interaction between spontaneous emission and stimulated emission in an optical gain medium operating under population inversion \cite{Siegman1986, Desurvire1994}. Unlike laser emission, which is seeded by an injected optical field, ASE is initiated by spontaneously emitted photons that arise randomly when excited atoms or ions decay to lower energy states. Each emitted photon possesses an intrinsically random emission time, phase, polarization, and propagation direction, reflecting the probabilistic nature of the underlying quantum process \cite{Loudon2000, schranz2019effects}. The amplification mechanism is governed by stimulated emission. Under population inversion, the number of particles in the excited state exceeds that in the lower state, so spontaneously emitted photons stimulate additional emissions while propagating through the gain medium. As a result, the initially weak spontaneous emission is exponentially amplified, producing a broadband optical field known as ASE \cite{Siegman1986, Desurvire1994}.

From the perspective of quantum optics, spontaneous emission originates from the interaction between excited emitters and the vacuum electromagnetic field \cite{Loudon2000, ScullyZubairy1997}. Consequently, the amplified fluctuations observed in ASE ultimately inherit their randomness from quantum vacuum fluctuations rather than from classical technical noise. This quantum origin makes ASE an attractive entropy source for high-speed quantum random number generators, since microscopic quantum fluctuations are converted into a macroscopic optical signal that can be directly detected using conventional photodetectors \cite{duda2023development, Yang2020quantification}.

The quantum origin of ASE is directly reflected in its statistical properties. Since spontaneous emission populates the optical modes independently and photons obey Bose–Einstein statistics, the emitted field exhibits photon bunching, meaning that photons tend to arrive in groups rather than independently. Consequently, the photon-number fluctuations are substantially larger than the shot-noise level expected for coherent laser light, providing a strong quantum signal that can be efficiently exploited for random number generation \cite{duda2023development, MandelWolf1995}.

For a single optical mode, the photon-number distribution follows the Bose–Einstein law,
\begin{equation}
   P(n)= \frac{\bar{n}^n}{(1+\bar{n})^{n+1}} \,,
\end{equation}
where $n$ is the detected photon number and $\bar{n}$ is the mean photon occupation number \cite{MandelWolf1995, Loudon2000}. This distribution has a variance larger than its mean, a direct consequence of photon bunching, in contrast to the Poisson statistics that characterize coherent laser fields \cite{duda2023development, Loudon2000}. As a result, the optical intensity exhibits large stochastic fluctuations whose amplitude is determined primarily by the amplified spontaneous emission process rather than by detector shot noise.

In practical ASE-QRNG, the detected signal results from the superposition of a large number of independent optical modes within the receiver bandwidth \cite{Williams2010, Martin2015, Qi2017incoherent, Wei2017SLED, Yang2020quantification, Guo2021, Huang2021thermalCVQKD}.
According to the central limit theorem, the electrical voltage generated after photodetection can therefore be accurately approximated by a zero-mean Gaussian random variable \cite{Goodman2015}. This Gaussian behavior considerably simplifies digitization using high-speed ADCs and enables efficient estimation of the available min-entropy prior to randomness extraction \cite{duda2023development, schranz2019effects, Guo2021}. The resulting macroscopic intensity fluctuations constitute the entropy source employed in most high-speed ASE QRNGs.

The quantum entropy carried by ASE is most commonly extracted through direct intensity measurements \cite{Williams2010, Argyris2012, Wei2017SLED}. In this configuration, the broadband ASE field is incident on a photodetector, which converts the optical intensity fluctuations into a stochastic electrical current. Since the detected intensity is proportional to the instantaneous photon flux, the amplified photon-number fluctuations are directly mapped onto voltage fluctuations that preserve the underlying quantum randomness of the optical field \cite{Williams2010, duda2023development}.



Several optical gain media have been employed to generate ASE for quantum random number generation, the most common being erbium-doped fiber amplifiers \cite{Williams2010, Martin2015}, semiconductor optical amplifiers (SOAs) \cite{Argyris2012}, and superluminescent light-emitting diodes (SLEDs) \cite{Wei2017SLED, Li2011SLED, Liu2013SLED}. Erbium-doped fiber amplifiers generate ASE by optical amplification in erbium-doped fibers, typically operating around the telecommunications wavelength of 1550 nm. Their broad optical bandwidth, high output power, and compatibility with standard fiber-optic components have made them one of the earliest and most widely adopted entropy sources for ASE QRNGs\cite{Williams2010, Martin2015}. In contrast, SOAs employ semiconductor gain media, providing a compact footprint, fast carrier dynamics, and straightforward integration with photonic integrated circuits. These characteristics have enabled the development of plug-and-play QRNGs and high-speed integrated architectures \cite{duda2023development, Guo2021}. Superluminescent light-emitting diodes constitute a third important class of ASE sources. Unlike conventional laser diodes, their cavity design strongly suppresses optical feedback, preventing laser oscillation while maintaining high optical gain. Consequently, SLEDs emit broadband, low-coherence ASE with relatively flat spectral profiles, making them particularly suitable for wavelength-division multiplexing and spectral-slicing techniques, where independent spectral channels are exploited for parallel random-number generation \cite{Guo2021, schranz2019effects}. Although all of them rely on the same physical principle, they differ in gain mechanism, spectral characteristics, integration capability, and achievable generation rates. 

The choice of the active medium ultimately depends on the target application. Erbium-doped fiber amplifier systems are preferred in fiber-based laboratory systems because of their high optical power and mature telecommunications technology, whereas SOAs and SLEDs are generally favored for compact and integrated implementations requiring high bandwidth and low power consumption. Regardless of the selected platform, the quantum entropy originates from the same spontaneous-emission process, while the active medium primarily determines the optical power, bandwidth, integration level, and scalability of the QRNG architecture \cite{duda2023development, Guo2021}.


\section{Entropy Estimation} \label{sec:Est_Ent}

Entropy plays an important role in the probability distributions of random variables \cite{cover1999elements}. In his seminal work, Shannon (1948) discussed the basis of information theory conceptualizing entropy as a metric for the unpredictability of a given random variable \cite{Shannon1948}, defined as

\begin{equation} \label{eq:shannon}
    H(X) = - \sum_{x \in \mathcal{X}} p_X(x) \log_2(p_X(x))\,,
\end{equation}
where $X$ is a random variable with distribution $p_X$, such that $p_X$ is the probability of $x$ being observed considering an alphabet $\mathcal X$ with cardinality $N$. In a more operational picture\footnote{A quantity is considered operational when defined by a real task of information processing. \cite{Konig2009}.}, entropy corresponds to the average number of bits necessary to describe the random variable $X$ \cite{cover1999elements}. 

Since Shannon entropy corresponds to an average metric in the number of events, it provides a worst-case bound only in the asymptotic limit \cite{tomamichel2015quantum, Tomamichel5319753}. For non-asymptotic scenarios, the Rényi entropy offers a natural generalization by defining a family of nonadditive functions parameterized by $\alpha$, 
\begin{equation}
H_{\alpha}(X) = \frac{1}{1-\alpha} \log_2 \left( \sum_{x \in \mathcal{X}} p_X(x)^\alpha \right) \,,
\end{equation}
where $\alpha$ recovers eq. \eqref{eq:shannon} in the limit $\alpha \rightarrow 1$ \cite{renyi1961measures}. This definition has operational relevance in the limit $\alpha \rightarrow \infty$, since it establishes a lower bound for the amount of randomness available in a probabilistic distribution, also called min-entropy \cite{Renner2004}. This implies that an eavesdropper could guess $x$ with probability $2^{- H_{\mathrm{min}}}$ (see Example \ref{ex:min_ent}). 

\begin{definition} \label{def:min_ent}
    (Min-entropy) Let $X$ be a random variable with a probability distribution $p_X(x)$, such that $p_X(x)$ is the probability of $x$ being observed in an alphabet $\mathcal{X} \equiv \{0, 1\}^n$. Its minimum entropy is given by    
    \begin{equation}
        H_{\mathrm{min}}= - \log_2 \left [\max_{x \in \mathcal{X}} \ p_X(x) \right ] \,.
    \end{equation}
\end{definition}

\begin{example}{Min-Entropy}{min_ent}
Let $X$ be a random variable that takes values in the alphabet $\mathcal{X} = \{x_1, x_2, x_3, x_4\}$, where:
\begin{align*}
x_1 &= 00, & p_X(x_1) &= 1/4 \\
x_2 &= 01, & p_X(x_2) &= 1/2 \\
x_3 &= 10, & p_X(x_3) &= 1/8 \\
x_4 &= 11, & p_X(x_4) &= 1/8
\end{align*}

Using Def. \ref{def:min_ent}, one can compute
 \begin{align*}
        H_{\mathrm{min}} & =  - \log_2 \left [\ p_X(x_2) \right ] \\
        & =  - \log_2 \left [ 1/2 \right ] = 1 \text{ bit} \,.
\end{align*}
    
The probability of an eavesdropper guessing the value of $x$ in the worst case is given by $2^{-H_{\mathrm{min}}(X)} = 2^{-1} = 0.5$. Therefore, it is possible to extract, at most, 1 bit of true randomness by symbols. 
\end{example}

Min-entropy is highly used in the context of secrecy instead of Shannon entropy, because a secret can have high Shannon entropy yet be highly vulnerable if one value dominates the distribution \cite{Smith2009}. Viewing secrecy as a resource that is gradually consumed by a system’s observable outputs, the average posterior vulnerability gives rise to the average min‑entropy leakage \cite{6957119, ESPINOZA201357}. This static measure is non‑negative, monotonic under channel cascades, and compositionally well‑behaved, making it a practical tool for quantifying information flow in probabilistic systems such as anonymity protocols or side‑channel attacks \cite{kawamoto2017compositionality}. In the quantitative information flow, the min-entropy relative to the QRNG outputs can be seen as the quantity of secrecy available for cryptographic tasks in information-theoretic channels. 

However, Definition \eqref{def:min_ent} does not account for practical imperfections in the quantum system. In realistic devices, the state cannot be prepared as a pure state, and the measurements are never perfect because of inefficient detectors \cite{frauchiger2013truerandomnessrealisticquantum}. Rather, the system is a mixed state, and the measurements are no longer projective but act like a Positive-Operator Valued Measure (POVM) \footnote{Neumark’s extention of the POVM provides an theoretical approach for treating a POVM as a projective measurement on a product space by considering an additional space \cite{Peres1990}.} \cite{DAriano_2005}. Therefore, the side-information must be considered in this metric. 


In a rigorous security scenario, Eve is limited only by quantum theory itself. Since unpredictability is guaranteed within all possible extensions of the theory \cite{Colbeck2011}, adopting this attack strategy ensures a general framework for entropy estimation \cite{Tomamichel2010}. This implies a quantum-side information approach, where Eve no longer holds a  set of classical bits but instead a purifying quantum system $E$ \footnote{For classical side information, the same approach can be adopted by replacing the density operator with probability distributions.}. Before applying the hash function, this situation can be characterized conveniently by a classical-quantum state 
\begin{equation}
    \rho_{XE} = \sum_{x \in \mathcal X} p_X (x) \ket{x} \bra{x} \otimes \rho_E^x \,,
\end{equation}
such that $\{\ket{x}\}_x$ is a family of mutual orthogonal vectors representing the classical values of $X$ and $\rho_{XE}$ is defined on the Hilbert space $\mathcal{H}_{XE} = \mathcal{H}_X \otimes \mathcal{H}_E$, where $X$ belongs to $\mathcal{H}_X$ and $\mathcal{H}_E$ is a space to which only $E$ has access. 

The conditional min-entropy, given in Def. \eqref{def:cond_min_ent}, provides a theoretical lower bound for the amount of independent randomness present in the quantum process by including the quantum side information in the state $\rho_E^x$. Also, note that even though it accounts for side information, there is no need to model the side information explicitly in this definition. This is relevant since providing a complete and explicit model for all side-information present in a realistic device can be challenging \cite{frauchiger2013truerandomnessrealisticquantum}. 

\begin{definition} \label{def:cond_min_ent}
    (Conditional min-entropy) Let $\sigma_E \in \mathcal{H}_E$ be a reduced state of $E$, the minimum entropy of $X$ conditioned on $E$ is then defined as
    \begin{equation}
        H_{\mathrm{min}}(X|E) = -\inf_{\sigma_E} H_{\mathrm{min}} (\rho_{XE} || \mathrm{id}_X \otimes \sigma_E) \,,
    \end{equation}
    or 
    \begin{equation}
        H_{\mathrm{min}}(X|E) = \sup_{\sigma_E} \log(\lambda) \,,
    \end{equation}
    where $\lambda$ is the smallest real number such that $\lambda \cdot \mathrm{id}_X \otimes \sigma_E - \rho_{XE}$ is non-negative. 
\end{definition}

Conditional min-entropy satisfies some important properties of information theory. For instance, the data processing inequality plays an important role in entropy estimation, as it can be used to show that no clever manipulation of data can improve the inferences that can be made from the data \cite{cover1999elements}. In other words, removing side information never reduces the entropy, such that for any two systems $E$ and $E'$ \cite{10.5555/2230996.2231000}, we have
\begin{equation}
    H_{\mathrm{min}}(X|EE') \leq H_{\mathrm{min}} (X|E)  \,.
\end{equation}

However, both min- and max-entropy are discontinuous: an arbitrarily small modification of the system's state may induce an arbitrarily large change in its entropy (see Example~\ref{ex:desc_min_ent}). This deficiency is addressed by the framework of smooth entropies, whereby the underlying entropy is optimized over a ball of states $\rho'$ that are $\epsilon$-close to $\rho$ in purifying distance~\cite{tomamichel2015quantum}. The smooth min-entropy is monotonically increasing in $\epsilon$ and inherits all the properties of the min-entropy, as established in Ref.~\cite{renner2006securityquantumkeydistribution}.

\begin{example}{Discontinuity at maximum entropy}{desc_min_ent}
    Let $X$ be a random variable that takes values in the alphabet $\mathcal{X} = \{0, 1, \dots, n-1\}$ with the following probability distribution:
    \begin{itemize}
        \item $P_X(0) = P_X(1) = \frac{1-\delta}{2}$.
        \item $P_X(x > 1) = \frac{\delta}{n-2}$, such that $\alpha > 0$. 
    \end{itemize}
    
    By definition
    \begin{equation*}
    H_{\max}(P_X) = \log n \,. 
    \end{equation*}
    
    If we slightly modify the distribution for $\bar{P}_X(x > 1) = 0$, so $H_{\max}(\bar{P}_X) = 1$. The distance between both distributions is small $\|P_X - \bar{P}_X\|_1 \leq \delta $, but the impact on the maximum entropy is significant:
    \begin{equation*}
    H_{\max}(P_X) \gg H_{\max}(\bar{P}_X) \quad \text{for } n \gg 1 \,.
    \end{equation*}
\end{example}

Therefore, to achieve higher security in estimating the number of genuinely random bits from an entropy source, the conditional minimum entropy should be replaced by its smoothed version \cite{Tomamichel2010}, as set out in Definition \ref{def:min_ent_sua}. 

\begin{definition} \label{def:min_ent_sua}
    (Conditional smooth min-entropy) Let $\rho_{XE}$ be a bipartite density operator and $\epsilon \geq 0$. The conditional smooth min-entropy of $X$ conditioned to $E$ is
    \begin{equation}
        H_{\mathrm{min}}^\epsilon (A|B)_\rho := \sup_{\rho'} H_{\mathrm{min}} (A|B)_{\rho'} \,,
    \end{equation}
    where $\rho' (A|B)$ is the set of states $\epsilon\text{-closed}$ of $\rho_{AB}$, such that
    \begin{equation}
        ||\rho_{AB} - \rho'_{AB}|| \leq \epsilon \\.
    \end{equation}
    
\end{definition}

In any case, the task of computing conditional (smooth) min-entropy is far from trivial. Indeed, estimating the min‑entropy of a distribution generated by an efficient classical sampler is proven to be SBP‑complete, implying that no polynomial‑time classical algorithm can approximate it to within an additive constant unless complexity classes collapse \cite{Watson2016}. Since the measurements in the entropy source are modeled by a classically sampled distribution, rigorously and tightly bounding the available (smooth) min‑entropy is computationally infeasible. Thus, practical QRNG implementations cannot rely on exact or finely resolved entropy estimates derived from such models, but must instead adopt provably conservative lower bounds. Computational methods try to avoid this complexity barrier by adopting practically feasible methods for relevant entropy estimation tasks. The use of semidefine programming (SDP) is a well-known example, where a convex optimization relative to some constraints is performed \cite{Brown2021, tomamichel2015quantum}. Similarly, efficient min‑entropy estimators based on collision entropy ($\alpha = 2$) are proposed \cite{9393401}. While not universal polynomial‑time algorithms in the worst‑case sense, these estimators perform very well on realistic random‑number sources. 

In general, the most direct way to compute min-entropy is by knowing the probability distribution of the measurements relative to a quantum process. Fortunately, quantum theory provides a well-known description of these distributions \cite{dirac1935principles}, such as gaussian quantum information \cite{Weedbrook2012} and discrete-event measurements \cite{nielsen2010quantum}. Also, there are theoretical frameworks that compute the min-entropy using the POVM relative to the measurements\cite{frauchiger2013truerandomnessrealisticquantum}. While this is a more direct approach for discrete-variable sources, it can be challenging in the continuous case.  Thus, practice QRNG implementations can obtain a good metric of independent randomness by leveraging these feasible techniques, even though a perfectly general classical estimator provably cannot exist.





\section{Randomness Certification} \label{sec:Cert_al}

The core advantage of QRNGs lies in the intrinsic probabilistic nature of quantum mechanics, which ensures that suitably chosen quantum measurements can produce outcomes that are intrinsically unpredictable. However, as discussed before, cryptographic RNGs face a critical problem of trust. 
For instance, historical precedents like the \textit{Dual Elliptic Curve Deterministic Random Bit Generator} (Dual--EC--DRBG) \cite{johnson2004x9} show that even standardized algorithms may contain cryptographic trapdoors that enable attackers with the required secret information to recover internal states and predict subsequent outputs from a small amount of observed data \cite{checkoway2014practical}. In response to security concerns and a loss of public confidence, NIST removed Dual--EC--DRBG from its random-number generation recommendations in 2014 and advised users to migrate to alternative approved algorithms \cite{nist2014dualec}.

At the hardware level, the threat of hardware trojans and malicious modifications such as dopants inserted into circuits can introduce hard-to-detect errors in real-world RNGs \cite{tehranipoor2010survey, becker2014stealthy}. Furthermore, physical components are subject to spontaneous failure or degradation, which may remain unnoticed while the device continues to produce output \cite{RevModPhys.89.015004}.

To mitigate these risks, certification of quantum randomness has emerged as a method to provide an experimental signature that guarantees the quantum origin and unpredictability of the data, moving beyond simple statistical suites like NIST SP 800-22 \cite{NIST_Website}. From a conceptual standpoint, certification methods can be organized into three distinct layers based on the level of trust assigned to the devices:

\begin{itemize}
    \item \textbf{Trusted device or device-dependent:} These terms are commonly used as equivalent terms to describe protocols whose security relies on a detailed and trusted model of the physical devices. These methods assume the hardware is perfectly calibrated and operates exactly according to the theoretical physical model. While practical and simple to implement, they are vulnerable to implementation loopholes or undetected hardware manipulations \cite{RevModPhys.89.015004,mannalatha2023review}. The downside is that such experiments are extremely sensitive and that all relevant loopholes must be properly addressed for the system to be considered reliable. Since the devices are fully trusted and characterized, randomness is not certified from the observed statistics in the sense discussed throughout this section.
    
    \item \textbf{Self testing or device-independent:} Although these terms are also sometimes used interchangeably in the QRNG literature, \emph{device-independent} specifies the trust model, whereas \emph{self-testing}, in its strict sense, refers to the stronger ability to infer the underlying quantum state and measurements from the observed correlations. This represents the most robust security tier, requiring no internal model of the devices. Certification is derived directly from the observation of nonlocality through Bell tests in spatially separated systems \cite{Pironio2010}. This ensures security even against an adversary who controls the hardware, though it demands sophisticated experimental infrastructure \cite{Acin2016}.

    \item \textbf{Semi-Self testing:} Similarly to the previous cases, the terms \emph{semi-device-independent} and \emph{semi-self-testing} are sometimes used as synonyms, although they emphasize different aspects of randomness certification. Semi-device-independent describes a trust model in which the devices are not fully characterized but must satisfy a weak physical assumption, such as a bound on dimension, energy, mean photon number, or state overlap. Since this term is also used for distinct architectures, including source device-independent and measurement device-independent, we adopt \emph{semi-self-testing} to emphasize protocols that use the observed input-output statistics to quantify randomness under an explicit physical assumption. With that said, the class of semi-self-testing occupy a middle ground between the other two. The idea is to analyze this feature modeled by a linear functional $W$ and check if a pre-defined classical bound $W^{(C)}$ is crossed or not (See Example \ref{ex:certification} to see the description of randomness certification based on dimension witness). 

\end{itemize}

\begin{example}{Randomness certification with dimension witness}{certification}
In a prepare-and-measure scenario, Alice sends a state $\rho_x$ and Bob measures it with setting $y$, obtaining outcome $b$, so that
\begin{equation}
    p(b|x,y)=\operatorname{Tr}\!\left(\rho_x M_{b|y}\right).
\end{equation}
Assuming the communicated system has dimension at most $d$, these probabilities define a dimension witness,
\begin{equation}
    W=\sum_{b,x,y}c_{bxy}\,p(b|x,y),
\end{equation}
and violating the classical bound, $W_{\mathrm{obs}}>W_d^{(C)}$, rules out any classical strategy of dimension $d$ \cite{brunner2008testing,gallego2010device}.

The certified randomness follows from maximizing the adversary's guessing probability over all dimension-$d$ quantum strategies compatible with $W_{\mathrm{obs}}$, giving the min-entropy
\begin{equation}
    H_{\min}(B|X,Y,E)
    =
    -\log_2 P_{\mathrm{guess}}^{\star}(W_{\mathrm{obs}},d).
\end{equation}
Hence $P_{\mathrm{guess}}^{\star}<1$ certifies private randomness with no full device characterization, provided the dimension assumption holds.
\end{example}

Note that there is a conceptual difference between certification and randomness resting with statistical validation. To validate statistically means, for example, showing that a data sequence is unpredictable when read bit by bit (see Section \ref{sec:Test}). However, if an attacker has prepared the list in advance, they are able to fully model the internal workings of the device generating it, then even if the list appears random, it will not be. We are considering an extreme case here, in which the eavesdropper can fully manipulate the QRNG or even be the very provider of the device itself. In this case, certification serves as an unequivocal way of verifying whether the system operates as claimed, \textit{i.e.}, whether it is, in fact, a quantum system.

Ultimately, the goal of these certification protocols is to establish a bound on the adversary's guessing probability ($P_{guess}$), which is then used to calculate the certified min-entropy \cite{Konig2009} and the bit rate generation (bit/s). Alongside well-established discrete-variable approaches, recent studies have investigated quantum randomness certification using continuous-variable and hybrid optical platforms, including displacement-based photodetection, homodyne and heterodyne detection. These measurement schemes offer complementary advantages and limitations in terms of efficiency, scalability, modeling assumptions, and experimental complexity. In the next parts of this section, we discuss both families and compare their main certification strategies.

\subsection{Discrete-variable setup}

The theoretical foundation for certifying randomness without trusting the internal workings of the devices began with the pioneering thesis of Roger Colbeck \cite{Colbeck2006}, who proposed the first self-testing randomness expansion protocol based on quantum nonlocality (see Section \ref{sec:enhance} for a more detailed description of randomness expansion). This conceptual breakthrough was experimentally realized in 2010 \cite{Pironio2010} by Pironio \textit{et al.}, who provided the first theoretical and experimental demonstration of a self testing QRNG certified by the violation of a Bell inequality using trapped ions: by heralding entanglement between two trapped $^{171}$Yb$^+$ ions and closing the detection loophole, they certified approximately 42 random bits directly from the observed violation.

Following these initial demonstrations, Vazirani and Vidick (2011) formalized the security of these protocols against general quantum adversaries, proving the possibility of exponential randomness expansion \cite{vazirani2011certifiable} . The introduction of the Entropy Accumulation Theorem by Arnon-Friedman \textit{et al.} (2018) provided a robust mathematical framework for composable security proofs in realistic scenarios \cite{arnon2018practical}. In the same year, Bierhorst \textit{et al.} \cite{bierhorst2018experimentally} executed a strictly loophole-free Bell test, closing both the detection and locality loopholes simultaneously, using a photonic setup with measurement stations separated by 187 meters, they extracted 1024 certified random bits from about 55 million experimental trials collected over roughly ten minutes of data acquisition.

Despite these interesting results, fully device-independent implementations remain experimentally demanding due to the strict requirement of closing all loopholes simultaneously and generating high-quality entangled states \cite{Acin2016,zhang2020experimental}. To bridge the gap between rigorous security and practical implementation, semi-self-testing models emerged around 2010--2011, driven by works such as Gallego \textit{et al.} \cite{gallego2010device} and Li \textit{et al.} \cite{Li2011}. These protocols typically operate in a Prepare-and-Measure (P\&M) scenario, replacing the stringent need for nonlocality with weaker, verifiable physical assumptions.

Within this scope, in the semi-self-testing regime, randomness is certified by evaluating a physical property that acts as a signature of quantum behavior. For instance, a dimension witness operates by restricting the dimension of the prepared quantum state (\textit{e.g.}, $d \leq 2$ for qubits)~\cite{brunner2008testing,gallego2010device,Li2011}. In a practical setup, Alice prepares a state from a finite set of options and Bob measures it. If the observed probability distribution yields a functional value $W$ that exceeds the classical bound $\beta$ ($W > \beta$), it proves that the correlations cannot be explained by any classical model restricted to that dimension. This violation directly certifies the presence of quantum randomness, bounding the adversary's guessing probability. A major advantage of dimension-bounded protocols is their compatibility with experimentally accessible discrete-variable systems, such as polarization and time-bin qubits. Such protocols can achieve certified generation rates in the kilobit-per-second regime using relatively simple optical setups \cite{lunghi2015self}.

However, strictly bounding the dimension of a system can be challenging in real physical systems. This has motivated the development of alternative frameworks based on experimentally accessible constraints, such as bounds on the energy or the overlap between prepared states \cite{VanHimbeeck2017,Brask2017,Rusca2019}. Instead of restricting the Hilbert-space dimension, these approaches impose a bound on a physical observable, such as the mean photon number of the optical pulses emitted by the source. Randomness can then be certified from the observed input-output correlations under this energy constraint. This assumption is naturally compatible with standard optical components, including attenuated lasers, photodetectors, and homodyne detectors. A subsequent homodyne-based implementation by Rusca \textit{et al.} (2020) achieved a certified random-bit rate of $145,5$ Mbit/s, illustrating the potential of energy-bounded protocols for practical high-rate generation \cite{Rusca2020}. Another experimentally relevant approach was introduced by Brask \textit{et al.} (2017), who certified randomness from the limited distinguishability of two nonorthogonal states. Their optical implementation achieved a certified generation rate of $16,5$ Mbits/s \cite{Brask2017}. These systems are discussed in more detail in the following subsection. 

In addition to property-bounded semi-self-testing, quantum steering offers an alternative intermediate trust model by introducing a one-sided device-independent architecture. In this asymmetric bipartite scenario, one party (\textit{e.g.}, Alice) is treated as an uncharacterized black box, while the other party (Bob) is trusted and capable of performing full state tomography. Quantum randomness is certified whenever Bob's measured states cannot be explained by any local hidden state model, demonstrating that Alice has genuinely steered his quantum system \cite{wiseman2007steering}. Theoretical foundations were established around 2014-2015 in \cite{law2014quantum, passaro2015optimal} where the authors formulated SDP frameworks to compute lower bounds on min-entropy under steering constraints. Experimental steering-based randomness certification was explored by Máttar \textit{et al.} (2017) using a multipartite photonic three-qubit state \cite{mattar2017experimental}. In 2022, Joch \textit{et al.} demonstrated the first generation of certified random bits in a steering scenario with the detection loophole closed \cite{joch2022certified}. In the continuous-variable regime, Ioannou \textit{et al.} (2022) certified randomness using squeezed states and homodyne measurements, estimating an extractable rate of approximately 70 kbit/s from experimental data and projecting that dedicated implementations could reach the Gbps regime \cite{ioannou2022steering}.

Complementing these methods, quantum contextuality provides a distinct mechanism for certifying randomness within a single quantum system without requiring bipartite entanglement or spatial separation. Based on the Kochen-Specker theorem \cite{kochen1967}, contextual certification relies on the impossibility of pre-assigning measurement outcomes independently of the choice of other compatible, co-measured observables. By evaluating non-contextual inequalities, such as the Klyachko-Can-Binicioglu-Shumovsky (KCBS) inequality \cite{klyachko2008simple}, on single multi-level systems (\textit{e.g.}, qutrits with $d \geq 3$), any experimental violation bounds the adversary's guessing probability ($P_{\text{guess}}$) directly.

Experimental feasibility was demonstrated in 2013 by two complementary implementations: Um \textit{et al.}\ used a single trapped $^{171}\mathrm{Yb}^{+}$ qutrit to close the detection-efficiency loophole in a contextuality-based randomness experiment, certifying $2.4\times10^{4}$ bits of min-entropy with $99\%$ confidence \cite{Um2013,Um2013Corrigendum}, while Deng \textit{et al.}\ employed a photonic qutrit to certify at least 5,246 net random bits \cite{Deng2013}. Later works pursued distinct theoretical and experimental directions. Um \textit{et al.}\ (2020) closed the detection-efficiency loophole with a modified KCBS inequality on a single trapped $^{138}\mathrm{Ba}^{+}$ ion, extracting $5.28\times10^{5}$ random bits at $270~\mathrm{bit/s}$ \cite{Um2020}; Pan (2021) proposed theoretical prepare-and-measure protocols based on Mermin's magic-square and magic-star proofs, certifying two and three bits of randomness, respectively \cite{Pan2021}; and Genzini \textit{et al.}\ (2026) implemented a KCBS-based semi-self-testing QRNG on two silicon-photonic chips with a heralded single-photon source, achieving an asymptotic certified rate of $21.7\pm0.5~\mathrm{bit/s}$ \cite{Genzini2026}.

Overall, quantum randomness certification encompasses several complementary approaches with distinct trust assumptions. Fully device-independent protocols infer randomness from Bell nonlocality, whereas semi-self-testing schemes rely on explicit physical constraints, such as dimension, energy, or state overlap. Steering-based protocols introduce an asymmetric trust model, in which only one party is characterized, while contextuality enables certification within a single quantum system through the violation of noncontextuality inequalities. These approaches therefore offer different trade-offs between security, experimental complexity, detection requirements, and generation rates, rather than forming a strict hierarchy of practical performance.

There is also the prospect of unifying these different frameworks through a common mathematical formulation, possibly based on quantum speed limits \cite{jones2026certified}. This line of research seeks to identify principles or symmetries shared by different physical constraints, such as dimension, energy, and distinguishability, allowing them to be described within a unified framework \cite{jones2026theory,manos2025extrapolation}.

\subsection{Continuous-variable setup}
\label{cv_qrnc}

Before discussing continuous-variable quantum random-number certification, an important distinction concerning the trust model must be made. A substantial fraction of the highest-rate certified CV-QRNG implementations follows the source-device-independent (Source-DI) approach. In this framework, the optical source may be uncharacterized or even controlled by an adversary, while the continuous-variable measurement device remains trusted and characterized \cite{Marangon2017,avesani2018source,PhysRevX.10.041048,bertapelle2025high,cizauskas202633}. 

Although Source-DI implementations represent prominent advances in CV-QRNGs, with some achieving Gbps generation rates, they are not the main focus of this section. As discussed previously, our primary interest lies in semi-self-testing protocols, in which the observed input-output statistics are used to quantify randomness under an explicit and limited physical assumption. In CV and hybrid optical platforms, such assumptions may involve the energy or mean photon number of the preparations, their distinguishability, or an effective restriction on the communicated system. We therefore emphasize protocols that combine homodyne, heterodyne, or displacement-based measurements with this operational form of self-testing.

Within this scope, continuous-variable semi-self-testing remains a comparatively recent and less explored direction, with only a few works identified in our mapping. One possible reason is that transferring Bell-based or dimension-bounded protocols to CV platforms is not straightforward. Although several Bell inequalities for continuous or infinite-dimensional systems have been developed \cite{Chen2002,Cavalcanti2007,Lee2009,Martin2016}, converting these constructions into practical randomness-certification protocols still requires suitable theoretical and experimental techniques. Similar difficulties arise for dimension witnesses because optical states generally occupy infinite-dimensional Fock spaces. Recent approaches have begun to address this issue by combining explicit physical constraints with discretized homodyne and displacement-based measurements \cite{alves2026semi}. However, such approaches still exhibit a hybrid character, combining continuous-variable optical measurements with discrete outcomes and finite-dimensional assumptions.

Some of the implementations we found in our mapping employ coherent-state preparations and quadrature measurements in prepare-and-measure configurations. Rusca \textit{et al.} (2020) and Avesani \textit{et al.} (2021) certify randomness under an upper bound on the energy of the preparations, using homodyne and heterodyne detection, respectively. These implementations achieved certified rates of $145.5~\mathrm{Mbit/s}$ and more than $113~\mathrm{Mbit/s}$, with the heterodyne scheme additionally enabling phase-drift compensation through post-processing \cite{Rusca2020,Avesani2021}. Wang \textit{et al.} (2023) considered a complementary trust model, in which the coherent preparations are characterized while the homodyne detector remains untrusted. Including finite-size effects, quantum side information, and non-IID measurement behavior, their implementation achieved a net randomness expansion rate of $4.98~\mathrm{kbit/s}$ over $10^{10}$ rounds \cite{Wang2023}.


For the methods based on quantum steering, Yi Li \textit{et al.}\ (2024) extended steering-based randomness certification to multipartite configurations and systems of arbitrary dimension, demonstrating the versatility of this approach for asymmetric quantum networks \cite{li2024randomness}. In 2025, Zhang \textit{et al.}\ experimentally implemented the continuous-variable route by distributing a Gaussian entangled state over a $2$-km fiber channel, achieving a certified randomness generation rate of $7.06~\mathrm{Mbit/s}$ with only the local station trusted \cite{zhang2025one}.

\section{post-processing} \label{sec:post}

Raw data produced by QRNGs inevitably contains not only the quantum entropy generated by the physical process, but also classical noise, bias, correlations, and other imperfections introduced by the experimental apparatus and measurement inefficiencies. These contributions may lead to deviations from true randomness, making post-processing an essential stage in any practical QRNG implementation. As illustrated in Fig.~\ref{fig:QRNG_Bulding_blocks}, post-processing aims to transform the raw experimental data into a sequence of bits that is statistically indistinguishable from a truly uniform distribution. This is achieved by suppressing residual bias, correlations, and classical patterns, while preserving the genuine quantum randomness contained in the raw data \cite{PhysRevA.87.062327}.

\subsection{Leftover Hash Lemma}\label{post_sub_A}

The task of post-processing is to extract from $X$ a $Z$ part that is completely unknown to the eavesdropper using the min-entropy as a conservative, worst-case bound for the amount of true randomness available. This requires classical processing, which is usually achieved by applying universal hash functions. The suitability of this technique, introduced in 1979 by Carter and Wegman \cite{10.1145/800105.803400}, stems precisely from its quasirandomness properties \cite{10.1007/3-540-48405-1_3}, which ensure that the extracted output is statistically close to uniform.

A $(D; N, M)$-hash family is defined as a set $\mathcal{F}$ of $D$ functions, where each $f \in \mathcal{F}$ maps $f : X \to Z$ \cite{stinson2001universal}. Such a family is said to be $\delta$-universal if, for any two distinct elements $x_1, x_2 \in X$, the number of functions $f \in \mathcal{F}$ that cause a collision by mapping both inputs to the same output, \textit{i.e.}, $f(x_1) = f(x_2)$, is at most $\delta D$. The parameter $\delta$, which is commonly called the collision probability, thus bounds the fraction of functions that fail to distinguish any given pair of distinct inputs. When $\delta := 1/|\mathcal{Z}|$, it is said that it is a particular case of a hash family, called two-universal hashing (see Definition \ref{def:two}). 

\begin{definition} \label{def:two}
(two-universal hashing) A family of functions $\mathcal{F} \subseteq \{ f \mid f : \mathcal{X} \to \mathcal{Z} \}$ is called {two-universal} if, for every pair of distinct inputs $x, x' \in \mathcal{X}$ and for a function $f$ chosen uniformly at random from $\mathcal{F}$, the probability of a collision satisfies 
\begin{equation}
\Pr_{f \in_R \mathcal{F}}\big[ f(x) = f(x') \big] \;\leq\; \frac{1}{|\mathcal{Z}|}.    
\end{equation}
The threshold $\delta = 1/|\mathcal{Z}|$ appearing on the right-hand side is exactly the collision probability obtained when $\mathcal{F}$ is taken to be the family of \emph{all} 
possible functions mapping $\mathcal{X}$ into $\mathcal{Z}$. Hence, a random hash function from the family must be no more  likely to cause a collision than a truly random function with the same range.
\end{definition}

In universal hash families, quaserandomness means that the application of a hash family $f: X \rightarrow Z$ take the values of $X$ to an almost uniform space $Z$ if the source has enough entropy \cite{10.1007/3-540-48405-1_3}.  In case the process consume the entropy source with $|Z| \leq |X|$, this is called randomness extraction \cite{NISAN1999148}. Since it is impossible to generate a perfect random sequence with a finite number of samples, we can make use of this properties to quantify the statistical distance between $Z$ and a uniform $u_Z$ \cite{stinson2001universal}. The key question to be answered in this task is whether it is possible to quantify this degree of non-uniformity for future corrections. The Lefotver Hash Lemma, initially discussed by Impagliazzo, Levin and Luby in 1989 \cite{10.1145/73007.73009}, answer this question by controlling the inputs and output of the extractor. 

\begin{definition} \label{def:leftover}
    (Leftover Hash Lemma) Let $X$ be a random variable over a set $\mathcal{X}$, let $E$ denote a quantum side-information system, and let $\mathcal{F}$ be a family of hash functions from $\mathcal{X}$ to $\{0,1\}^\ell$ that is two-universal. If we choose a function $f$ uniformly at random from $\mathcal{F}$ and define the extracted output as $Z := f(X)$, then the average (over the random choice of $f$) of the conditional trace distance between $Z$ and the uniform distribution on $\{0,1\}^\ell$, given $E$, is bounded by
    \begin{equation}
        \epsilon = \frac{1}{2} \sqrt{2^{\ell - H_{\mathrm{min}}(X|E)}}
    \end{equation}
        where $H_{\min}(X|E)$ denotes the conditional min-entropy of $X$ given $E$.
\end{definition}

The leftover hash lemma \eqref{def:leftover} implies that for any fixed joint distribution of $X$ and $E$, there exists a fixed function $f$ that extracts nearly uniform randomness. Indeed, for any $\epsilon > 0$, a suitable function $f$ can be found such that the extracted output satisfies the claimed guarantee
\begin{equation}
        m \leq H_{\mathrm{min}}(X|E) - 2\log{ \left (\frac{1}{\epsilon} \right )} \,.
    \end{equation}

At the end of the randomness extraction, one may expect that the correlation between the results and a possible eveasdropper is small. This situation can also be generalized using a classical-quantum state

\begin{equation}
    \rho_{ZE} = \sum_{z \in \mathcal Z} p_Z (z) \ket{z} \bra{z} \otimes \rho_E^z \,,
\end{equation}
where the raw data is uniformly conditioned in $E$ if $\rho_E^z $ has a product form $\omega_z \otimes \rho_E$,  with $\omega_z = \mathbb{I}_Z/|Z|$ being the maximally mixed state on $\mathcal{H}_Z$. This concept generalizes the definition of random distribution and it is useful to introduce the definition of $\epsilon\text{-truly}$ random \eqref{def:eTrue}.

\begin{definition} \label{def:eTrue}
    ($\epsilon\text{-truly}$ random) $Z$ is called $\epsilon\text{-truly}$ random if there exist a state $\sigma_E$ on $E$ for which the trace distance between $\rho_{ZE}$ and $\omega_Z\otimes \sigma_E$ is at most $\epsilon$. This can be express as
    \begin{equation}
        \frac{1}{2} ||\rho_{ZE} - \omega_z \otimes \rho_E|| \leq \epsilon \, ,
    \end{equation}
    where $||\cdot|| = \emph{tr}(|\cdot|)$ denotes the trace norm. 
\end{definition}

The power of the leftover hash lemma is in its capacity of turn an abstract measure of randomness into something operational, using only information about the entropy of the source and the input and output of the extractors \cite{stinson2001universal}. Indeed, even the approaches that do not use this lemma for security proofs, usually tries to find an equivalent proofs \cite{Wu2026}.  



\subsection{Deterministic Extractors}\label{post_sub_B}

A deterministic extractor is defined as follows.

\begin{definition} \label{def:ExD}
    (Deterministic Extractor) 
    Let $\mathcal{C}$ be a class of probability distributions over $\{0, 1\}^n$. A function $\mathrm{Ext} : \{0, 1\}^n \to \{0,
    1\}^m$ is an $\epsilon$-extractor for $\mathcal{C}$ if, for every distribution $X \in \mathcal{C}$, the output distribution
    $\mathrm{Ext}(X)$ is $\epsilon$-close to the uniform distribution $U_m$.
\end{definition}

In general, deterministic extractors cannot exist for arbitrary weak sources: for any fixed function $\mathrm{Ext}: \{0,1\}^n \to \{0,1\}^m$, one can always construct a
distribution $X$ with high min-entropy $H_\infty(X) \geq n-1$ such that $\mathrm{Ext}(X)$ is completely fixed and therefore not close to uniform \cite{Vadhan12, ChorGoldreich88, SanthaVazirani86, Zuckerman96}. This impossibility result implies that deterministic extraction is only feasible when the source belongs to a restricted, well-characterized class $\mathcal{C}$ \cite{ChorGoldreich88, Shaltiel11}. For this reason, each deterministic extractor must be paired with an explicit statement of the source model it assumes. 
In this review, we distinguish between two main source models: an \emph{IID-bit source} and an \emph{independent-bit source}. An {IID-bit source} is written as $X = (X_1, X_2, \ldots, X_n)$, where all bits are mutually independent and follow the same Bernoulli distribution \cite{vonNeumann51, Elias72, Peres92}. 
That is, there exists a constant bias $0 < p < 1$, possibly unknown, such that $\Pr[X_i = 1] = p$ for every $i$. 
Alternatively, an {independent-bit source} allows the bits to remain mutually independent while their biases may vary from position to position \cite{Vazirani87, ChorGoldreich88, SanthaVazirani86}. 

In this case, for some constant $\delta > 0$, each bit satisfies $\delta \leq \Pr[X_i = 1] \leq 1-\delta$, but the values $p_i = \Pr[X_i = 1]$ are not required to be the same for all $i$.
Note that every IID-bit source is an independent-bit source, but not vice versa: the IID class additionally requires that all bits share the same bias $p$. This distinction is not merely taxonomic, but it determines which extractors are applicable, as detailed below for the Von Neumann \cite{vonNeumann51, Elias72, Peres92} and XOR \cite{Vazirani86thesis, Goldreich11xor} constructions.

\subsubsection{Von Neumann Extractor} 

The origin of deterministic extractors traces back to the pioneering work of John von Neumann in the 1950s \cite{von195113}, who proposed a foundational method to eliminate bias from a sequence of independent coin tosses with an unknown bias \cite{Vadhan12}. The Von Neumann extractor is designed specifically for IID-bit sources: it requires that all input bits share an unknown bias $p$ and be mutually independent. The requirement of identical bias across bits is essential to the method's correctness, since the symmetry argument on which it relies breaks down when consecutive bits have different biases \cite{Vadhan12}.

\begin{example}{Von Neumann Extractor}{von_extractor}
Consider the $12$-bit sequence:
$$X = (1,\; 1,\; 0,\; 1,\; 0,\; 0,\; 1,\; 0,\; 1,\; 1,\; 0,\; 1).$$

The sequence is split into non-overlapping pairs, and each pair is processed as follows:

\begin{center}
    \begin{tabular}{cccl}
    \toprule
    Pair & Bits & Output & Reason \\
    \midrule
    1 & $(1,1)$ & --- & Identical, discarded \\
    2 & $(0,1)$ & $0$ & Distinct, emit first bit \\
    3 & $(0,0)$ & --- & Identical, discarded \\
    4 & $(1,0)$ & $1$ & Distinct, emit first bit \\
    5 & $(1,1)$ & --- & Identical, discarded \\
    6 & $(0,1)$ & $0$ & Distinct, emit first bit \\
    \bottomrule
    \end{tabular}
\end{center}

The resulting output sequence is $(0, 1, 0)$, perfectly unbiased.
\end{example}
In practical scenarios, this framework models physical sources of randomness that exhibit a fixed but unknown systematic bias while maintaining independent trials \cite{QRNGInfocommunications}. To extract unbiased bits, the Von Neumann extractor partitions the incoming bitstream into non-overlapping pairs. For each pair, if the
two bits differ, the extractor outputs the first bit; if the bits are identical, the entire pair is discarded. This operational procedure is detailed in Example~\ref{ex:von_extractor}.

By producing completely unbiased and independent bits, the Von Neumann extractor serves as a $0$-extractor (\textit{i.e.}, with zero statistical distance from uniform) for IID-bit sources. Its correctness relies on the fact that for two independent bits drawn from the same distribution with $\Pr[X_i=1] = p$, the two asymmetric outcomes are equiprobable:
\begin{align*}
    P[X_i=1 \wedge X_{i+1}=0] &= p \cdot (1-p), \\
    P[X_i=0 \wedge X_{i+1}=1] &= (1-p) \cdot p.
\end{align*}

Since $p \cdot (1-p) = (1-p) \cdot p$ for \emph{any} $p$, the extractor completely eliminates the source bias without requiring knowledge of $p$. Note, however, that this symmetry holds only when both bits in the pair share the same bias. If the biases differ across positions, as in an independent-bit source, the argument breaks down, and the Von Neumann extractor no longer produces a perfectly unbiased output \cite{Vadhan12, An_Introduction}.

\subsubsection{XOR Extractor} 

The XOR extractor \cite{Jennewein2000,vazirani1987strong}, also known as the parity extractor, addresses a strictly broader source class than the Von Neumann extractor: it is designed for independent-bit sources, where the individual biases $p_i = \Pr[X_i = 1]$ may differ across bits, provided all bits remain mutually independent and none is too biased \cite{Vadhan12}. This generalization over the IID assumption makes the XOR extractor applicable to a wider range of physical sources, at the cost of producing only a single output bit per application. The definition of an extractor is 
\begin{equation}
    \mathrm{Ext}(X) \;=\; X_1 \oplus X_2 \oplus \cdots \oplus X_n
    \,,
\end{equation}
where $\oplus$ denotes addition modulo~2 (see Ex. \ref{ex:xor_extractor}). 

A notable limitation of the XOR extractor is that it maps an $n$-bit input to a single output bit, resulting in a compression ratio of $1/n$. More critically, both the Von Neumann and XOR extractors share a fundamental constraint: neither can handle sources whose bits exhibit correlations, since both proofs rely on mutual independence \cite{An_Introduction, Vadhan12}. Raw bits from physical QRNG sources often violate this assumption due to electronic crosstalk and detector after pulsing, making subsequent extraction stages necessary in practice \cite{RevModPhys.89.015004}.

\begin{example}{XOR Extractor}{xor_extractor}
Consider the $12$-bit sequence :
\begin{equation*}
    X = (1,\; 1,\; 0,\; 1,\; 0,\; 0,\; 1,\; 0,\; 1,\; 1,\; 0,\; 1),
\end{equation*}

Partitioning $X$ into three non-overlapping blocks of four bits and applying the XOR extractor to each:
\begin{align*}
    B_1 &= (1, 1, 0, 1) \;\Rightarrow\; \mathrm{Ext}(B_1) = 1\oplus 1\oplus 0\oplus 1 = 1, \\
    B_2 &= (0, 0, 1, 0) \;\Rightarrow\; \mathrm{Ext}(B_2) = 0\oplus 0\oplus 1\oplus 0 = 1, \\
    B_3 &= (1, 1, 0, 1) \;\Rightarrow\; \mathrm{Ext}(B_3) = 1\oplus 1\oplus 0\oplus 1 = 1.
\end{align*}
The resulting output sequence is $(1, 1, 1)$: three unbiased bits
extracted from twelve, at a rate of one output bit per block.
\end{example}

\subsection{Random extractors}\label{post_sub_C}

The impossibility of a efficient deterministic extraction from arbitrary weak sources motivates a more general approach in which the extractor is allowed to use an additional input. This input, usually called the \emph{seed}, is a short string of uniformly random bits that is independent of the raw source (see Definition \ref{def:ExA}). Operationally, the seed selects one function from a family of deterministic maps. This small amount of trusted randomness makes it possible to extract nearly uniform bits from the broad class of all sources with sufficiently large min-entropy, rather than only from restricted source models such as IID-bit or independent-bit sources \cite{An_Introduction}.

\begin{definition}\label{def:ExA}(Seeded Randomness Extractor)
    A function 
    $$ Ext : \{0, 1\}^n \times \{0, 1\}^d \to \{0, 1\}^m $$
    is a $(k,\epsilon)$-extractor if, for every random variable $X$ over $\{0,1\}^n$ satisfying $H_\infty(X)\geq k$, and for an independent uniform seed $Y=U_d$, the output distribution $\mathrm{Ext}(X,Y)$ is $\epsilon$-close to the uniform distribution $U_m$ over $\{0,1\}^m$. 
\end{definition}

The role of the seed is to provide the extractor with a small amount of independent randomness, which greatly broadens the class of sources that can be treated. Instead of requiring detailed knowledge of the source structure, it is sufficient to establish a reliable lower bound $k$ on the min-entropy of the raw data \cite{Bennett1988PrivacyAB}. Provided that the seed is uniform and independent of $X$, the extractor produces an output distribution that is $\epsilon$-close to uniform \cite{stinson2001universal}. Importantly, the seed does not create randomness by itself; it only enables the extraction of the entropy already contained in the raw source.

For cryptographic post-processing, especially in QRNG and QKD applications, it is often important that the extracted output remains compatible with Definition \eqref{def:eTrue} when the seed is known.
This requirement defines a strong $(k,\epsilon)$-extractor: for every source $X$ with $H_\infty(X)\geq k$ and for an independent uniform seed $Y$, the pair $(Y,\mathrm{Ext}(X,Y))$ must be $\epsilon$-close to $(Y,U_m)$, that is,
$\Delta\big((Y,\mathrm{Ext}(X,Y)),(Y,U_m)\big)\leq \epsilon$.
This property is especially useful when the seed is public or transmitted together with the extracted string. Universal hashing constructions, such as Toeplitz hashing, are widely used in this setting because the Leftover Hash Lemma \eqref{def:leftover} guarantees that they can act as efficient strong extractors \cite{CARTER1979143,wegman1981new,stinson2001universal,PhysRevA.87.062327}.

\

\subsubsection{Toeplitz Extractor} 

A Toeplitz extractor is a seeded extractor based on universal hashing \cite{xu2012ultrafast, wegman1981new}. It maps a raw block of $n$ bits into an $m$-bit output by multiplying the input vector by a binary Toeplitz matrix, as detailed in Def. \ref{def:toeplitz}. The seed specifies the Toeplitz matrix, and therefore determines the particular hash function used in that extraction instance \cite{math12142243}.

\begin{definition}\label{def:toeplitz}(Toeplitz Matrix)
    A matrix $T\in\mathbb{R}^{r\times c}$ is called a \emph{Toeplitz matrix} if its entries are constant along each diagonal parallel to the main diagonal. Equivalently, there exists a function $t:\mathbb{Z}\to\mathbb{R}$ such that 
    $$T_{ij}=t(i-j), \qquad i=1,\ldots,r,\quad j=1,\ldots,c.$$
\end{definition}

All operations are performed over the finite field $\mathbb{F}_2$. Hence, multiplication corresponds to logical AND, while addition corresponds to XOR. In this sense, Toeplitz extraction can be implemented as a binary matrix-vector multiplication, as detailed in Ex. \ref{ex:toeplitz_extractor}.

\begin{example}{Toeplitz Extractor}{toeplitz_extractor}
Consider a raw block of $n = 4$ bits with min-entropy $k = 2$, and suppose we wish to extract $m = 2$ nearly uniform bits. The seed specifies a Toeplitz string $ts$ of length $n + m - 1 = 5$:
\begin{equation*}
    X = (1,\; 0,\; 1,\; 1), \qquad ts = (1,\; 0,\; 1,\; 0,\; 1).
\end{equation*}
The Toeplitz matrix $T \in \{0,1\}^{2 \times 4}$ is constructed by taking consecutive length-$n$ substrings of $ts$ as rows:
\begin{equation*}
    T =
    \begin{pmatrix}
        ts[0:3] \\[2pt] ts[1:4]
    \end{pmatrix}
    =
    \begin{pmatrix}
        1 & 0 & 1 & 0 \\
        0 & 1 & 0 & 1
    \end{pmatrix}.
\end{equation*}
Each output bit $y_i = \mathrm{row}_i(T) \cdot X \pmod{2}$ is computed via AND then XOR:
\begin{align*}
    y_0 &: (1,0,1,0)\cdot(1,0,1,1) = 1\oplus 0 \oplus 1 \oplus 0 = 0,\\
    y_1 &: (0,1,0,1)\cdot(1,0,1,1) = 0\oplus 0 \oplus 0 \oplus 1 = 1.
\end{align*}
The extracted output is $y = (0,\; 1)$.
\end{example}

A key advantage of Toeplitz hashing is its compact seed representation. While an arbitrary binary matrix in $\{0,1\}^{m\times n}$ would require $mn$ bits to be fully specified, a Toeplitz matrix is determined by only $n+m-1$ bits. This reduction is particularly important in high-throughput QRNG post-processing, where the extractor must process large raw data blocks efficiently and with limited memory overhead. For this reason, Toeplitz hashing is widely used in practical and hardware-oriented implementations of randomness extraction \cite{RevModPhys.89.015004, PhysRevA.87.062327, Guo:19, PRXQuantum.4.010330,math12142243}.

From a theoretical perspective, Toeplitz hashing belongs to a two-universal family of hash functions \eqref{def:two}. Therefore, its extraction guarantee follows from the Leftover Hash Lemma \eqref{def:leftover}. In this framework, the output length $m$ is selected according to the estimated min-entropy of the raw source and the target security error $\epsilon$. The extractor should thus be understood as a compression mechanism: it does not generate new randomness, but transforms the entropy already present in the raw sequence into a shorter output whose deviation from uniformity is bounded by the adopted security parameter \cite{CARTER1979143,wegman1981new,stinson2001universal}.

\

\subsubsection{Trevisan Extractor} 

Beyond universal hashing, Trevisan proposed a fundamentally different construction of a strong seeded extractor \cite{Trevisan2001}. 
Its central idea is to apply a strong one-bit extractor like XOR and Von Neumann repeatedly to the same raw input $x$, each time using a different subset of the seed bits, and to concatenate the resulting bits into the final output (see Ex. \ref{ex:trevisan_extractor}).
The seed subsets $S_0, S_1, \dots, S_{m-1} \subseteq [d]$ are chosen according to a
combinatorial structure called a \emph{weak design}, which controls the overlap between consecutive subsets and thereby limits the correlation between successive output bits \cite{trevisanReducing}.

\begin{definition}\label{def}(Trevisan)
    The extractor is a concatenation function,
    \begin{equation}
        \mathrm{Ext}(x, y)
        \;=\;
        \mathrm{Ext}_1(x, y_{S_0})
        \;\|\;
        \cdots
        \;\|\;
        \mathrm{Ext}_1(x, y_{S_{m-1}}),
    \end{equation}
    where $y_{S_i}$ denotes the bits of the seed $y$ indexed by $S_i$, and $\|$ denotes concatenation.
\end{definition}

Two properties make Trevisan's extractor particularly relevant for QRNG post-processing. First, it has been proven to remain secure even when the adversary holds quantum side information about the raw source \cite{de2012trevisan}, a property not guaranteed by
construction for most classical extractors. Second, the seed length grows only polylogarithmically with the input size, $\mathcal{O}(\log^3 n)$ bits suffice to extract all of the min-entropy \cite{trevisanReducing}, in sharp contrast with Toeplitz hashing,
which requires a seed of length $n + m - 1$. This seed efficiency makes Trevisan's construction attractive in resource-constrained settings where trusted randomness is scarce.

\begin{example}{Trevisan Extractor}{trevisan_extractor}
    Let $n = 6$, $m = 3$, and consider the raw input and seed:
    \begin{equation*}
        x = (1,\, 0,\, 1,\, 1,\, 0,\, 1), \\
        y = (0,\, 1,\, 1,\, 0,\, 1,\, 0).
    \end{equation*}
    The weak design specifies three disjoint index subsets of size $t = 2$:
    \begin{equation*}
        S_0 = \{0, 1\}, \qquad S_1 = \{2, 3\}, \qquad S_2 = \{4, 5\}.
    \end{equation*}
    The corresponding seed substrings are extracted from $y$:
    $y_{S_0} = (y_0, y_1) = (0, 1) ; y_{S_1} = (y_2, y_3) = (1, 0); y_{S_2} = (y_4, y_5) = (1, 0).$
    Using the XOR function as the one-bit extractor, each sub-seed
    selects the corresponding positions of $x$ and computes their
    parity:
    \begin{align*}
        \mathrm{Ext}_1(x,\, y_{S_0})
            &= x_0 \oplus x_1 = 1 \oplus 0 = 1, \\
        \mathrm{Ext}_1(x,\, y_{S_1})
            &= x_2 \oplus x_3 = 1 \oplus 1 = 0, \\
        \mathrm{Ext}_1(x,\, y_{S_2})
            &= x_4 \oplus x_5 = 0 \oplus 1 = 1.
    \end{align*}
    Concatenating the three output bits yields $\mathrm{Ext}(x, y) = (1,\, 0,\, 1)$.
\end{example}

\subsubsection{ Other extractor}

Beyond Toeplitz and Trevisan, a few other seeded constructions have been used or proposed for QRNG post-processing. Most of them are two-universal families, so their security also follows from the Leftover Hash Lemma. They differ mainly in seed length and implementation cost.

Random binary matrix is the simplest two-universal family uses an arbitrary matrix $M\in\{0,1\}^{m\times n}$, at the cost of the $mn$-bit seed discussed above.
Because it requires only AND/XOR logic, it was adopted in early commercial and hardware QRNGs \cite{troyer2012quantis, frauchiger2013true}. The matrix is usually fixed and pre-generated by another QRNG. Examples include an FPGA extractor for a 1.25-GHz QKD system~\cite{martin2015qrng}, an image-sensor QRNG \cite{amri2016qis}, and an extractor integrated on a SPAD-array chip \cite{regazzoni2021onchip}.

Several families reduce the seed while keeping $\mathcal{O}(n\log n)$ complexity: The modified Toeplitz hash, with an $(n-1)$-bit seed, is widely used for privacy amplification in QKD~\cite{hayashi2016dual}; The Circulant extractor, with an $(n+1)$-bit seed, is available in the open-source Cryptomite library and is applied to the output of a commercial QRNG \cite{foreman2025cryptomite,foreman2024statistical} and multiplication in $\mathrm{GF}(2^n)$ and polynomial hashing \cite{carter1979universal}.

In Ref.~\cite{li2024improved}, it was shown that raw QRNG data often forms a \emph{reverse block source}. They proposed hashing each block with a seed taken from the output of the previous block, which reduces the initial seed to $\mathcal{O}(\log N)$ or $\mathcal{O}(1)$ bits, where $N$ is the total input length; an FPGA simulation reached about 300~Gbit/s~\cite{li2024improved}.
Sampled sub-block hashing likewise makes Toeplitz extraction of very large inputs feasible in hardware \cite{ng2024subblock}.

Keyed cryptographic conditioners based on the AES block cipher, with a secret 128-bit key or initialization vector as the seed, have been used in vacuum-fluctuation QRNGs \cite{haw2015maximization, haylock2019multiplexed, samsonov2020multimode, pervushin2021vacuum}.
AES-CMAC, for instance, is a vetted conditioning component in NIST SP~800-90B~\cite{nist80090b}. These conditioners provide only computational security and are not proven secure against quantum side information.
Software implementations have also been targeted by side-channel attacks~\cite{kuznetsov2019side}.
 
Finally, the seed can be replaced by a second, independent weak source through two-source extractors, such as the inner-product construction \cite{chor1988unbiased, bouda2012improving} or the construction of Dodis \cite{dodis2004improved}. This approach has been applied in real-time seedless QRNG post-processing \cite{li2024seedless, haylock2019multiplexed}.

\section{Randomness Enhancement} \label{sec:enhance}

Intrinsic randomness can be regarded as a physical resource associated with the fundamental unpredictability of quantum measurement outcomes. In the context of randomness expansion and amplification, while deterministic procedures cannot generate true randomness, quantum protocols can be leveraged to process and increase the amount or improve the quality of certified randomness available. This constitutes the core paradigm of quantum physical extractors for randomness enhancement: utilizing quantum protocols either to expand a short uniform random seed into a longer certified random string or to generate nearly uniform and private randomness from a weak randomness source. The former corresponds to \emph{randomness expansion} protocols, whereas the latter is addressed by \emph{randomness amplification} protocols. In this section, we briefly discuss the central principles behind these two processes.


\subsection{Randomness Expansion}

Randomness expansion enables expanding an initial source of private random bits into a larger string of certified random bits. The concept was initially introduced by Colbeck (2007) as a protocol that was not yet optimized for rate efficiency and lacked a full security proof against general quantum attacks \cite{colbeck2011quantumrelativisticprotocolssecure}. Later, Colbeck and Kent (2011) formalized the task of private randomness expansion using untrusted devices \cite{Colbeck_2011}. The key idea is to allow an honest user (Bob), operating within a single secure laboratory, to generate new private randomness by certifying Bell-nonlocal correlations produced by untrusted quantum devices supplied by a potential adversary (Eve).

The protocol relies on the following central assumptions \cite{Colbeck_2011}:
\begin{enumerate}
    \item Bob’s laboratory is secure. In particular, secret messages cannot be sent from Eve’s devices, once within his laboratory, to the outside world, and Eve cannot probe his laboratory from the outside.
    \item Bob can isolate any devices in his laboratory, preventing them from sending any signal outside an isolated region.
    \item Bob has secure classical information processing devices, with secure authenticated classical communication between them within his laboratory. In particular, Eve’s devices are unable to spoof classical communications within Bob’s laboratory; they can output to Bob’s classical devices only via prescribed channels.
    \item Eve is constrained by the laws of quantum theory.
    \item All communication channels and devices operate noiselessly.
\end{enumerate}

In this protocol, three untrusted devices are deployed to perform Greenberger-Horne-Zeilinger (GHZ) nonlocality tests \cite{greenberger1989going}. Bob partitions his initial private seed $X$ into two parts: a substring $X_1$ used to choose the measurement inputs ($P_i$ or $Q_i$) for each device, and a seed $R$ reserved for privacy amplification \cite{Bennett1988PrivacyAB}. Each device receives only its assigned input without knowledge of the full input string. From the measurement outcomes, Bob verifies whether the product of outcomes satisfies the GHZ nonlocality relations; if an invalid outcome is observed, the protocol aborts. Upon passing all tests, Bob forms a raw output string $\tilde{X}$ and concatenates it with the input string to form $X' = (X_1, \tilde{X})$. Finally, Bob applies privacy amplification using two-universal hash functions selected using $R$ to compress $X'$ into a shorter, $\delta$-private string $S$. By applying privacy amplification to $X_1$ alongside $\tilde{X}$, the protocol eliminates vulnerabilities to abort-conditioning attacks, thereby achieving universally composable security against quantum adversaries.

Bell nonlocality is the central resource underlying device-independent private randomness expansion. Beyond early formulations based on Greenberger-Horne-Zeilinger (GHZ) tests, subsequent research demonstrated that non-classical correlations derived from bipartite games, such as the Clauser-Horne-Shimony-Holt (CHSH) inequality \cite{CHSH1969}, can similarly certify entropy against potential eavesdroppers \cite{Pironio2010, VaziraniVidick2012}. A major theoretical breakthrough was achieved by Miller and Shi (2017), who proved full quantum security for spot-checking protocols under general nonlocal games, raising the noise tolerance threshold for the CHSH game to 10.3\% \cite{MillerShi2014}. Moreover, by leveraging sequential CHSH game rigidity, adaptive schemes demonstrated that a constant number of non-signaling quantum devices could perform unbounded, infinite randomness expansion \cite{CoudronYuen2014, MillerShi2014, ChungShiWu2014}.

To provide tight, practical rates against general quantum adversaries, modern security analyses combine single-round entropy bounds, often obtained through semidefinite-programming relaxations, with the Entropy Accumulation Theorem (EAT) to derive finite-size expansion rates secure against quantum side information. \cite{Acin2012, NietoSilleras2014, ArnonFriedman2018}. These theoretical advancements enabled the first successful experimental realizations of device-independent quantum randomness expansion \cite{Pironio2010, Christensen2013}. Utilizing photonic loophole-free Bell tests and spot-checking protocols grounded in either quantum probability estimation or the EAT, recent experiments demonstrated net expansion gains exceeding hundreds of millions of certified random bits secure against quantum side information \cite{singh2024localcontextualitybasedselftestssufficient, bierhorst2018experimentally, zhang2020experimental, Li2021, Shalm2021}.

To overcome the severe experimental demands and modest bit generation rates of full Bell nonlocality, complementary approaches have explored semi-device-independent protocols in all it extent of possibilities, from measurement and souce device-independence to semi-self testing platforms \cite{Ma2016, RevModPhys.89.015004}. Paradigms such as source-independent and measurement-device-independent randomness expansion trade full device independence for realistic, physically verifiable assumptions placed on either the state preparation or the measurement apparatus \cite{Ma2016, Zhang2026, Cao2016}. In particular, P\&M protocols leveraging discrete-modulated coherent states coupled with continuous-variable homodyne detection allow for certified randomness expansion secure against quantum side information without requiring Bell nonlocality or detector calibration \cite{Wang2023, Zhang2026}. Crucially, homodyne detection operates at room temperature and avoids the detector-blinding vulnerabilities and efficiency constraints of single-photon detectors \cite{Wang2023}. The recent realization of self-testing MDI randomness expansion on fully integrated silicon photonic chips \cite{Zhang2026} marks a major milestone, proving that provably secure, composable quantum randomness expansion can be successfully implemented on scalable, miniaturized photonic platforms \cite{Wang2023, Zhang2026}.

\subsection{Randomness Amplification}

While randomness expansion generates a longer string of certified random bits from a short, perfectly uniform private seed, {randomness amplification} addresses a fundamentally distinct and more challenging task: transforming a single weak, imperfect source of randomness into a string of certified random bits without assuming access to any prior uniform seed \cite{ColbeckRenner2012, Brandao2016}. Weak randomness is conventionally modeled by a Santha--Vazirani ($\epsilon$-SV) source \cite{SanthaVazirani1986}, where each bit generated has a bounded bias $\epsilon \in [0, 1/2)$ conditioned on all past outcomes and external side information $E$. In a seminal classical result, Santha and Vazirani (1986) proved a strict no-go theorem: no deterministic classical algorithm can extract or amplify a single $\epsilon$-SV source into a less biased or uniform output \cite{SanthaVazirani1986}. Consequently, classical randomness extraction requires at least two independent weak sources, making device-independent certification impossible from a single imperfect classical source. 

Quantum nonlocality provides a physical mechanism to overcome this fundamental classical barrier \cite{ColbeckRenner2012}. Colbeck and Renner (2012) demonstrated the first quantum randomness amplification protocol, showing that by feeding weak choices from an $\epsilon$-SV source into spatially separated quantum devices performing chained Bell tests, nonlocal correlations certify fresh randomness that can amplify sources with initial bias $\epsilon < 0.058$ into arbitrarily pure, uncorrelated random bits under the no-signalling principle \cite{ColbeckRenner2012}. Subsequently, Gallego \textit{et al.} (2013) solved the open question of amplifying arbitrarily weak sources ($\epsilon \to 1/2$) by exploiting five-party Mermin Bell inequalities and GHZ nonlocality \cite{Gallego2013}. Further theoretical studies established exact thresholds for bipartite chain correlations \cite{Grudka2014}, single-device tripartite GHZ configurations under min-entropy constraints \cite{PleschPivoluska2014}, and physical extractors for general min-entropy sources beyond the SV-model \cite{ChungShiWu2014}.

Early theoretical schemes faced severe practical bottlenecks, as they required either an infinite number of non-signalling devices or discarding devices after each single measurement round \cite{Brandao2016}. To render quantum randomness amplification more experimentally feasible, Brand\~ao \textit{et al.} (2016) introduced robust, noise-tolerant protocols that reuse a small, constant number of devices (four or eight non-signalling components) to amplify arbitrarily weak $\epsilon$-SV sources against general no-signalling adversaries \cite{Brandao2016}. Concurrently, Ramanathan \textit{et al.} (2016) established a minimal two-device architecture capable of amplifying any non-fully deterministic SV source under realistic noise \cite{Ramanathan2016}. Their scheme combined a two-party Bell test with a partial tomographic verification procedure, proving that certified min-entropy can be extracted even when nonlocality only guarantees weak, partial randomness in specific input--output channels \cite{Ramanathan2016}.

Recently, Kulikov \textit{et al.} (2026) achieved the first experimental realization of device-independent quantum randomness amplification, demonstrating quantum advantage over classical computation \cite{kulikov2026amplification}. Implemented on a solid-state circuit quantum electrodynamics platform featuring two superconducting transmon qubits connected via a 30-meter cryogenic link, the setup executed a loophole-free CHSH test with high multi-day stability, attaining an observed Bell value $S_{\text{obs}} = 2.271$. By feeding inputs from physical laser-phase-diffusion random number generators possessing an input bias $\mu \le 0.75\%$, and processing the observed Bell outcomes with a two-source extractor, the experiment generated $m = 4.5 \times 10^7$ certified, unbiased random bits ($\mu' = 0$) with a composable failure probability $\epsilon_{\text{sound}} = 10^{-12}$ \cite{kulikov2026amplification}. This breakthrough confirms that quantum devices can perform physically certified randomness amplification in realistic, scalable experimental environments.

\section{Randomness Testing} \label{sec:Test}

Randomness testing aims to determine whether a given dataset can be regarded as effectively random under specified conditions \cite{Mengdi_2021}. In essence, such tests seek to identify patterns, regularities, or structures in finite sequences that may indicate that the underlying generation process is not truly random. Randomness tests are statistical procedures applied to finite data sequences with the purpose of assessing the degree of confidence in their random behavior \cite{article_417502}. These procedures differ from theoretical approaches, which are typically concerned with infinite sequences, and are fundamental for evaluating the quality of random number generators, whether physical or computational \cite{RevModPhys.89.015004, PhysRevA.87.062327}.

The notion of randomness does not have a universally accepted definition and may be interpreted from different perspectives \cite{philosophy_statis, Bera_2017}. Nevertheless, several approaches share common elements, such as the presence of uncertainty and the idea that perfect randomness is an ideal and absolute property. Although a perfectly random sequence is expected to exhibit maximal complexity in an idealized sense, many practical applications only require sequences that are sufficiently random for a given purpose. Therefore, a finite set of tests and an acceptance threshold are usually defined according to the context of application \cite{10.1145/3447773}. For instance, the randomness requirements for a charity raffle are considerably less stringent than those required for selecting components in a nuclear reactor \cite{philosophy_statis}.

Several practical approaches are available for assessing the randomness of a sequence, particularly in the binary case. The main categories include:
\begin{itemize}
\item \textbf{Statistical tests:} these tests evaluate deviations from properties expected in random sequences \cite{6104067}.
\item \textbf{Spectral transformations:} these methods use tools such as Fourier or Hadamard transforms to detect hidden periodicities \cite{philosophy_statis}.
\item \textbf{Complexity-based tests:} these tests assess the compressibility of a sequence or the difficulty of describing it concisely \cite{testealeat}.
\end{itemize}

According to the National Institute of Standards and Technology\footnote{NIST is a U.S. federal agency responsible for developing technical standards, measurements, and guidelines in science, technology, and information security, including the widely recognized NIST Cybersecurity Framework \cite{NIST_Website}.} (NIST), randomness should be understood as a probabilistic property. A sequence is regarded as random when the results of the statistical tests applied to it do not provide sufficient evidence to reject the null hypothesis ($H_0$), which assumes that the data originate from a truly random process. Thus, each randomness test is formulated to evaluate $H_0$ and may lead to two possible outcomes. If $H_0$ is rejected, the sequence is considered to exhibit statistically significant deviations from the expected behavior of a random sequence and is therefore classified as non-random. Conversely, if $H_0$ is not rejected, there is insufficient statistical evidence to discard the randomness hypothesis, and the sequence is considered compatible with random behavior \cite{Neyman_Pearson_1933, Casella_Berger_2002}.

In this context, two types of statistical error may occur in the test decision. A type I error, denoted by $\alpha$, occurs when the test rejects the null hypothesis ($H_0$) even though the sequence is, in fact, random. The parameter $\alpha$ is known as the significance level of the test and represents the probability of incorrectly concluding that a random sequence is non-random  \cite{ Casella_Berger_2002}. In cryptographic applications, typical values of $\alpha$ range from $0.001$ to $0.01$. For example, when $\alpha = 0.01$, approximately one out of every 100 truly random sequences may be incorrectly rejected \cite{Rukhin2001NIST}. A type II error, denoted by $\beta$, occurs when the test fails to reject $H_0$ even though the sequence is non-random. In this case, the testing procedure fails to detect patterns, biases, or structural deviations in the data. Unlike $\alpha$, the value of $\beta$ is not fixed, since it depends on the specific way in which the sequence violates randomness; different forms of non-randomness lead to different probabilities of type II error. For a detailed discussion on null hypotheses and Type I and II errors, we refer the reader to \cite{Montgomery_Runger}.

The result of each statistical test is expressed by means of a $p$-value, which represents the probability, under the null hypothesis, of obtaining a test statistic at least as extreme as the one observed. Therefore, $p$-values smaller than the significance level ($p < \alpha$) lead to the rejection of $H_0$, indicating statistical evidence of non-randomness \cite{NISTteste}. On the other hand, $p$-values satisfying $p \geq \alpha$ indicate that there is not enough evidence to reject the null hypothesis, and the sequence is considered statistically compatible with randomness.

The Diehard Test Suite is another case of widely known randomness test batteries \cite{Marsaglia1996Diehard}, being one of the earliest and most influential collections of statistical tests designed to evaluate the quality of PRNG. Its main purpose is to detect non-random patterns that may not be revealed by simpler tests, thereby assessing whether the generated sequences exhibit statistical properties close to those expected from a truly random source. It operates by applying a set of statistical tests that evaluate properties such as uniformity, independence, absence of spatial patterns, and regularity in the distribution of bits  \cite{sys2022bad}. Each test returns statistical measures, particularly $p$-values, which indicate whether the observed behavior is consistent with the randomness hypothesis. Extremely small or extremely large $p$-values may suggest that the generator fails to produce sufficiently random sequences. Although Diehard was a pioneering test suite, it does not cover all possible defects in PRNGs and does not guarantee cryptographic security.

Over time, more modern and comprehensive suites have been developed, such as TestU01 \cite{Lecuyer2007TestU01} and the NIST SP 800-22 test suite \cite{Rukhin2001NIST}. These suites provide broader sets of statistical tests and more flexible methodologies for evaluating random and pseudorandom sequences. The Dieharder suite \cite{Brown2006Dieharder} is a more recent and extensive implementation that includes the original Diehard tests, additional tests, a modern interface, and support for different random number sources.

Being the most used test, the NIST Statistical Test Suite was developed by NIST and consists of 15 main statistical tests for binary sequences \cite{NIST_Website}. In this suite, the parameter ($n_{\min}$) denotes the minimum length required for a binary sequence for the corresponding statistical test to be valid and produce reliable results \cite{zaman2012reviewstudyniststatistical}. This requirement is important because, when the analyzed sequence has a length smaller than ($n_{\min}$), the test statistics may no longer follow the theoretical distribution assumed by the test, thereby compromising the reliability of the randomness assessment.

\section{Mainstream implementations} \label{sec:mainstream}

The previous sections discussed the fundamental building blocks of QRNG, including the underlying entropy sources, randomness extraction methods, and security considerations. Building upon this background, this section reviews representative implementations that have shaped the development of the field. Rather than focusing on the physical origin of entropy or on specific post-processing techniques, the discussion highlights notable demonstrations reported in the literature and the technological directions they established. 

To provide a broad perspective on the state of the art, a representative body of literature was surveyed, covering QRNG implementations developed across different platforms and application contexts. The selected works are organized according to their primary contribution to the advancement of the field, namely high-generation-rate systems, integrated photonic realizations, dedicated hardware architectures, and cryptographic applications. This structure allows the identification of the main technological trends, performance milestones, and deployment strategies that have driven the evolution of QRNG from laboratory demonstrations to practical security solutions.

\subsection{Hight-generation-speed implementations}

The generation rate is one of the main performance indicators of a QRNG, but its interpretation requires distinguishing the raw-data rate from the rate effectively obtained after entropy estimation and randomness extraction. The final performance depends not only on the bandwidth of the quantum process, but also on the detection efficiency, the digitization rate and resolution, the min-entropy per sample, and the capacity of the post-processing stage \cite{RevModPhys.89.015004, mannalatha2023review}. In addition, results obtained offline, in real time, in a single channel, or through the aggregation of multiple channels are not directly equivalent. 

\begin{figure*}
    \centering
    \includegraphics[width=0.8\linewidth]{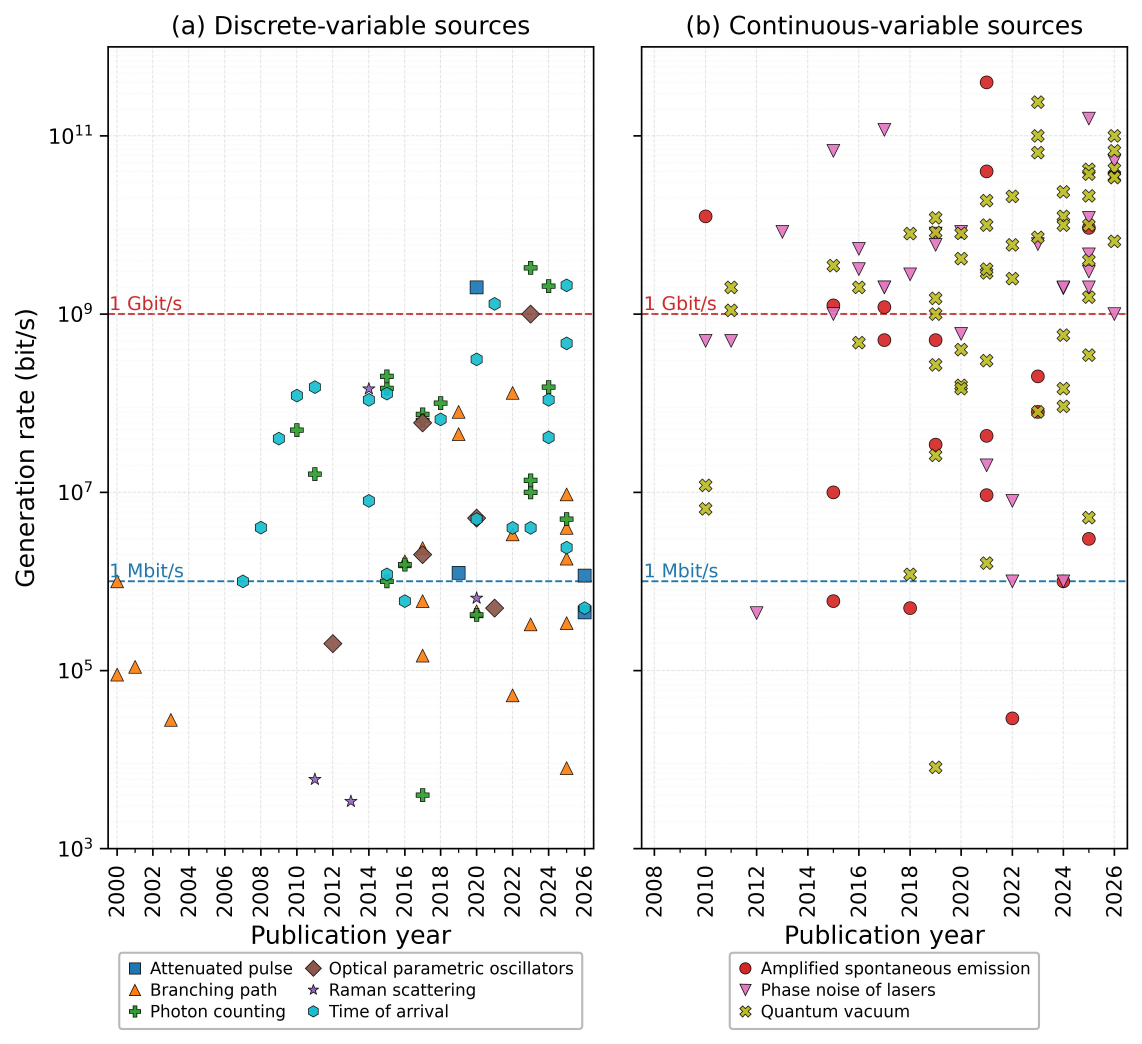}
    \caption{Generation rates of surveyed DV-QRNGs (left) and CV-QRNGs (right), grouped by entropy source.}
    \label{fig:speed}
\end{figure*}

Under these criteria, the evolution of high-speed QRNGs can be understood as a transition from DV-architectures, limited by the detection of individual events, to CV-sources and parallel systems capable of exploiting substantially larger bandwidths. This is because real-valued measurements are mapped onto a discrete space with a cardinality of $2^n$, where advances in microelectronics have enabled the use of ADCs with $n > 12$. To illustrate these distinctions, Figure \ref{fig:speed} presents the evolution of the generation rate of recent DV-QRNG (left) and CV-QRNG (right).  Henceforth, we discuss the main results, highlighting the highest generation speed related to every entropy source. 

Early DV-optical sources exploited either the random path selection or the detection time of individual photons, providing a direct association between each quantum event and the resulting binary sequence. In photon-path systems, generation rates evolved from approximately 1 Mbit/s in 2000 \cite{Jennewein2000} to 9.49 Mbit/s after extraction in an integrated implementation in 2025 \cite{du2025source}, the highest value confirmed among the dated studies surveyed for this source. In this architecture, optical losses, dark counts, multiphoton events, and detector dead time keep the net rate below the nominal capacity in several implementations \cite{dudek2025optical,francis2017quantum,soubusta2001quantum}. Photon time-of-arrival measurements exhibited greater scalability, progressing from approximately 1 Mbit/s in 2007 in an architecture employing a restartable clock \cite{stipcevic2007quantum} to 2.1 Gbit/s in 2025 by combining the temporal and spatial information associated with each detection event \cite{wang2025efficient}. This evolution shows that multidimensional encoding and detector parallelism can mitigate the limitations inherent in generating only one bit per event.

In photon-counting QRNG, the increase in generation rate resulted primarily from extracting more information from each acquisition window or detection event and from detector parallelism. An integrated LED--PMT system achieved approximately 50 Mbit/s in 2010 by exploiting detector dead time to reduce bias without post-processing \cite{Furst:10}, whereas photon-number-resolving detection produced two bits per pulse and reached 16 Mbit/s in 2011 \cite{10.1063/1.3613952}. An integrated SPAD-based system reached an aggregate output rate of 3.3 Gbit/s in 2023 by operating two arrays of quantum random flip-flops in parallel, the highest value among the dated photon-counting studies surveyed \cite{10129008}. In 2024, a \(32 \times 32\)-pixel CMOS-SPAD array combined with temporal resolution enabled the experimental extraction of up to 20 bits per event, achieving 2.067 Gbit/s \cite{li2024efficient}. In attenuated-pulse sources, an early demonstration in 2009 exploited the balance between detection and no-detection events and indicated theoretical scalability toward Gbit/s rates \cite{Wei:09}. Subsequently, a self-testing implementation based on an energy bound achieved 1.25 Mbit/s in 2019 \cite{PhysRevA.100.062338}. In both approaches, the generation rate remains constrained by the repetition frequency, detector dead time, afterpulsing, and optical-intensity stability.

Sources based on Raman scattering and optical parametric oscillators have remained less competitive in terms of absolute generation rate. In stimulated Raman scattering, an early demonstration in 2011 quantized the random phase of the Stokes field into 64 intervals, corresponding to six bits per sample and a nominal raw rate of up to 6~kbit/s for the 1~kHz acquisition limit \cite{Bustard2011}, whereas a waveguide implementation achieved up to 145~Mbit/s in 2014 \cite{10.1063/1.4864095}, the highest Raman generation rate among the reviewed studies. In degenerate OPOs, an early demonstration in 2012 achieved all-optical generation at 10~kbit/s without post-processing by exploiting the spontaneous selection between two phase states \cite{Marandi2012}. In nondegenerate OPOs, measuring the intensity fluctuations of twin beams enabled the simultaneous extraction of two sequences at 60~Mbit/s each in 2017 using Toeplitz hashing; this result corresponds to a CV implementation and the highest experimental rate among the OPO studies considered in this survey \cite{Zhang_2017}. Subsequently, in 2023 temporal multiplexing and cavity elongation enabled the generation of five-bit sequences in a femtosecond OPO, while configurations pumped at 1 GHz were indicated as a route toward rates approaching 1~Gbit/s \cite{Li:23}. In these sources, scalability remains constrained by intracavity-field reset requirements and optical-cavity complexity.

CV-QRNG enabled a pronounced increase in generation rates by digitizing analog measurements  into multibit words, shifting the limit from individual-event detection to receiver and ADC bandwidth \cite{ng2023240,bai202118}. In vacuum-fluctuation architectures, balanced homodyne detection converts quantum quadratures into electrical samples while suppressing common-mode classical noise. The earliest implementations in the survey, published in 2010, reported generation rates of 12~Mbit/s \cite{Shen2010} and 6.5~Mbit/s \cite{Gabriel2010}. A photonic integrated circuit achieved 18.8~Gbit/s in real time in 2021 \cite{bai202118}, whereas a wideband integrated homodyne implementation reported 240~Gbit/s in 2023 with offline Toeplitz extraction, the highest rate for this source among the reviewed studies \cite{ng2023240}. In 2026, 9-tap PFIR equalization in the ADC evaluation board and adaptive Toeplitz hashing in an FPGA enabled an aggregate real-time rate of 100.3~Gbit/s across two channels \cite{bomhals2026100}. These results show that the highest generation rates depend on the co-design of the optical source, receiver, digitization stage, and randomness extractor.

Entropy sources based on quantum phase fluctuations have also achieved some of the highest single-channel generation rates by converting laser phase diffusion into intensity variations through interferometry \cite{Xu2012}. The earliest implementation in this category in the survey was published in 2010 and demonstrated a generation rate of 500~Mbit/s \cite{Qi2010}. In 2015, an implementation reported a min-entropy-based rate estimate of 70.4~Gbit/s and a final rate of 68~Gbit/s after offline Toeplitz extraction, using a 15-GHz photodetector and sampling at 10~GSa/s \cite{Nie2015}. Published online in 2025, a dual-CW-laser implementation combined 4--24~GHz digital filtering with downsampling to 40~GSa/s, yielding 3.962 bits of conditional min-entropy per sample and 156~Gbit/s after offline Toeplitz extraction; the same work achieved 10~Gbit/s in real time \cite{Yang2026}. In an integrated SOI implementation presented in 2020, a tunable unbalanced Mach--Zehnder interferometer produced 12.5~Gbit/s after extraction with a \(5/8\) ratio \cite{imran2020quantum}. These results highlight the trade-off between absolute generation rate, integration, stability, and reconfigurability.

Amplified spontaneous emission has become particularly suitable for ultrahigh-rate generation because it combines broad optical bandwidth with macroscopic fluctuations produced by the amplification of spontaneous emission, favoring rapid sampling and spectral parallelization \cite{Guo2021}. An early implementation demonstrated a generation rate of 12.5~Gbit/s in 2010 \cite{Williams2010FastPhysical}. In 2021, 40~Gbit/s were demonstrated in a single channel through optical sampling with a terahertz optical asymmetric demultiplexer (TOAD), circumventing bandwidth and jitter limitations associated with conventional electronic conversion \cite{Guo2021}. Also in 2021, spectral slicing into 42 channels and the use of orthogonal polarizations enabled 400~Gbit/s per channel and an aggregate rate of 16.8~Tbit/s \cite{Cao2021}. However, this aggregate value is not directly equivalent to the throughput of a single processing chain, as it corresponds to the sum of the rates of multiple parallel channels. Toward integration, InGaN micro-LEDs demonstrated 12.5~Gbit/s in 2026 using offline self-subtraction and five-LSB retention, extending the range of compact spontaneous-emission sources, which differ physically from amplifier-based ASE \cite{hsiao2026microled}.

Table~\ref{tab:qrng_generation_rates} summarizes the highest random bit generation rates reported for each entropy source among the selected studies, indicating the publication year and operating regime. Taken together, the reported results show that the highest absolute generation rates have been achieved with CV-QRNG, particularly vacuum fluctuations, laser phase noise, and ASE, owing to the combination of high bandwidth, multibit digitization, and parallelization \cite{ng2023240,bomhals2026100,Yang2026,Cao2021}. Discrete-variable sources based on single-photon detection have also progressed into the Gbit/s range through SPAD arrays, temporal resolution, and multidimensional encoding, although they remain more directly constrained by detector efficiency, dead time, and afterpulsing \cite{wang2025efficient,li2024efficient}. Regardless of the source, the final secure rate depends on the acquisition rate, the extractable min-entropy per sample or event, the security margin, and the throughput of the extractor \cite{RevModPhys.89.015004,PhysRevA.100.052318,bruynsteen2023100}. Therefore, raw, aggregate, or offline rates should not be equated with the final real-time single-channel throughput. Sustaining high generation rates in real time benefits from dedicated hardware capable of performing digital conditioning and hashing concurrently with data acquisition, preventing post-processing from becoming the QRNG throughput bottleneck \cite{bomhals2026100}.

\begin{table*}[!t]
\caption{Highest reported random bit generation rates among the selected records of the survey.}
\label{tab:qrng_generation_rates}
\begin{ruledtabular}
\begin{tabular}{llll}
Entropy source &
Generation rate &
Year &
Operating regime \\
\hline
Photon path \cite{du2025source} &
9.49~Mbit/s &
2025 &
Not specified \\

Photon arrival time \cite{wang2025efficient} &
2.1~Gbit/s &
2025 &
Not specified \\

Photon counting \cite{10129008} &
3.3~Gbit/s (aggregate) &
2023 &
Real-time \\

Attenuated pulses \cite{Rusca2019} &
1.25~Mbit/s &
2019 &
Not specified \\

Raman scattering \cite{10.1063/1.4864095} &
145~Mbit/s &
2014 &
Not specified \\

Optical parametric oscillators \cite{Zhang_2017} &
60~Mbit/s (per sequence) &
2017 &
Not specified \\

Vacuum fluctuations \cite{ng2023240} &
240~Gbit/s &
2023 &
Offline \\

Laser phase noise \cite{Yang2026} &
156~Gbit/s &
2025 &
Offline \\

Amplified spontaneous emission \cite{Cao2021} &
400~Gbit/s (per channel) &
2021 &
Offline \\
\end{tabular}
\end{ruledtabular}
\end{table*}

\subsection{Integrated photonic implementations} 
The integration of QRNG into photonic integrated circuits (PICs) has become one of the main technological directions for translating laboratory demonstrations into practical devices. Compared with bulk-optical and fiber-based implementations, integrated photonic platforms substantially reduce the system footprint while improving mechanical stability and reproducibility by eliminating free-space optical alignment. Furthermore, semiconductor fabrication technologies enable wafer-scale manufacturing, facilitating the deployment of QRNGs in applications requiring compact and low-power hardware, including quantum communication systems, embedded cryptographic modules, and Internet-of-Things (IoT) devices \cite{imran2020quantum,imran2021chip}.

Several photonic platforms have been investigated for integrated QRNGs, each providing different trade-offs between optical performance, fabrication complexity, and integration density. Silicon-on-insulator (SOI) has become one of the most widely adopted platforms owing to its compatibility with mature CMOS fabrication processes and the availability of low-loss passive photonic components \cite{raffaelli2018generation,bai202118}. Since silicon is an indirect-bandgap semiconductor, efficient light sources generally require external lasers or heterogeneous integration with III--V materials \cite{imran2021chip,wang2024compact}. In contrast, indium phosphide (InP) enables the monolithic integration of optical sources, modulators, passive waveguides and photodetectors operating in the telecommunications band \cite{roger2019real,chrysostomidis2023long}. Other technologies have also been explored for specific applications, including CMOS-SPAD processes for DV-QRNG and plasmonic waveguides, which provide subwavelength optical confinement at the expense of increased propagation losses \cite{Khanmohammadi2015,acerbi2018robust,francis2017quantum,strydom2024quantum}.
%





Although the physical entropy source varies across implementations, integrated QRNGs generally comprise three functional blocks: an optical source, a photonic circuit responsible for manipulating the optical field, and a photodetection stage. Continuous-variable implementations typically employ Mach--Zehnder interferometers or balanced homodyne receivers to convert vacuum fluctuations or laser phase noise into measurable intensity fluctuations \cite{raffaelli2018generation,roger2019real,bai202118,li2024chip}. Discrete-variable implementations, on the other hand, commonly rely on integrated SPAD arrays combined with on-chip timing electronics for measuring photon arrival times or photon-counting statistics \cite{Khanmohammadi2015,acerbi2018robust,stanco2020efficient}. Since the operating principles of these entropy sources have already been discussed in Section \ref{sec:entropy_sources}, photonic integration primarily seeks to increase the level of hardware integration while reducing optical losses and environmental perturbations.

Different integration strategies have consequently emerged. Fully monolithic implementations integrate both active and passive components within a single photonic platform, an approach commonly adopted in InP technologies and CMOS-SPAD devices \cite{roger2019real,chrysostomidis2023long,acerbi2018robust}. Hybrid integration combines passive silicon photonic circuits with III--V lasers or high-performance photodetectors through heterogeneous integration or flip-chip bonding, taking advantage of the mature silicon photonics ecosystem while preserving the optoelectronic performance of compound semiconductors \cite{bai202118,wang2024compact,huang2025hybrid}. Simpler architectures also remain relevant, particularly for proof-of-concept demonstrations, where only the passive photonic circuit is integrated while lasers and detection electronics remain external to the chip \cite{raffaelli2018generation,imran2021chip}.

Recent developments indicate an increasing adoption of CV-integrated architectures capable of operating at generation rates exceeding 100~Gbps by combining broadband photonic circuits with high-speed balanced photodetectors and ADC \cite{bruynsteen2023100,li2024chip}. In parallel, source-independent and semi-device-independent implementations have recently been demonstrated using programmable photonic circuits incorporating high-speed phase modulators for real-time entropy estimation under reduced trust assumptions \cite{kincaid2023source,li2024chip,bertapelle2025high,du2025source}. Meanwhile, CMOS-SPAD arrays continue to improve the performance of DV-QRNG through spatial and temporal multiplexing, increasing the attainable generation rate while preserving single-photon sensitivity \cite{stanco2020efficient,li2024efficient}.

Despite the substantial progress achieved during the last decade, several challenges remain before fully integrated QRNGs become widely deployable. Optical coupling and propagation losses directly reduce the available quantum entropy, particularly in vacuum-state and source-independent implementations \cite{bertapelle2025high,leone2023generation}. Monolithic integration also introduces electrical crosstalk between high-frequency driving electronics and low-level photodetection signals, requiring careful electromagnetic isolation during circuit design \cite{chrysostomidis2023long}. In addition, thermal fluctuations modify the optical phase accumulated in interferometric structures, requiring active stabilization to preserve the operating point of balanced receivers \cite{imran2021chip,chrysostomidis2023long}. Finally, although integrated photonic hardware is now capable of generating raw random data at hundreds of gigabits per second, real-time randomness extraction remains a significant computational bottleneck, making the co-design of photonic circuits and digital post-processing hardware an important direction for future integrated QRNG systems \cite{zhang2016note,bai202118,li2024chip}.
The principal characteristics of these photonic integration platforms are summarized in Table~\ref{tab:PIC_platforms}.

\begin{table*}[t]
\caption{Comparison of the principal photonic integration platforms employed for QRNGs.}
\label{tab:PIC_platforms}
\begin{ruledtabular}
\begin{tabular}{llll}
Platform &
Entropy source &
Integration strategy &
Main limitation \\
\hline
SOI &
Vacuum fluctuations, phase noise &
Hybrid or external source &
Requires external or heterogeneous light source \\

InP &
Vacuum fluctuations, phase noise &
Monolithic &
Complex fabrication \\

CMOS-SPAD &
Photon arrival time &
Monolithic &
Detector dead time \\

Plasmonics &
Path superposition &
Passive photonic integration &
Propagation loss \\
\end{tabular}
\end{ruledtabular}
\end{table*}

\subsection{Dedicated hardware implementations} 
The development of hardware for post-processing in QRNGs focuses on overcoming real-time processing limitations \cite{10484939, zhang2016note}. Because randomness extraction is mathematically analogous to privacy amplification in QKD protocols, post-processing strategies frequently serve a dual purpose: compressing raw entropy and ensuring information-theoretic security. Currently, hardware implementations are divided between the use of field-programmable gate arrays (FPGAs) to manage high transfer rates and the integration into silicon components to obtain compact devices \cite{park2020qec, 10484939, zhang2016note}.

For systems requiring high-speed processing, the use of FPGAs enables continuous and real-time data extraction \cite{10484939, zhang2016note}. A system based on vacuum fluctuations, integrated with photonic-electronic detectors, demonstrated a generation rate of 100.3 Gbps using digital processing on an FPGA \cite{bomhals2026100}. Another implementation reported a rate of 3.2 Gbps employing a Virtex-6 family chip to execute the randomness extraction \cite{zhang2016note}. In configurations where sampling rates are limited by measuring the arrival time of photons in SPADs, FPGAs such as the Spartan-6 or Zynq-7010 are frequently employed to digitize the time intervals \cite{park2020qec, Khanmohammadi2015}. In these architectures, the components typically operate at clock frequencies between 50 MHz and 100 MHz \cite{park2020qec, stanco2020efficient}.

With the aim of enabling miniaturization, monolithic integrated solutions in CMOS technology have been developed to accommodate the physical generation and initial processing on the same substrate \cite{park2020qec, Khanmohammadi2015}. The use of 0.35 $\mu$m CMOS technology allowed the fabrication of a QRNG where a silicon-based photon emitter and a SPAD detector coexist on the same die \cite{Khanmohammadi2015}. Additionally, entropy sources based on radioactive decay were produced in 180 nm CMOS technology, resulting in silicon active areas of 1.4 mm $\times$ 1.4 mm \cite{park2020qec}. Due to their small size, these components can act as Intellectual Property (IP) blocks embedded directly into microcontroller units \cite{park2020qec}.

In the context of Application-Specific Integrated Circuits (ASICs) and System-on-Chip (SoC) designs, post-processing architectures are specifically tailored to balance silicon area, power consumption, and throughput. To execute the aforementioned dual-purpose strategies, universal hashing functions, particularly those based on Toeplitz matrix multiplication, represent a standard approach \cite{10484939, kincaid2023source}. To accommodate the stringent resource constraints of ASICs while sustaining multi-gigabit-per-second throughputs, these Toeplitz extractors are typically mapped into highly parallelized XOR-AND trees or optimized linear feedback shift registers (LFSRs). Furthermore, embedding lightweight cryptographic primitives, such as AES cores or Hash-DRBGs, directly into the ASIC silicon provides a compact mechanism to simultaneously compress raw data and mask physical vulnerabilities against side-channel attacks \cite{park2020qec}.

The evolution of post-processing hardware for QRNGs converges into complementary development trajectories dictated by specific application requirements. For high-speed systems based on laser phase or vacuum fluctuations, real-time architectures operate in the multi-gigabit-per-second regime by employing FPGAs integrated with FIR equalizers to mitigate temporal correlations, subsequently feeding parallelized Toeplitz matrix extraction blocks \cite{10484939, zhang2016note}. Conversely, for detector arrays coupling SPADs with time-to-digital converters (TDCs), the hardware executes extraction algorithms, such as those by Peres and Zhou-Bruck, which are tailored to compensate for TDC digitization non-linearities, detector dead time, afterpulsing effects, and cross-correlation between pixels \cite{stanco2020efficient}. Finally, in single-chip System-on-Chip (SoC) architectures targeted at low-power devices, the processing is streamlined through the direct selection of the least significant bits (such as LSB 2 and 3) from the analog-to-digital converter. This embedded technique mitigates classical noise directly at the physical circuit level, ensuring a quantum min-entropy exceeding 0.98 per bit without the need for external randomness extraction modules \cite{gras2021quantum}.

\subsection{Cryptographic applications}

The use of random numbers finds applications in various fields, most notably in cryptography \cite{goldreich1999modern, katz2007introduction}. In classical digital cryptography, randomness is an indispensable primitive that underpins Kerckhoffs' principle \cite{kerckhoffs1883cryptographie}, ensuring that cryptosystem security relies solely on secret parameters rather than algorithmic secrecy \cite{marton2010randomness}. In this scenario, QRNG can be applied to cryptographic key generation for both symmetric and asymmetric ciphers, where secret keys must possess high entropy, independence of values, uniform distribution, and unpredictability to withstand cryptanalytic attacks \cite{Saini2022, BLANCOROMERO2024110877, Sharma2025, Mohamed2024}.  Beyond cryptographic key generation, QRNGs can also provide entropy for CSPRNGs, ephemeral encapsulation secrets, signature nonces, and side-channel countermeasures based on randomized masking techniques \cite{RevModPhys.89.015004}.

In Quantum Key Distribution, QRNGs play a fundamental role across both the quantum transmission stage and the subsequent classical post-processing procedures, ensuring that security assumptions are rooted in genuine unpredictability rather than computational complexity \cite{Renner2008}. During the quantum communication phase, random numbers are required at both the sender (Alice) and receiver (Bob) nodes to determine state preparations, basis selections, and measurement settings, thereby preventing adversaries from exploiting deterministic patterns in the protocol execution \cite{pirandola2020advances,zhang2023continuous, laudenbach2018continuous, RevModPhys.74.145}. As modern QKD systems operate at clock rates reaching several gigahertz, high-throughput QRNGs, particularly those based on CV-measurements, are often integrated directly into the hardware platform to provide entropy at a rate compatible with real-time operation \cite{bomhals2026100}. Randomness remains equally important during the classical post-processing stage, where secret random seeds are employed in tasks such as privacy amplification and parameter estimation, enabling the extraction of provably secure keys even in the presence of partial information leakage to an eavesdropper \cite{Yang2020, Kiktenko_2016}. Consequently, the security and performance of practical QKD systems are intimately linked to the quality, throughput, and trustworthiness of the underlying random number generation process.


In Post-Quantum Cryptography (PQC), QRNGs provide a source of physical entropy for security-critical operations within quantum-resistant cryptographic schemes \cite{chen2016report, Bernstein2017}. As modern standards increasingly adopt lattice-based constructions, such as CRYSTALS-Kyber \cite{Bos2018} and CRYSTALS-Dilithium \cite{ducas2018crystals}, as well as code-based approaches including Classic McEliece \cite{Overbeck2009}, high-quality randomness becomes essential for key generation, parameter sampling, noise-vector generation, and other probabilistic procedures underlying these primitives \cite{Bavdekar2023}. In lattice-based cryptography, in particular, the security of schemes relying on Module Learning With Errors (MLWE) assumptions depends on the correct generation of statistically sound random values, as deficiencies in entropy sources may weaken practical security and facilitate cryptanalytic attacks \cite{marton2010randomness}. Consequently, robust random number generation remains a fundamental requirement for the secure implementation and deployment of post-quantum cryptographic systems. 

\begin{figure*}
    \centering
    \includegraphics[width=0.9\linewidth]{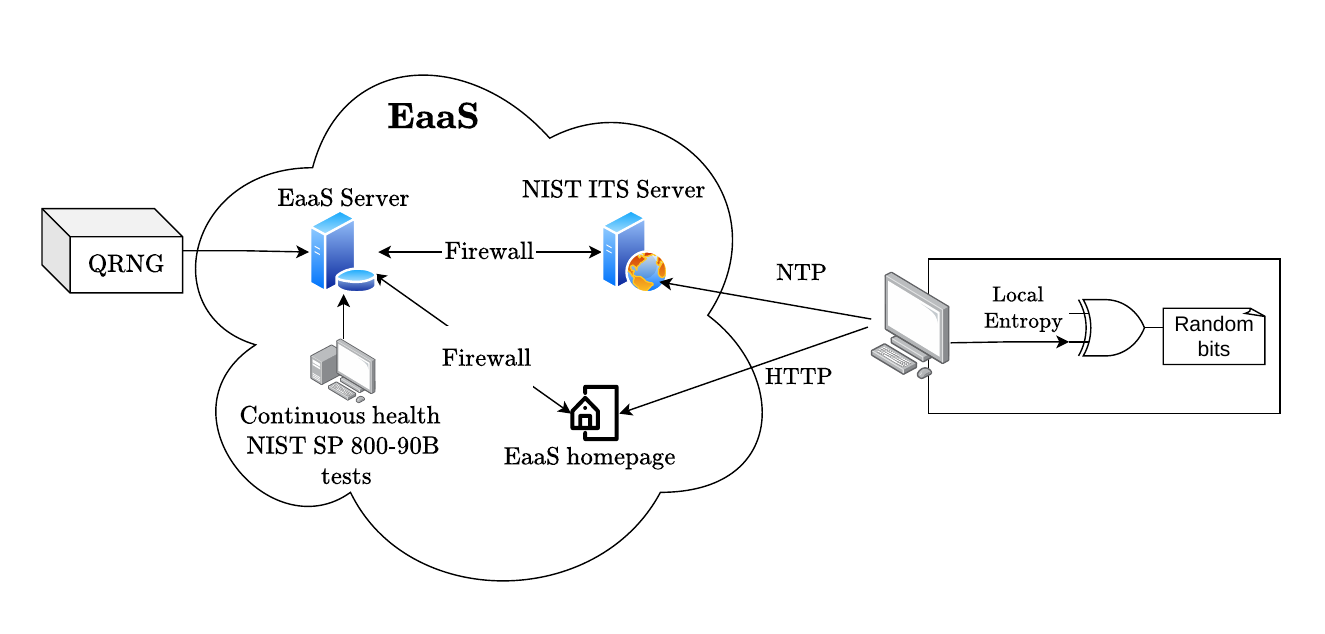}
    \caption{\justifying Simplified EaaS architecture. True random numbers are produced by a QRNG and sent to an EaaS server. The EaaS server is subjected to continuous health tests in accordance with NIST SP 800-90B and is connected to an EaaS homepage, which a user accesses via HTTP. As an intermediate step, both the EaaS server and the user must connect to a NIST ITS server to synchronize their clocks, using the NTP protocol. On the user side, the true random numbers can be combined with a local entropy generator. To compromise this established provisioning model and take control of a deployed device, an attacker would have to penetrate factory security, record every device key issued, and monitor every interaction between the device and EaaS. Missing even one such interaction would thwart the attack. This sets a high security bar for attackers.}
    \label{fig:EaaS}
\end{figure*}

In modern cloud computing and Internet of Things (IoT) infrastructures, Entropy as a Service (EaaS) has emerged as an approach to mitigate entropy shortages in virtualized, embedded, and resource-constrained environments \cite{Apostol2016}. Virtual machines instantiated from identical images, containers deployed at scale, and headless devices lacking traditional sources of environmental randomness often struggle to accumulate sufficient entropy during system initialization, potentially weakening the security of cryptographic operations \cite{Heninger2012, Vassilev2014}. To address this limitation, EaaS platforms provide remote access to high-quality entropy generated by centralized physical random number generators, increasingly including quantum random number generators (QRNGs), through secure network interfaces, as illustrated in Fig. \ref{fig:EaaS}.

In typical deployments, entropy servers distribute random data to client systems using authenticated and encrypted communication channels, enabling secure integration with local deterministic random bit generators and operating-system entropy pools \cite{nist80090b}. The externally supplied entropy can then be combined with locally collected randomness through cryptographic conditioning mechanisms, enhancing the robustness and availability of random number generation across cloud services, edge-computing platforms, and large-scale IoT ecosystems \cite{Apostol2016, Romero2026, Ullah2020, Blanco2025}. As a result, EaaS extends the benefits of hardware-based and quantum-generated randomness to systems where dedicated entropy generation may be impractical due to cost, power, or deployment constraints.

\section{Summary and outline} \label{sec:summary}

The development of quantum random number generators illustrates how the intrinsic unpredictability of quantum measurements can be transformed into a practical resource for information processing and cryptographic applications. The literature examined in this survey reveals that the performance and security of a QRNG cannot be attributed solely to its physical entropy source. Rather, they emerge from the combined operation of quantum-state preparation, measurement, entropy estimation, randomness extraction, and hardware implementation. By examining these components within a common framework, this survey identifies the relationships between physical entropy generation, security assumptions, and achievable output rates, providing a basis for interpreting the technological advances and limitations reported across different QRNG architectures.

The surveyed implementations reveal distinct trade-offs between discrete-variable and continuous-variable architectures. DV approaches exploit individual quantum events, including photon-path selection, arrival times, and photon-counting statistics, providing a direct correspondence between measurement outcomes and discrete symbols. Their performance is primarily constrained by detection efficiency, dead time, afterpulsing, and acquisition electronics, although detector arrays and multidimensional encoding have enabled substantial improvements in generation rates. Conversely, CV architectures based on vacuum fluctuations, laser phase noise, and amplified spontaneous emission exploit broadband optical signals and multibit digitization to achieve higher reported throughputs. Their performance depends critically on quantum-to-classical noise separation, receiver bandwidth, sampling characteristics, and digital processing capacity. Photonic integration and parallel acquisition have expanded the implementation possibilities of both approaches. Nevertheless, differences in experimental conditions, extraction procedures, and operating regimes prevent generation rate alone from establishing a general hierarchy among QRNG architectures.

From a security perspective, the quantum origin of the measured fluctuations does not independently guarantee the unpredictability of the final bit sequence. Practical security requires conservative entropy bounds that account for classical noise, device imperfections, and potentially accessible side information. Conditional min-entropy, including its smooth formulation for finite-size security analyses, provides the operational connection between the raw measurements and the randomness available for extraction. The surveyed certification frameworks further demonstrate that security guarantees depend on explicitly defined trust assumptions, ranging from characterized devices to semi-device-independent and fully device-independent protocols. Randomness extractors can subsequently produce nearly uniform outputs with quantified security errors, provided their entropy and seed assumptions are satisfied. Randomness expansion and amplification extend these principles to distinct information-processing tasks, although their experimental requirements and achievable rates remain dependent on the adopted security model. Statistical testing provides complementary evidence of output quality but cannot replace entropy estimation or cryptographic randomness certification.

Despite advances in generation speed and device integration, several challenges remain. Reliable operation requires accounting for finite-size effects, temporal correlations, classical noise, component drift, and imperfections that may invalidate the assumed entropy model. At the implementation level, photonic integration introduces constraints associated with optical losses, thermal stability, electrical crosstalk, and the integration of high-speed acquisition with real-time extraction. These limitations motivate further development of experimentally verifiable entropy models, continuous device monitoring, and coordinated photonic-electronic processing architectures. Greater emphasis on finite-size security analysis and reduced-trust certification is also warranted, particularly where practical throughput must be reconciled with stronger security assumptions. Moreover, meaningful comparisons across implementations require consistent reporting of raw and extracted generation rates, entropy bounds, security parameters, and operating conditions, explicitly distinguishing real-time operation from offline processing and single-channel throughput from aggregated rates.

Taken together, the surveyed literature indicates that progress in QRNG technology depends on balancing the physical generation of quantum entropy with its secure characterization and efficient conversion into usable random bits. For applications involving quantum key distribution, post-quantum cryptography, and distributed entropy provisioning, the relevant performance criterion is therefore not merely the rate of quantum measurements, but the rate at which randomness can be delivered under clearly specified security and implementation assumptions. This relationship provides a common basis for evaluating existing architectures and guiding further developments toward integrated, scalable, and cryptographically dependable QRNG systems.

\begin{acknowledgments}
This work has been partially funded by the project QRNG: Post-processing and Experimental supported by QuIIN - Quantum Industrial Innovation, EMBRAPII CIMATEC Competence Center in Quantum Technologies, with financial resources from the PPI IoT/Manufatura 4.0 of the MCTI grant number 053/2023, signed with EMBRAPII. VLS and JBSJ thanks Coordenação de Aperfeiçoamento de Pessoal de Nível Superior - Brasil (CAPES) - Finance Code 001. VLS tanks CNPQ support through grant no. 308518/2026-8.
\end{acknowledgments}

\bibliography{sample}

\end{document}